\documentclass[journal=jacsat,manuscript=article]{achemso}

\usepackage{chemformula} % Formula subscripts using \ch{}
\usepackage[T1]{fontenc} % Use modern font encodings
\author{Conor A Nixon}
\email{conor.a.nixon@nasa.gov}
\phone{+1 301 286-6757}
\affiliation[NASA GSFC]
{Planetary Systems Laboratory, NASA Goddard Space Flight Center \\ 8800 Greenbelt Road, Greenbelt, MD 20771}

\title[An \textsf{achemso} demo]
  {The Composition and Chemistry \\ of Titan's Atmosphere
  \footnote{Accepted for publication in ACS Earth and Space Chemistry}}

\abbreviations{IR,NMR,UV}
\keywords{American Chemical Society, \LaTeX}

\newcommand{\methane}{CH$_4$}
\newcommand{\ethane}{C$_2$H$_6$}
\newcommand{\acet}{C$_2$H$_2$}
\newcommand{\acetylene}{C$_2$H$_2$}
\newcommand{\benzene}{c-C$_6$H$_6$}
\newcommand{\diacet}{C$_4$H$_2$}
\newcommand{\hydrogen}{H$_2$}

\newcommand{\propyne}{CH$_3$CCH}
\newcommand{\propadiene}{CH$_2$CCH$_2$}

\newcommand{\cpld}{c-C$_3$H$_2$}
\newcommand{\ethylene}{C$_2$H$_4$}
\newcommand{\propane}{C$_3$H$_8$}
\newcommand{\propargyl}{C$_3$H$_3$}
\newcommand{\propene}{C$_3$H$_6$}
\newcommand{\cyclohexane}{c-C$_6$H$_{12}$}

\newcommand{\butane}{C$_4$H$_{10}$}

\newcommand{\phosphine}{PH$_3$}
\newcommand{\hsulfide}{H$_2$S}

\newcommand{\formaldehyde}{H$_2$CO}
\newcommand{\methanol}{CH$_3$OH}
\newcommand{\water}{H$_2$O}
\newcommand{\coo}{CO$_2$}

\newcommand{\nitrogen}{N$_2$}
\newcommand{\ammonia}{NH$_3$}
\newcommand{\methanimine}{CH$_2$NH}

\newcommand{\pyrimidine}{c-C$_4$H$_4$N$_2$}
\newcommand{\pyridine}{c-C$_5$H$_5$N}
\newcommand{\adenine}{C$_5$H$_5$N$_5$}

\newcommand{\cyanogen}{C$_2$N$_2$}
\newcommand{\dicyanoacet}{C$_4$N$_2$}
\newcommand{\cyanoacet}{HC$_3$N}
\newcommand{\methylcyn}{CH$_3$CN}
\newcommand{\acetonitrile}{CH$_3$CN}
\newcommand{\cyanopropyne}{CH$_3$C$_3$N}
\newcommand{\butynenitrile}{CH$_3$C$_3$N}
\newcommand{\vinylcyn}{C$_2$H$_3$CN}
\newcommand{\acrylonitrile}{C$_2$H$_3$CN}
\newcommand{\ethylcyn}{C$_2$H$_5$CN}
\newcommand{\propionitrile}{C$_2$H$_5$CN}

\newcommand{\eg}{$e.g.$}

\newcommand{\cm}{cm$^{-1}$}
\newcommand{\micron}{$\mu$m}
\newcommand{\dg}{$^\circ$}

\begin{document}


\begin{abstract}
In this article I summarize the current state of knowledge about the composition of Titan’s atmosphere, and our current understanding of the suggested chemistry that leads to that observed composition. I begin with our present knowledge of the atmospheric composition, garnered from a variety of measurements including {\em Cassini-Huygens}, the {\em Atacama Large Millimeter/submillimeter Array (ALMA)}, and other ground and space-based telescopes. This review focuses on the typical vertical profiles of gases at low latitudes, rather than global and temporal variations. The main body of the paper presents a chemical description of how complex molecules are believed to arise from simpler species, considering all known ‘stable’ molecules – those that have been uniquely identified in the neutral atmosphere. The last section of the paper is devoted to the gaps in our present knowledge of Titan's chemical composition and how further work may fill those gaps. 
\end{abstract}

\section{Introduction}

Titan, Saturn’s largest moon, was first observed in 1655 by the Dutch astronomer Christiaan Huygens. This important discovery that Saturn, in addition to Jupiter, had its own satellite helped to consolidate the Copernican worldview: that the Earth was no longer to be considered the center of the Solar System, but rather one of several planets orbiting the Sun, and possessing a natural satellite of its own. In his excitement at this new finding, Huygens would have little suspected what would transpire 350 years later: that a machine devised and launched into the heavens by humanity and bearing his name would traverse an unimaginable void, and then land softly on his new world, finding it stranger and more alien than even the machine’s designers had anticipated\cite{lebreton05}.

\begin{figure}
\includegraphics[scale=0.72]{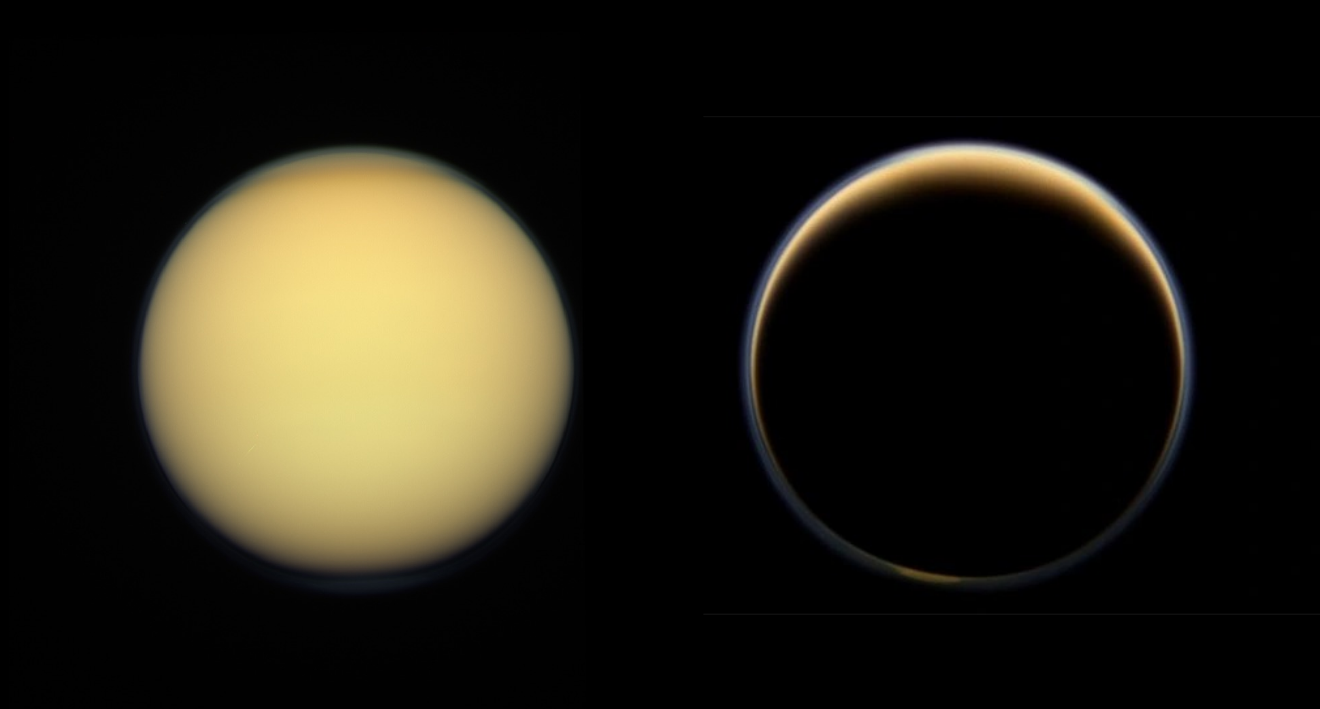}
\caption{The atmosphere of Titan seen by {\em Cassini}'s Imaging Science Subsystem (ISS). 
The dayside image (December 16th 2011) shows Titan's ubiquitous golden haze, opaque to visible light. A darker haze collar is seen around the north pole, along with fainter haze hoods over both the north and south poles, thought to be created by the interaction between chemical and dynamical processes.
The nightside image (PIA14924, June 6th 2012, range: 216,000 km) clearly shows the detached atmospheric haze surrounding the entire limb and an elevated stratospheric condensate cloud over the south pole, thought to be composed at least partly of HCN.
Image credit: NASA/JPL-Caltech/Space Science Institute/CICLOPS, with reprocessing by Kevin M. Gill.}
\label{fig:daynight}
\end{figure}

Over the 13 years from 2004 to 2017, the {\em Cassini-Huygens} mission\cite{lebreton02, matson02} was able to significantly reveal Titan, both to our eyes and to our minds.  Between the successful landing of the ESA-built {\em Huygens} probe carrying six scientific suites in January 2005, and the 127 flybys of the NASA-built {\em Cassini} spacecraft with its own twelve science instruments, our knowledge of Titan now is vastly greater than before the mission arrived. So, in our privileged position of hindsight, what do we now know about Titan, almost four centuries since its discovery? 

\begin{figure}
\includegraphics[scale=0.15]{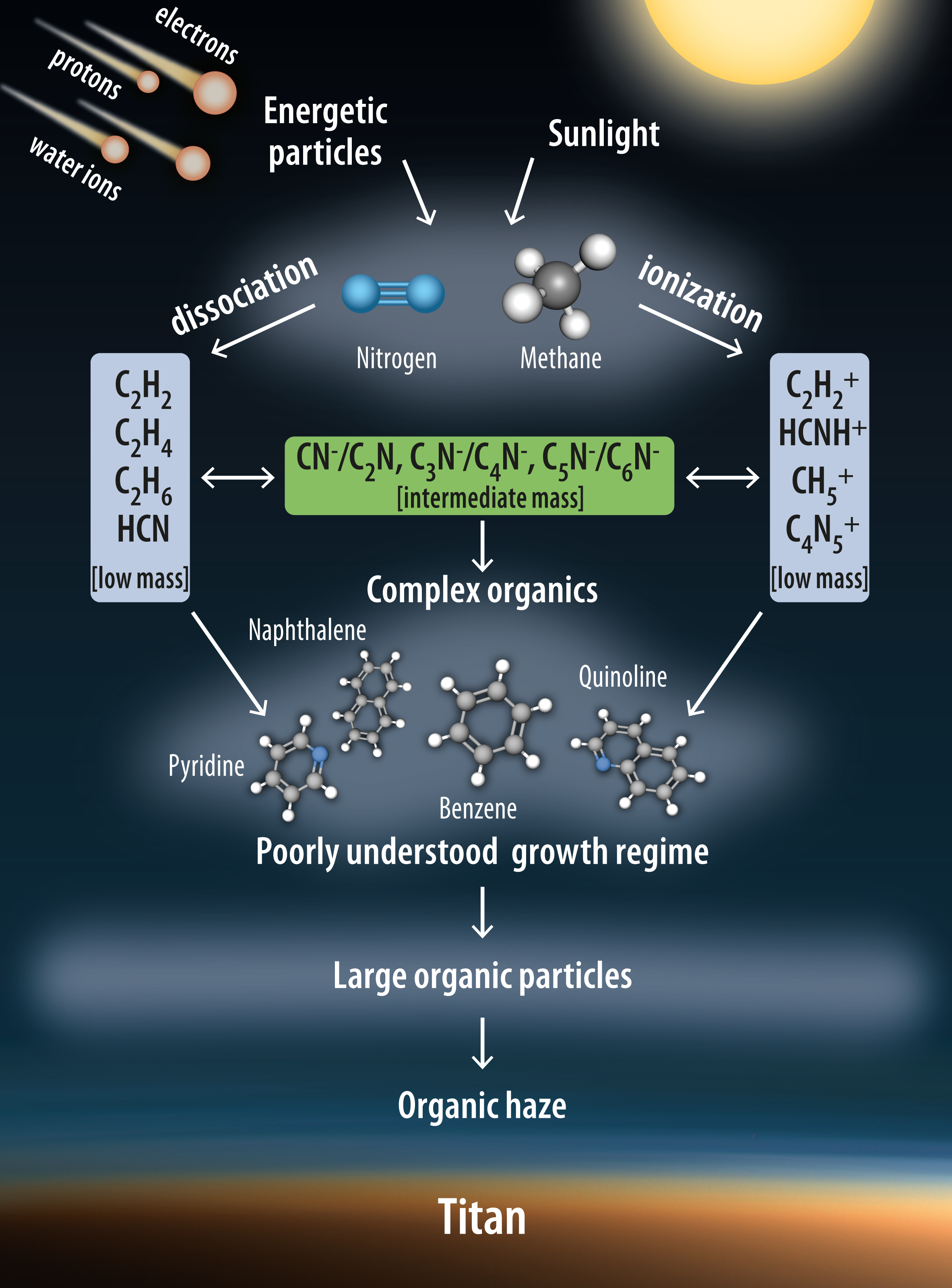}
\caption{Simplified portrayal of Titan's atmospheric photochemistry, showing the presumed progression from simpler molecules (CH$_4$, N$_2$) to more complex molecules, and eventually haze particles that can sediment out from the atmosphere. Image adapted from ESA graphic. }
\label{fig:chemistry}
\end{figure}

First and foremost, it is a moon with a dense atmosphere (Fig.~\ref{fig:daynight}), the only such body known in our Solar System. Also that this atmosphere, composed primarily of molecular nitrogen and methane, is a largely anoxic environment, with little oxygen to cause the termination of complex organic reactions. The result is a chemical wonderland, with a breathtaking array of complex organic molecules, of which we presently have only the most rudimentary understanding. Fig.~\ref{fig:chemistry} shows a schematic overview of the presumed chemistry that occurs in Titan's upper atmosphere, where the `raw ingredients' of its photochemical reactions, \nitrogen\ and \methane , are broken apart and recombined into successively larger molecules, and finally haze particles.

Titan is a world that is today both tantalizingly more known, and more unknown than ever before. Our direct investigations of its atmosphere by the {\em Huygens} lander and {\em Cassini} spacecraft have both increased our understanding of Titan enormously and also multiplied our questions. Significant outstanding questions include:

\begin{itemize}
\item
Why is Titan the only moon in the Solar System with a significant atmosphere, and how did it come to be in its present state?
\item
Is the atmosphere today in a steady state, or is it growing or shrinking or changing in some way?
\item
To what degree has the atmosphere interacted with and shaped the surface and subsurface?
\item
What degree of chemical complexity is reached in Titan's atmosphere, and are precursor biomolecules among the products?
\item
Are there other moons with atmospheres similar to Titan elsewhere in the galaxy?
\end{itemize}

Achieving a better understanding of Titan's atmosphere and its chemistry is important both for the sake of Titan science itself, and because of its potential to inform us about other environments. This includes the present-day Earth, since Titan and Earth are the only objects in the Solar System today to have a hydrological cycle of evaporation, condensation and precipitation, and associated rivers, lakes and seas \cite{hueso06, lunine08a, witek15, hayes16}. Like the Earth, Titan also experiences seasons, due to orbiting close to Saturn's equatorial plane, which is tilted $\sim$27\dg\ to the ecliptic. Titan therefore experiences summer and winter in each hemisphere, seasons that last $\sim$7.4$\times$ longer than on Earth, with transitional equinox periods of equal daylight at all latitudes.

We also note the relevance to the early Earth, which likely had a much more chemically reducing atmosphere in its distant past \cite{kasting93, mckay99, tian05, trainer06, he14a, zahnle20a}, before the Great Oxidation Event \cite{sessions09, gumsley17, zahnle20b}. Finally we can surmise the likely relevance to exoplanets, which vastly outnumber the planets in our own Solar System, and more likely than not include Titan-like bodies somewhere in our galaxy \cite{lunine10,bourgalais21, woitke21}.

In this review paper, I attempt to lay out a simple picture of the known characteristics of Titan’s atmosphere, with a focus on the composition and chemistry of the dense lower atmosphere. By necessity, this review will not cover, except in passing, many related areas: the origin of the atmosphere and possible replenishment mechanisms by internal or external sources; isotopic composition and time evolution of isotopic ratios; winds and dynamics; condensates and meteorology; and the chemical composition of large particulates (haze particles). All of these topics have been covered extensively in review articles and chapters elsewhere\cite{hayes16,horst17,nixon18}, and in two books written about the results of the {\em Cassini-Huygens} mission \cite{brown10book, mueller-wodarg14}.

In presenting a simple overview and summary focusing on the chemistry and composition of the neutral atmosphere, it is hoped that I will do sufficient justice to this one area to make this article a useful primer for undergraduate or graduate students, or others new to the field, to quickly gain a basic understanding of Titan’s bulk atmospheric composition, and why it is that way - at least at the present era. 

The paper is organized as follows: I first review basic knowledge about Titan's atmospheric temperature structure and gas composition. The main section of the paper contains an exposition on the chemistry of the 24 known molecules in the neutral, lower atmosphere. This is followed by a detailed discussion of future research directions in Titan atmospheric composition studies, followed by a Summary and Conclusions.

%%%%%%%%%%%%%%%%%%%%%%%%%%%%%%%%%%%%%%%%%%%%%%%%%%%%%%%%%%%%%%%%%%%%%
\section{Atmospheric Composition and Structure}
%%%%%%%%%%%%%%%%%%%%%%%%%%%%%%%%%%%%%%%%%%%%%%%%%%%%%%%%%%%%%%%%%%%%%

\subsection{Atmospheric Composition}

Titan's atmosphere is largely composed of two gases: N$_2$ and CH$_4$. The vertical profile of methane comes from measurements by instruments on the {\it Cassini-Huygens} space mission, primarily the {\it Huygens} GCMS (Gas Chromatograph and Mass Spectrometer) \cite{niemann02} from 0--146 km, the {\it Cassini} Visual and Infrared Mapping Spectrometer (VIMS)\cite{brown04} (50--850~km), the {\it Cassini} UVIS (Ultraviolet Imaging Spectrometer) \cite{esposito04} (400--1650~km), and the {\it Cassini} INMS (Ion and Neutral Mass Spectrometer) \cite{waite04} (900--1500~km). Their results have been reported in publications from the mission \cite{waite05, shemansky05, bellucci09, niemann10}.

\begin{figure}
\includegraphics[scale=0.80]{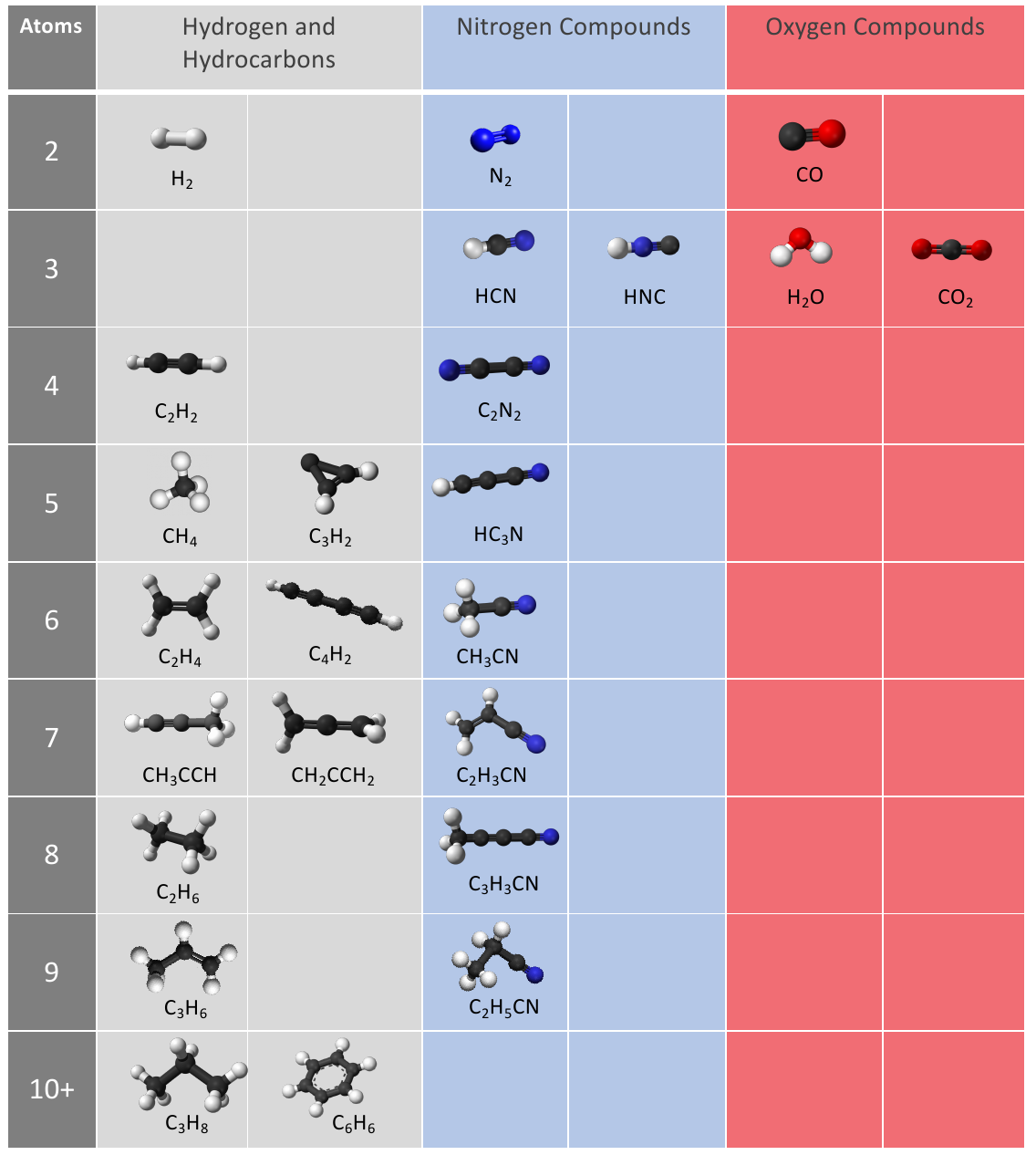}
  \caption{Molecules detected in Titan's neutral atmosphere sorted by number of atoms and composition. Hydrocarbons are the most abundant and complex species type, followed by nitriles. No complex oxygen-bearing molecules, including organics, have been detected on Titan to date.}
  \label{fig:table-fig}
\end{figure}

Aside from noble gases ($^{36}$Ar, $^{40}$Ar and $^{22}$Ne) \cite{niemann05,  niemann10}, 22 molecular species other than \nitrogen\ and \methane\ have been definitively detected in Titan's atmosphere at the time of writing (see Fig.~\ref{fig:table-fig}): ten hydrocarbons (\acetylene, \ethylene, \ethane, \cpld, \propadiene, \propyne, \propene , \propane, \diacet, \benzene ), eight cyanides\footnote{Also referred to as `nitriles' when -CN occurs in an organic molecule.} (HCN, HNC, \cyanoacet , \cyanogen , \methylcyn, \vinylcyn , \ethylcyn , \cyanopropyne ), three oxygen-bearing species (CO, \coo , \water ) plus \hydrogen . These gases were originally detected by a variety of astronomical and remote sensing techniques from the ground and space.

%, depicted in Fig.~\ref{fig:discoveries}.

%\begin{figure}
%\includegraphics[scale=0.86]{telescope-discoveries.pdf}
%  \caption{Observatories responsible for first detections of gases in Titan's neutral atmsosphere.\cite{kuiper44, gillett75, trafton72, gillett73, lutz83, broadfoot81, hanel81, maguire81, kunde81, samuelson81, lombardo19b, bezard92, coustenis98, coustenis03, nixon13a, moreno11, cordiner15, palmer17, thelen20, nixon20}
% The first flyby of Titan by {\em Voyager 1} in 1980 contributed the most new discoveries regarding Titan's atmospheric chemical composition to date, although more recently the ALMA (Atacama Large Millimeter/submillimeter Array) observatory has increased the pace of new discoveries once more.}
%\label{fig:discoveries}
%\end{figure}

All of the major types of hydrocarbons have been detected (alkanes, alkenes, alkynes, a carbene, an aromatic ring). However, major chemical families of nitrogen-bearing molecules (including amines, imines, azines, and N-heterocyclic rings) and oxygen-bearing molecules (such as aldehydes, ketones, alcohols, ethers) are possible ingredients of the atmosphere but remain undetected - a subject we will return to in a later section.

Oxygen has yet to be detected on Titan in an organic molecule such as methanol (CH$_3$OH) or formaldehyde (H$_2$CO), being found so far only in the simple inorganic molecules CO, \coo , and \water . This limits the presently confirmed scope of astrobiological molecules ({\em i.e.} those with the elements CHON in a variety of functional groups) - at least in the atmosphere. At the surface and in the subsurface - where hydrocarbons are thought to be readily hydrolyzed as seen in laboratory experiments \cite{khare86, neish10, ramirez10} - the astrobiological potential may be much greater \cite{raulin08, raulin10, lunine20}.

The reaction pathways that lead between these molecules have been compiled into computational models of the atmospheric chemistry, which have largely been successful at replicating the observed gas abundances. Models pre-dating the {\rm Cassini-Huygens} mission  \cite{strobel74, strobel82,yung84, yung87, toublanc95, english96, lara94, lara96, wilson04} primarily focused on replicating the observed neutral gas abundances as measured by {\it Voyager} \cite{coustenis89a} and the {\em Infrared Space Observatory} (ISO) \cite{coustenis03}. 
However, some models were also developed for the ionosphere \cite{keller92, fox97, galand99, mullerwodarg00, banaszkiewicz00}. 
During the {\it Cassini-Huygens} mission and since, new information collected by the spacecraft, especially from direct sampling of the ionosphere,\cite{waite05, szego05, wahlund05, hartle06, cravens06, mullerwodarg06, coates07, waite07, mullerwodarg08, cravens08, wahlund09, cui09a, cui09b, rymer09, crary09, coates09a, coates12, aagren12, westlake11, westlake12, snowden13, shebanits13, teolis15, cui16, chatain21}
has prompted many new and revised models of Titan's atmosphere \cite{lavvas08a, lavvas08b, horst08, krasnopolsky09, krasnopolsky10, krasnopolsky12, krasnopolsky14, hebrard05, hebrard07, hebrard09, robertson09, hebrard12, hebrard13, vuitton08, vuitton09, vuitton12, bell10a, lara14, dobrijevic14, loison15, lic15, willacy16, dobrijevic16, loison19, vuitton19}. 

At the opposite end of the size scale, molecular growth by covalent bonding and agglomeration results in macro-molecular haze particles \cite{khare02,sagan92, wilson03, trainer04, imanaka04, trainer06, waite07, sekine08, imanaka12, cable12, horst18}, composed of thousands to millions of individual atoms \cite{lavvas10}. As these particles reach a size of $\sim$1~$\mu$m, they begin to sediment (or form the nuclei for condensate growth) and are removed from the atmosphere \cite{lorenz93,lorenz95,mckay01,karkoschka09}, apparently forming vast dune fields on the surface \cite{lorenz06, radebaugh08, mastrogiuseppe14}.

\subsection{Atmospheric Temperature Structure}

\begin{figure}
\includegraphics[scale=0.85]{ %titan-temperatures-smoothed_wht_bg.pdf
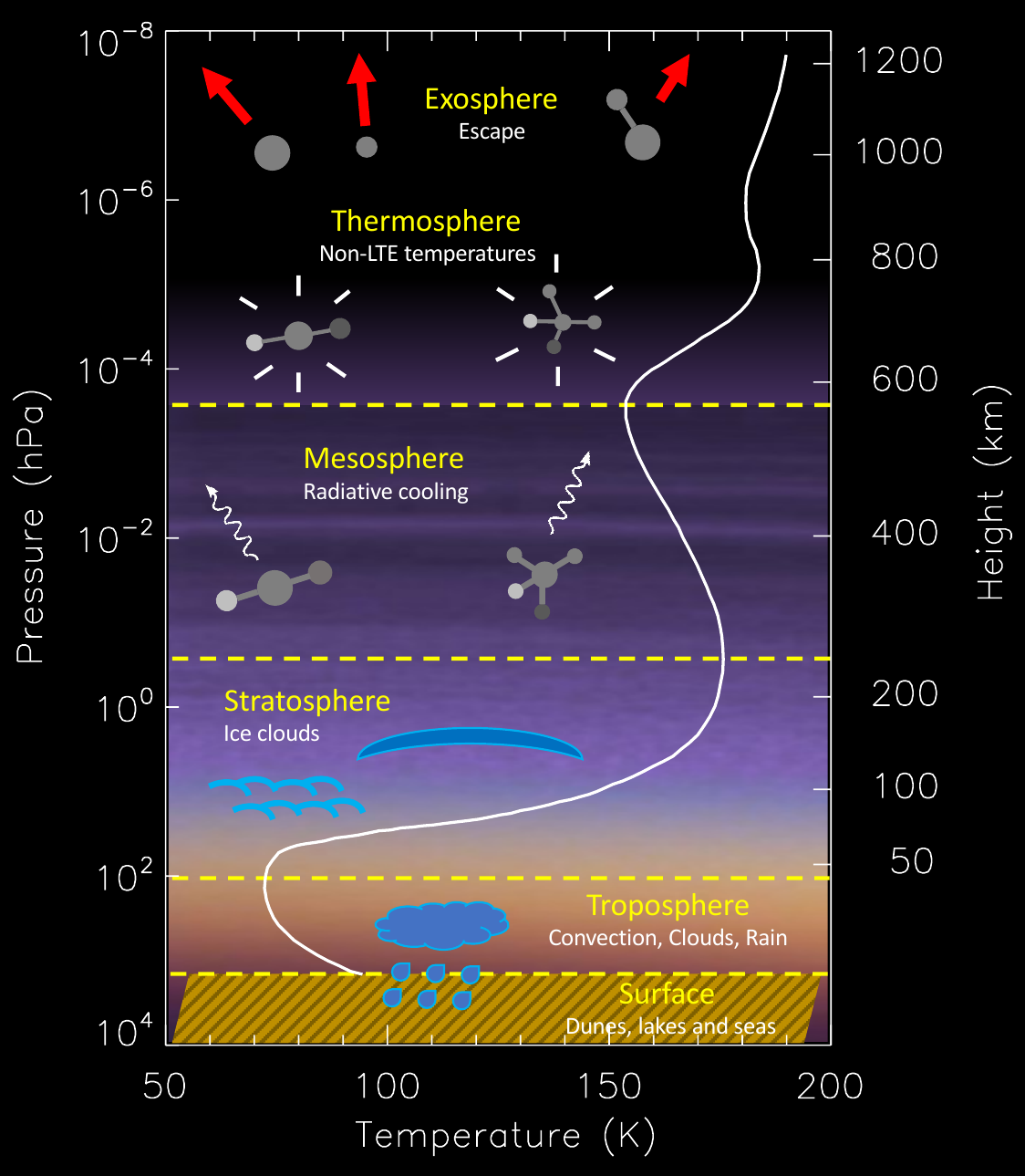}
  \caption{Typical low-latitude atmospheric temperature structure of Titan, composited from multiple measurement sources\cite{fulchignoni05, schinder11a, teanby19, lellouch19}, showing processes responsible for creating each layer. }
  \label{fig:temps}
\end{figure}

Titan's atmospheric temperature structure (Fig.~\ref{fig:temps}) is a result of the competing heating and cooling processes that take precedence at different altitudes. In the dense lower atmosphere, convection driven by surface heating leads to a vertically decreasing temperature profile, as warm air rises and adiabatically cools. A temperature minimum is reached at $\sim$45~km, the tropopause, as confirmed by direct measurements\cite{fulchignoni05} from the {\em Huygens} Atmospheric Structure Experiment \cite{fulchignoni02} (HASI) and occultations by the {\em Cassini} Radio Science Subsystem\cite{kliore04} (RSS) \cite{schinder11a,schinder12,schinder20}. 

Above the tropopause, heating by absorbed solar energy, primarily by atmospheric haze particles, causes temperatures to rise again in the stratosphere \cite{yelle91}. Stratospheric temperatures have been measured by {\em Cassini} HASI \cite{fulchignoni05} and by the {\em Cassini} Composite Infrared Spectrometer  \cite{flasar04b} (CIRS) \cite{flasar05, achterberg08a}, ALMA \cite{serigano16, thelen18, lellouch19}, as well as through radio occultations for the lower stratosphere \cite{schinder11a,schinder12,schinder20}. At around 250--400~km, a temperature maximum, the stratopause, is reached at $\sim$180~K \cite{strobel09}. The exact altitude (pressure) and temperature of the stratopause varies with both latitude and season \cite{achterberg11, teanby12, teanby19} over the course of Titan's long year (29.46 Earth years), being higher and warmer (by $\sim $20~K) over the winter pole. This somewhat counter-intuitive result can be understood as resulting from adiabatic compression of air in the descending branch of the global stratospheric Hadley cell.

Temperatures fall throughout the next layer, the mesosphere, as haze becomes thin, and radiative cooling by gases such as HCN and \acetylene\ becomes increasingly important \cite{yelle91,strobel09}. Titan's mesopause is reached at $\sim$600~km \cite{lellouch19}, above which altitude temperatures rise again. This is primarily due to methane UV absorption \cite{lellouch90}, blocking of outgoing IR radiation by \ethane , and far-infrared HCN rotational lines \cite{yelle91}. This is the thermosphere, a region where gas collisions are rare, and molecules must wait to spontaneously emit a photon to lose energy. 

The temperature structure of the upper atmosphere is highly variable\cite{galand99}. Thermal oscillations of significant amplitude were inferred by {\em Huygens} HASI\cite{fulchignoni05} above 500~km, while \latin{in situ} measurements of electron temperature (by the {\em Cassini} Radio and Plasma Wave Spectrometer - RPWS \cite{gurnett04}) and density and composition (by INMS) have shown significant time variability on diurnal \cite{cui09b, agren09, bell14} and longer timescales depending on the level of solar activity \cite{edberg13a} as well as the position of Titan within Saturn's magnetosphere \cite{garnier09,snowden11a, snowden11b, edberg13b}. More recently ALMA measurements are now able to probe the thermal structure of the upper atmosphere as well, providing the ability to monitor secular changes over time.\cite{lellouch19, thelen22}

Above the four layers of the bound atmosphere is the exosphere, beginning at the exobase ($\sim$1500~km \cite{cui08}), a region where gases can freely escape to space. These five regions mirror the temperature structure of the Earth's atmosphere, but with a substantially larger scale height (approximately $\times$5) due to the lower surface gravity. Overlapping the upper thermosphere is the ionosphere (${\sim}z > 1000$~km), defined as the region where "significant numbers of free thermal ($<$1~eV) electrons and ions are present." \cite{schunk04}

It is important to note that the vertical profiles of temperature, minor gas abundances and haze density all vary with both latitude and season. Titan - like the Earth and other planets with atmospheres - exhibits one or more convection cells in the middle atmosphere. Near to the equinoxes, air rises at mid-latitudes and flows to both poles, where it descends and then returns equator-ward \cite{hourdin95, lebonnois01, hourdin04, rannou04, lebonnois09}. However, close to the solstices the circulation more closely resembles a single cell with flow from the summer to winter hemisphere. These cells act to redistribute thermal energy, trace gases and hazes in both altitude and latitude. This topic has been the subject of extensive measurements and modeling in the literature (e.g. \cite{teanby19, lora19, vinatier20, mathe20, sylvestre20, cordiner20, coustenis20, sharkey21}, and references therein). In this review I will not further discuss latitudinal or longitudinal variations in atmospheric structure and focus only on the vertical chemistry variations typical of mean conditions at low latitudes.

\subsection{Gas Vertical Profiles}

\begin{figure}
\includegraphics[scale=0.65]{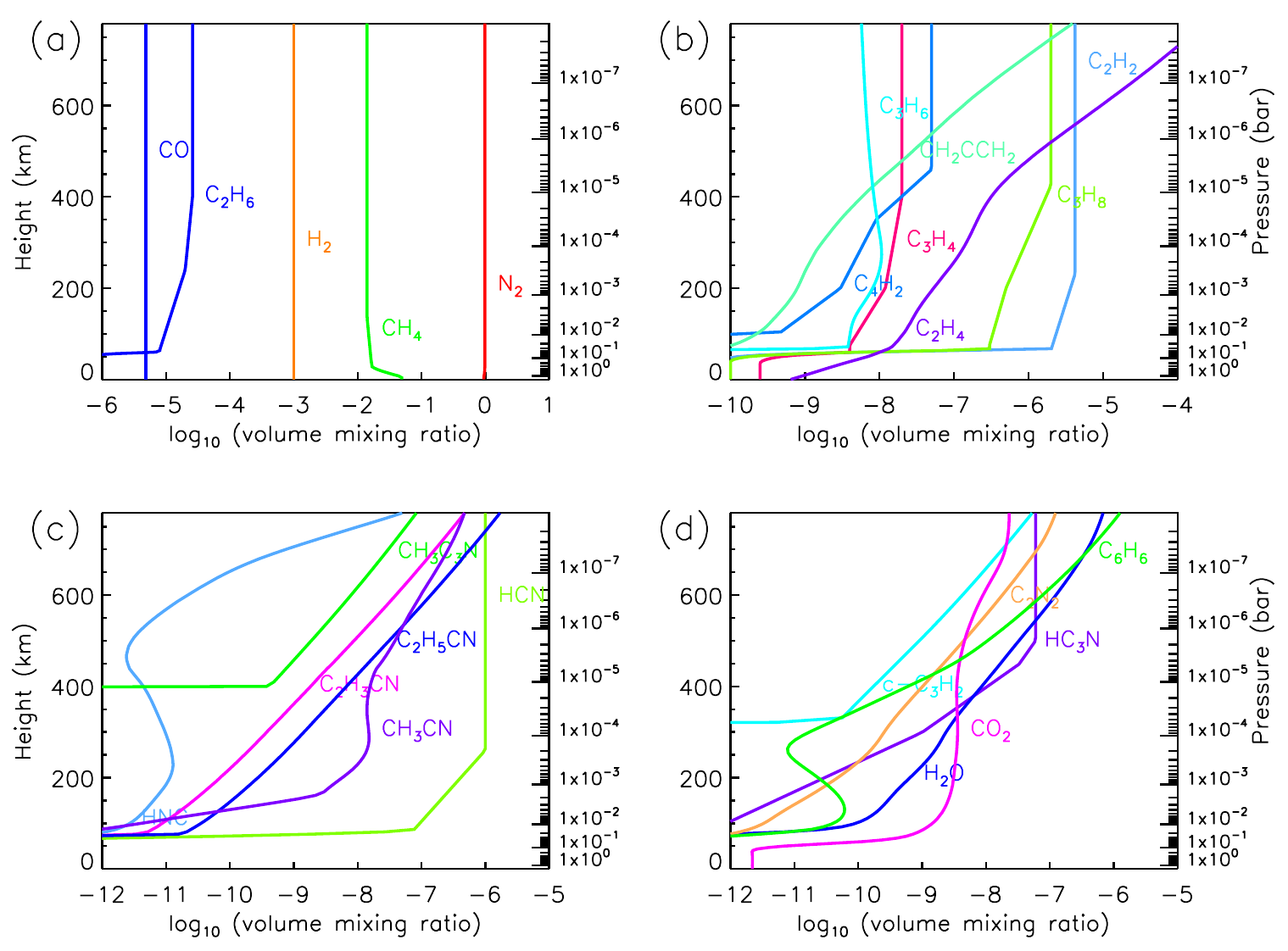}
  \caption{Typical vertical profiles of gases in Titan's atmosphere at low latitudes. 
  Sources: \nitrogen , \methane , \hydrogen\  - Niemann et al., (2010); \cite{niemann10}
  CO - Serigano et al. (2016); \cite{serigano16}
  \acet , HCN - Teanby et al. (2007);  \cite{teanby07}
  \ethane , \propyne , \propane , \diacet\ - Vinatier et al. (2007); \cite{vinatier07a}
  \cyanoacet\ - Marten et al. (2002); \cite{marten02}
  \water , \cyanogen\ - Loison et al. (2015); \cite{loison15}
  \cpld\ - Nixon et al. (2020); \cite{nixon20}
  \vinylcyn\ - Palmer et al. (2017); \cite{palmer17}
  \ethylcyn\ - Cordiner et al. (2015); \cite{cordiner15}
  \butynenitrile\ - Thelen et al. (2020); \cite{thelen20}
  \methylcyn\ - 2015 profile from Thelen et al. (2019a)\cite{thelen19b} with extensions to troposphere from Marten et al. (2002)\cite{marten02} and mesosphere from Loison et al. (2015); \cite{loison15}
  \propene\ - Lombardo et al. (2019a); \cite{lombardo19a}
  \propadiene\ - Lombardo et al. (2019b); \cite{lombardo19b}
  HNC - combination of Lellouch et al. (2019)\cite{lellouch19} and Dobrijevic et al. (2016); \cite{dobrijevic16}
   \ethylene , \benzene, \coo\ - photochemical models from Vuitton et al. (2019). \cite{vuitton19}
  } 
\label{fig:profiles}
\end{figure}

Fig.~\ref{fig:profiles} shows typical vertical profiles of the 24 known molecular species at low latitudes, compiled from a combination of ground and space-based measurements, and  some photochemical model profiles constrained by observations. These include two pairs of structural isomers: HCN and HNC, \propyne\ and \propadiene . 

Some important trends can be noted. Methane is an unusual outlier, with a greater abundance in the troposphere ($z \leq 45$~km) than above. This is due to the `cold trap' effect, where it reaches saturation as the tropospheric temperature drops with altitude, and therefore its mixing ratio is reduced as it forms clouds at $\sim$15-30~km. The gases \nitrogen\ and CO are well-mixed, having approximately uniform profiles throughout the atmosphere due to long photochemical lifetimes. Two other gases: \hydrogen\ and \ethylene\ also do not condense at the `cold trap' - the coldest part of the atmosphere around the tropopause at 45~km ($\sim$70~K). The profile of \hydrogen\ is shown here as constant at the 0.1\% value typical of the lower stratosphere, since measurements of its vertical profile remains uncertain. 

All other gas species show profiles that typically decrease downwards from the upper atmosphere, due to having a source due to photochemistry at high altitudes, and then becoming diluted as they are mixed downwards into the denser part of the atmosphere. In many cases the actual measured profiles are still rudimentary, constrained by only a few data points and with even less knowledge of meridional and temporal variations. Some gases may exhibit increases again towards the stratosphere, due to either secondary production peaks (e.g. due to cosmic ray deposition) or due to redistribution by atmospheric circulation.

%%%%%%%%%%%%%%%%%%%%%%%%%%%%%%%%%%%%%%%%%%%%%%%%%%%%%%%%%%%%%%%%%%%%%
%%%%%%%%%%%%%%%%%%%%%%%%%%%%%%%%%%%%%%%%%%%%%%%%%%%%%%%%%%%%%%%%%%%%%
\section{Atmospheric Photochemical Processes}
%%%%%%%%%%%%%%%%%%%%%%%%%%%%%%%%%%%%%%%%%%%%%%%%%%%%%%%%%%%%%%%%%%%%%
%%%%%%%%%%%%%%%%%%%%%%%%%%%%%%%%%%%%%%%%%%%%%%%%%%%%%%%%%%%%%%%%%%%%%

\begin{figure}
\includegraphics[scale=0.65]{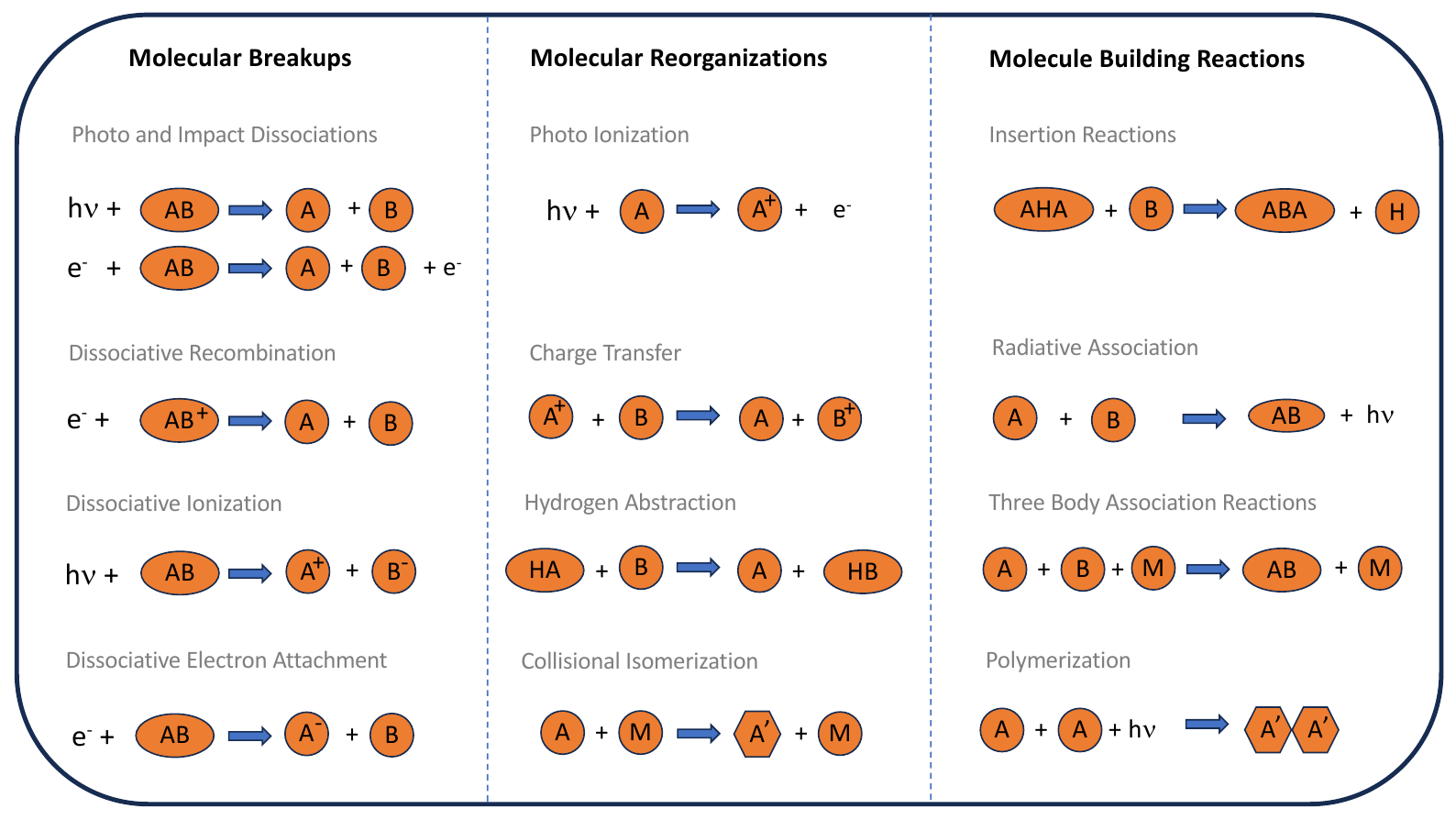}
 \caption{Principal processes and reaction types relevant to the chemistry in Titan's atmosphere. Plus (+) and minus (-) superscripts indicate ions, while $A$ and $A^{\prime}$ are molecules that have re-arranged into a different structure without change of composition. }
\label{fig:reactions}
\end{figure}

In this section I present a brief overview of the types of reactions that occur in Titan’s atmosphere, as a prelude to the discussion of the chemistry of individual molecules in the following section (see also Table~5 of Vuitton et al. \cite{vuitton19} and description therein). Different chemical processes become important at different levels of the atmosphere, due to the altitude variation of temperature, density and penetration depths of charged particles and photons that affect the reactions. For example, Saturn magnetospheric electrons are stopped high up in the ionosphere, while solar photons penetrate to varying depths depending on wavelength.\cite{vuitton19} High energy cosmic rays are the deepest-penetrating rays in the atmosphere, peaking in energy deposition at around 100-150~km altitude.\cite{vuitton19} 
The various processes, depicted in Fig.~\ref{fig:reactions}, are now examined in more detail.

%A schematic view of important physical and chemical processes in various altitude ranges is shown in Fig.~\ref{fig:processes}.

%\begin{figure}
%\includegraphics[scale=0.85]{titan-chemistry_wht_bg.pdf}
%  \caption{Important physical and chemical processes in Titan's atmosphere by altitude range.}
%  \label{fig:processes}
%\end{figure}

%%%%%%%%%%%%%%%%%%%%%%%%%%%%%%%%%%%%%%%%%%%%%%%%%%%%%%%%%%%%%%%%%%%%%
\subsection{Photodissociation and electron impact dissociation}

Dissociation (break-up, or fragmentation) of a molecule occurs when a sudden influx of energy breaks molecular bonds. In Titan’s atmosphere this happens readily in the upper atmosphere ($z > \: \sim$700~km), due to both energetic UV solar photons (h$\nu$) and by the impact of fast-moving electrons ($e^-$) trapped in Saturn’s magnetic field \cite{cravens05, lavvas11b} (Fig.~\ref{fig:chemistry}). 

%\begin{eqnarray}
%{\rm AB} + h\nu & {\longrightarrow} & {\rm A + B} \\
%{\rm AB} + e^- & {\longrightarrow} & {\rm A + B} + e^- 
%\end{eqnarray}

Dissociation often leads to neutral (uncharged) molecular fragments, which may either be in the ground state or excited states, e.g.:\cite{mount77, erwin05, lee83}
\begin{eqnarray}
{\rm CH_4} + h\nu & {\longrightarrow} & {\rm ^1CH_2} + {\rm H_2} \\
{\rm CH_4} + e^- & {\longrightarrow} & {\rm ^3CH_2} + 2{\rm H} 
\end{eqnarray}

\noindent where ${\rm ^3CH_2}$ is  ground state methylene and  ${\rm ^1CH_2}$ is an excited state. Note that other possible fragmentation products are possible, the examples given above are only one possibility in each case. Also note that for simplicity, electronic states of molecules are usually omitted in this paper, unless required to distinguish between two otherwise identical reagents that have significantly different properties.

The molecular fragments are often radicals – i.e. highly reactive atomic or molecular species that have unpaired electrons, such as H, CH, and N. These radicals are quick to react with other radicals, or with neutral species.

A secondary peak of dissociation is expected to occur in the deep atmosphere at $\sim$100--150~km \cite{capone76, capone83, molina-cuberos99a, molina-cuberos99b} due to extremely high-energy cosmic rays (h$\nu$).

%%%%%%%%%%%%%%%%%%%%%%%%%%%%%%%%%%%%%%%%%%%%%%%%%%%%%%%%%%%%%%%%%%%%%
\subsection{Ionization}

Instead of breaking up (dissociation), a molecule may instead become ionized (positively or negatively charged), typically by losing an electron to become a cation, e.g.:
\cite{aoto06, shaw92, shemansky05}

\begin{eqnarray}
{\rm N_2} + h\nu & {\longrightarrow} & {\rm N_2^+} + e^-  \\
{\rm N_2} + e^- & {\longrightarrow} & {\rm N_2^+} + 2e^- 
\end{eqnarray}

However dissociative ionization may also occur, e.g.:\cite{mitsuke91, ruscic90}

\begin{eqnarray}
{\rm CH_4} + h\nu & {\longrightarrow} & {\rm CH_3^+ + H^-} \\
{\rm C_2H_2} + h\nu & {\longrightarrow} & {\rm H^+ + C_2H^-} 
\end{eqnarray}

Ionization of neutrals may also occur through dissociative electron attachment, {\eg}\cite{rawat08, stamatovic70}: 

\begin{eqnarray}
{\rm CH_4} + e^- & {\longrightarrow} & {\rm CH_2^- + H_2} \\ 
{\rm CO} + e^- & {\longrightarrow} & {\rm O^- + C}
\end{eqnarray}

For larger molecules, radiative electron attachment is also important:\cite{vuitton09, herbst08}

\begin{eqnarray}
{\rm C_6H} + e^- & {\longrightarrow} & {\rm C_6H^- + h\nu } 
\end{eqnarray}

%%%%%%%%%%%%%%%%%%%%%%%%%%%%%%%%%%%%%%%%%%%%%%%%%%%%%%%%%%%%%%%%%%%%%
\subsection{Ion reactions}

Dissociative recombination is the process whereby a positive ion reunites with an electron, and in the process breaks apart. An example in Titan’s atmosphere is:
\cite{kaminska10, semaniak98}

\begin{eqnarray}
{\rm CH_5^+} + e^- & {\longrightarrow} & {\rm CH_3 + 2\: H} 
\end{eqnarray}

%\begin{eqnarray}
%{\rm A + B}   & {\longrightarrow} & {\rm AB} + h\nu 
%\end{eqnarray}

Radiative association is a reaction whereby two species combine, shedding excess energy via a photon.
These reactions typically occur only in rarefied environments ($p < 10^{-5}$~mbar\cite{klippenstein96} where a metastable intermediary complex has time to form and then stabilize by emission of a photon. Despite being impossible to observe in laboratory conditions due to the long lifetimes of the intermediate states\cite{gerlich92,luca02}, such reactions are thought to occur in interstellar clouds\cite{herbst80, herbst85, herbst91, herbst21}, as well as in Titan's upper atmosphere. Radiative ion-neutral association reactions on Titan may include: \cite{kaiser02b, vuitton19} 

\begin{eqnarray}
{\rm C_4H_3^+ + C_2H_2} & {\longrightarrow} & {\rm C_6H_5^+ + h\nu } \\
{\rm C_3H_3^+ + H_2} & {\longrightarrow} & {\rm C_3H_5^+ + h\nu } 
\end{eqnarray}

\noindent In charge transfer reactions \cite{lindinger00, vuitton19}, at least two products result:

\begin{eqnarray}
{\rm CH_4^+ + HCN} & {\longrightarrow} & {\rm HCNH^+ + CH_3}  \\
{\rm N^+ + C_2H_2} & {\longrightarrow} & {\rm C_2H_2^+ + N } 
\end{eqnarray}

\noindent
Ions may also react with each other, leading to neutral products, although \citeauthor{vuitton19}\cite{vuitton19} argued that positive-negative ion recombination rates are too small to compete with ion-neutral reaction pathways.

%%%%%%%%%%%%%%%%%%%%%%%%%%%%%%%%%%%%%%%%%%%%%%%%%%%%%%%%%%%%%%%%%%%%%
\subsection{Radical reactions}
Radicals (molecules with an unpaired electron) react with other radical and non-radical species in multiple ways. Radicals may react with each other in association reactions \cite{wilson04}:

\begin{eqnarray}
{\rm ^3CH_2 + CH_3} & {\longrightarrow} &  {\rm C_2H_4 + H}
\end{eqnarray}

\noindent
Radicals may attack neutral molecules, for example in the case of hydrogen abstraction, which is a major loss process for methane \cite{yung84, opansky96, vuitton06b, vuitton19}:

\begin{eqnarray}
{\rm C_2H + CH_4} & {\longrightarrow} &  {\rm C_2H_2 + CH_3}
\end{eqnarray}

\noindent
Other examples include substitutions of terminal atoms or groups, typically at carbon-carbon double or triple bonds \cite{herbert92, sims93, gannon07}:

\begin{eqnarray}
{\rm CN + C_2H_2} & {\longrightarrow} &  {\rm HC_3N + H} \\
{\rm CN + C_2H_4} & {\longrightarrow} &  {\rm C_2H_3CN + H} 
\end{eqnarray}

\noindent
and insertions, which allow heavier molecules to be built from simpler ones \cite{mckee03, berman82, thiesemann01}:

\begin{eqnarray}
{\rm CH + C_2H_4}   & {\longrightarrow} & {\rm CH_3CCH + H}
\end{eqnarray}

% of the type:
%\begin{eqnarray}
%{\rm A + B + M}   & {\longrightarrow} & {\rm AB +M} 
%\end{eqnarray}

\noindent
Three-body association reactions occur when two reactants meet to form a metastable, intermediate complex that is then stabilized by collision with a third, non-reacting body that carries away energy, allowing the metastable complex to stabilize. The two steps are:

\begin{eqnarray}
{\rm A + B }   & {\longrightarrow} & {\rm AB^* } \\
 {\rm AB^* + M }   & {\longrightarrow} & {\rm AB + M^* }
\end{eqnarray}

\noindent 
where the asterisk is used to denote an excited state. In this paper I typically simplify such reactions to a single step:

\begin{eqnarray}
 {\rm A+ B + M }   & {\longrightarrow} & {\rm AB + M ^*}
\end{eqnarray}

Such reactions are of critical importance to the formation of many hydrocarbons, especially alkanes, e.g.\cite{slagle88, brouard89, kaiser02b}:

\begin{eqnarray}
{\rm CH_3 + H + M} & {\longrightarrow} &  {\rm CH_4 + M^* } \\
{\rm CH_3 + CH_3 + M} & {\longrightarrow} &  {\rm C_2H_6 + M^* }
\end{eqnarray}

\noindent
Note that three-body reactions are limited to Titan's dense, lower atmosphere where there is a sufficiently high collision rate to allow the collisional stabilization to occur.\cite{vuitton12}

Radiative association reactions may also occur between radicals. Vuitton et al.\cite{vuitton12} studied the effect of radiative associations on Titan's chemistry and proposed that reactions such as:

\begin{eqnarray}
{\rm C_4H_2 + H } & {\longrightarrow} &  {\rm C_4H_3 + h\nu} \\
{\rm C_4H_3 + H } & {\longrightarrow} &  {\rm C_4H_4 + h\nu} 
\end{eqnarray}

\noindent may occur in Titan's atmosphere. 

%%%%%%%%%%%%%%%%%%%%%%%%%%%%%%%%%%%%%%%%%%%%%%%%%%%%%%%%%%%%%%%%%%%%%
\subsection{Other reactions}

Molecules may also re-organize their structure to become more stable, such as in collisional isomerization. Two important known instances of this on Titan are the conversion between propadiene (allene) and propyne by H atoms,\cite{yung84,lic15} with a barrier of 65.1~kcal/mol:\cite{alnama07}

\begin{eqnarray}
{\rm CH_2CCH_2 + H}   & {\longrightarrow} & {\rm CH_3CCH + H} 
\end{eqnarray}

\noindent
and HNC to HCN\cite{willacy16} with a barrier of 30.2~kcal/mol:\cite{herbst00}

\begin{eqnarray}
{\rm HCN + H}   & {\longrightarrow} & {\rm HNC + H}
\end{eqnarray}

%\noindent or ring closure, such as from hexa-1,3-diene-2-yne to benzyne:

%\begin{eqnarray}
%{\rm C_6H_4 }   & {\longrightarrow} & {\rm c{\text -}C_6H_4}
%\end{eqnarray}

Polymerization is the process where multiple similar, unsaturated hydrocarbons join together form linear chains:

\begin{eqnarray}
n\rm{ (C_2H_4) } + h\nu  & {\longrightarrow} & {\rm CH_3}{\text -}{\rm (C_2H_4 )}_{n-2}{\text -}{\rm CH_3} \: + \: {\rm C_2H_2}
\end{eqnarray}

\noindent
Polymerization is thought to be one of the principal mechanisms leading to the formation of larger polyyne compounds, which may be a significant component of Titan's haze particles.\cite{courtin91, cabane93, clark97, lara99, mckay01, khare02, lebonnois02, tran03a, tran03b, wilson03, imanaka04, lavvas08b, perrin21}

%%%%%%%%%%%%%%%%%%%%%%%%%%%%%%%%%%%%%%%%%%%%%%%%%%%%%%%%%%%%%%%%%%%%%
%%%%%%%%%%%%%%%%%%%%%%%%%%%%%%%%%%%%%%%%%%%%%%%%%%%%%%%%%%%%%%%%%%%%%
\section{Chemistry of the Neutral Atmosphere}
%%%%%%%%%%%%%%%%%%%%%%%%%%%%%%%%%%%%%%%%%%%%%%%%%%%%%%%%%%%%%%%%%%%%%
%%%%%%%%%%%%%%%%%%%%%%%%%%%%%%%%%%%%%%%%%%%%%%%%%%%%%%%%%%%%%%%%%%%%%

\begin{figure}
\includegraphics[scale=0.65]{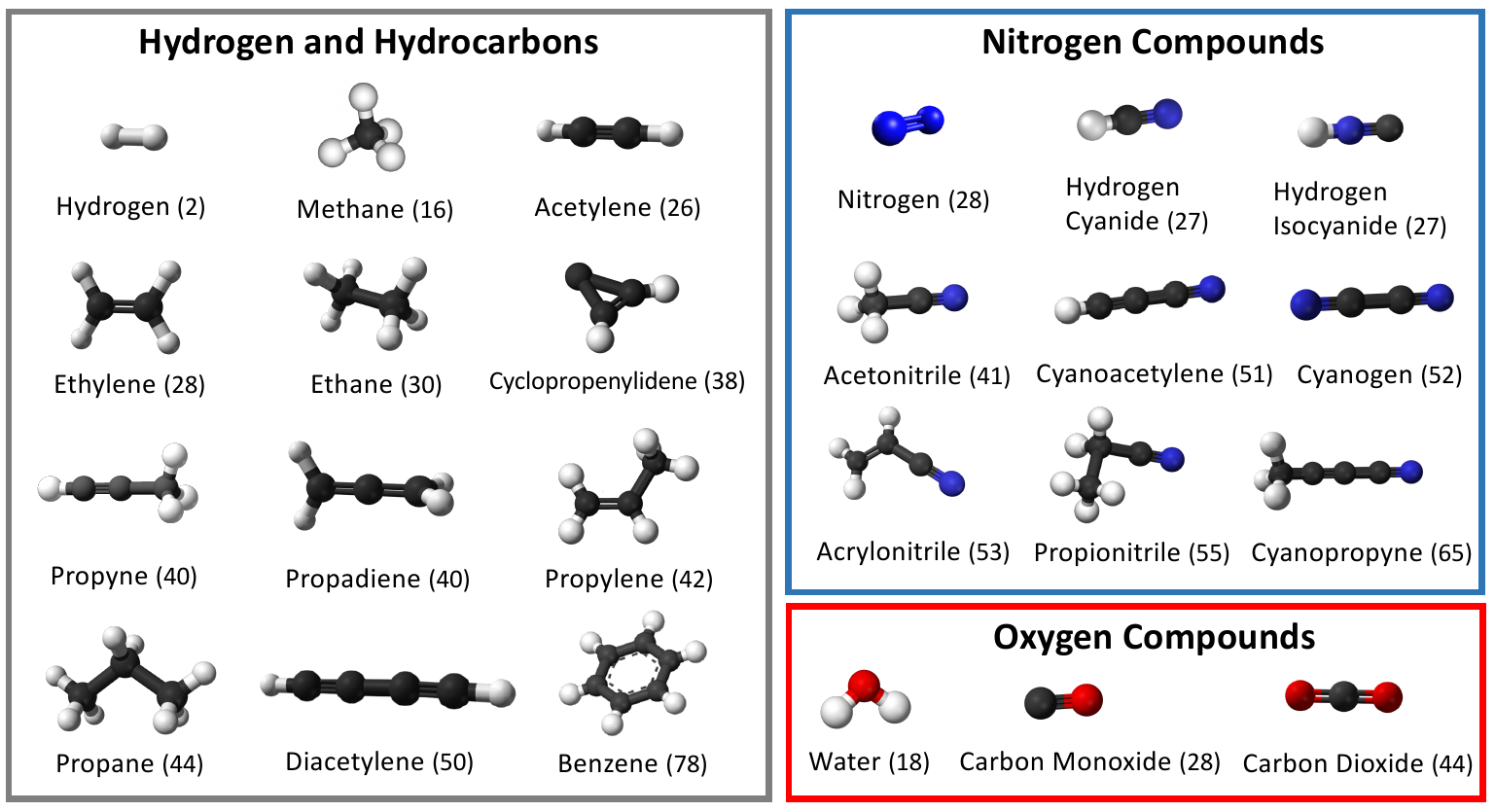}
  \caption{Molecules detected in Titan's neutral atmosphere, along with the mass of the most abundant isotopologue in Da.}
  \label{fig:molecules}
\end{figure}

In this review, I focus on the 24 molecules detected in Titan's dense, neutral atmosphere (Fig.~\ref{fig:molecules}). Many other neutral species have been inferred from ion and neutral mass spectroscopy in Titan's upper atmosphere \cite{waite05, cravens06, vuitton07, waite07, magee09, cui09a}, through photochemical models
\cite{lavvas08a, lavvas08b, horst08, krasnopolsky09, krasnopolsky10, krasnopolsky12, krasnopolsky14, hebrard05, hebrard07, hebrard09, hebrard12, hebrard13, vuitton08, vuitton09, vuitton12, bell10a, lara14, dobrijevic14, loison15, lic15, willacy16, dobrijevic16, loison19, vuitton19} and via laboratory experiments (see review by \citeauthor{cable12}\cite{cable12}). 
The choice to limit the discussion to the chemistry and composition of the 24 definitively identified molecules of the neutral atmosphere was made for several reasons: (i) this set of molecules includes the most easily detectable and likely the most abundant molecules of the atmosphere, which therefore provide a good overview of bulk chemistry and composition; (ii) other molecules inferred only through single-stage mass spectroscopy do not have robust structural identifications, since this technique cannot typically distinguish between isomers having the same chemical formula, occurring at the level of complexity of three carbon atoms and beyond; (iii) to allow for a more detailed discussion of the molecules that have been unambiguously detected; and (iv) because the composition of the neutral, lower atmosphere is the most important chemical inventory for consideration of other processes such as condensation and meteorology, sedimentation to the surface, and astrobiology at Titan's surface and interior.

{\em Note on chemical names:} the topic of chemical nomenclature remains eternally problematic. For example, the relatively simple compound \acrylonitrile\ has been commonly referred to as `vinyl cyanide' in most 20th century astronomical literature, though preference has shifted more recently towards the simpler, single-word name `acrylonitrile' (also with `methyl cyanide' to `acetonitrile'). In fact multiple valid names for \acrylonitrile\ exist, including `2-propenenitrile', `cyanoethene/ cyanoethylene' and `propenenitrile', however it must be noted that the preferred official (IUPAC) name is actually the rather cumbersome `prop-2-enenitrile'. 

 IUPAC molecular nomenclature certainly has its place. However, for the purposes of discourse in planetary atmospheric chemistry, dominated by small molecules, the formulaic names for molecules can be not only inconvenient, but an actual obstacle to reading and digesting information. Therefore, in this work I have followed a naming convention based on a combination of modernity, modified in places for greater simplicity, i.e.: (a) single word names are generally preferred over multiple word names (e.g. `propyne' over the older standard `methyl acetylene'); (b) avoiding numbers in names of small molecules except where necessary, and (c) accepting that some names are close enough to be interchangeable (e.g. there is little confusion engendered by using either `ethylene' and  `ethene'; `propylene' and `propene' etc). 

While I run the risk of offending practicing chemists, I hope that the terminology is consistent enough for readers to understand what molecule is being referred to, and convenient enough for simplified writing in typical planetary science usage.

%%%%%%%%%%%%%%%%%%%%%%%%%%%%%%%%%%%%%%%%%%%%%%%%%%%%%%%%%%%%%%%%%%%%%
\subsection{Hydrocarbons and Hydrogen}

\begin{figure}
\includegraphics[scale=0.60]{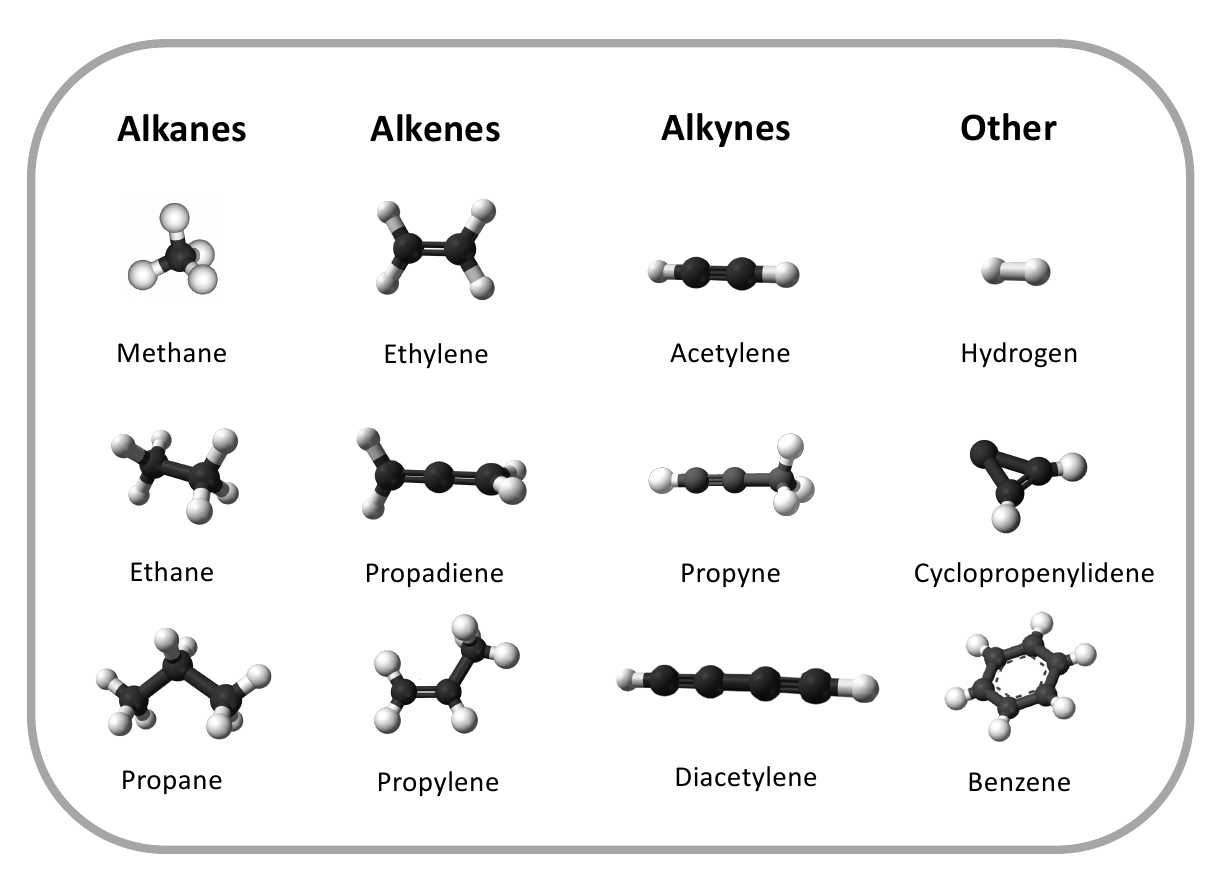}
  \caption{Hydrocarbon molecules detected on Titan.}
  \label{fig:hydrocarbons}
\end{figure}

Hydrocarbons are molecules formed from atoms of only hydrogen and carbon. Due to the four-fold valency of carbon, many bonding configurations are possible. 
The most common families of aliphatic (acyclic) hydrocarbons include (i) the alkanes - where carbon is saturated having four single-bonds; and the two unsaturated types - (ii) alkenes, featuring carbon-carbon double bonds, and (iii) alkynes with carbon-carbon triple bonds. Once there are four or more carbon atoms, mixed types become possible (see Fig.~\ref{fig:hydrocarbons}).

Cyclic hydrocarbons occur where there are one or more closed rings of carbon atoms (at least three are needed to make a ring, or cycle). Cyclic molecules that have a pi bond of delocalized electrons are known as `aromatic' (small aromatics are typically volatile at room temperature), although not all rings are aromatic. For example the unsaturated six-carbon ring benzene (\benzene ) is a well-known aromatic, whereas its saturated cousin cyclohexane (\cyclohexane ) is a non-aromatic cycle. Carbenes are reactive molecules where carbon has two unbonded but paired electrons, such as methylene (CH$_2$).

Hydrocarbon molecules are created through the breakup and recombination of fragments of methane. A prime example is the formation of ethane from two methyl radicals:\cite{walter91,cody03}

\begin{eqnarray}
{\rm CH_4} + h\nu & {\longrightarrow} &  {\rm CH_3 + H} \\
{\rm CH_3 + CH_3 + M} & {\longrightarrow} &  {\rm C_2H_6 + M^* }
\label{reaction:ethane}
\end{eqnarray}

Note that, in the process, significant amounts of molecular hydrogen will form from the photo-dissociated hydrogen atoms \cite{baulch05}:

\begin{eqnarray}
{\rm 2H + M}  & {\longrightarrow} & {\rm H_2 + M^*}
\end{eqnarray}

\noindent
leading to the trace amounts of hydrogen ($\sim$0.1\%) found in Titan’s lower atmosphere. Note that this termolecular reaction is an important process leading to the creation of \hydrogen\ in the Interstellar Medium (ISM).

Significant loss of hydrogen to space is thought to occur,\cite{strobel82, yelle06, cui08, strobel10a, strobel12, tucker13, strobel22} preventing methane from being recycled, as occurs on the giant planets, leading to gradual depletion of methane in Titan’s atmosphere in the absence of outgassing or other replenishment.\cite{yung84, tobie05, mandt09, mandt12, nixon12b}

\begin{figure}
\includegraphics[scale=0.60]{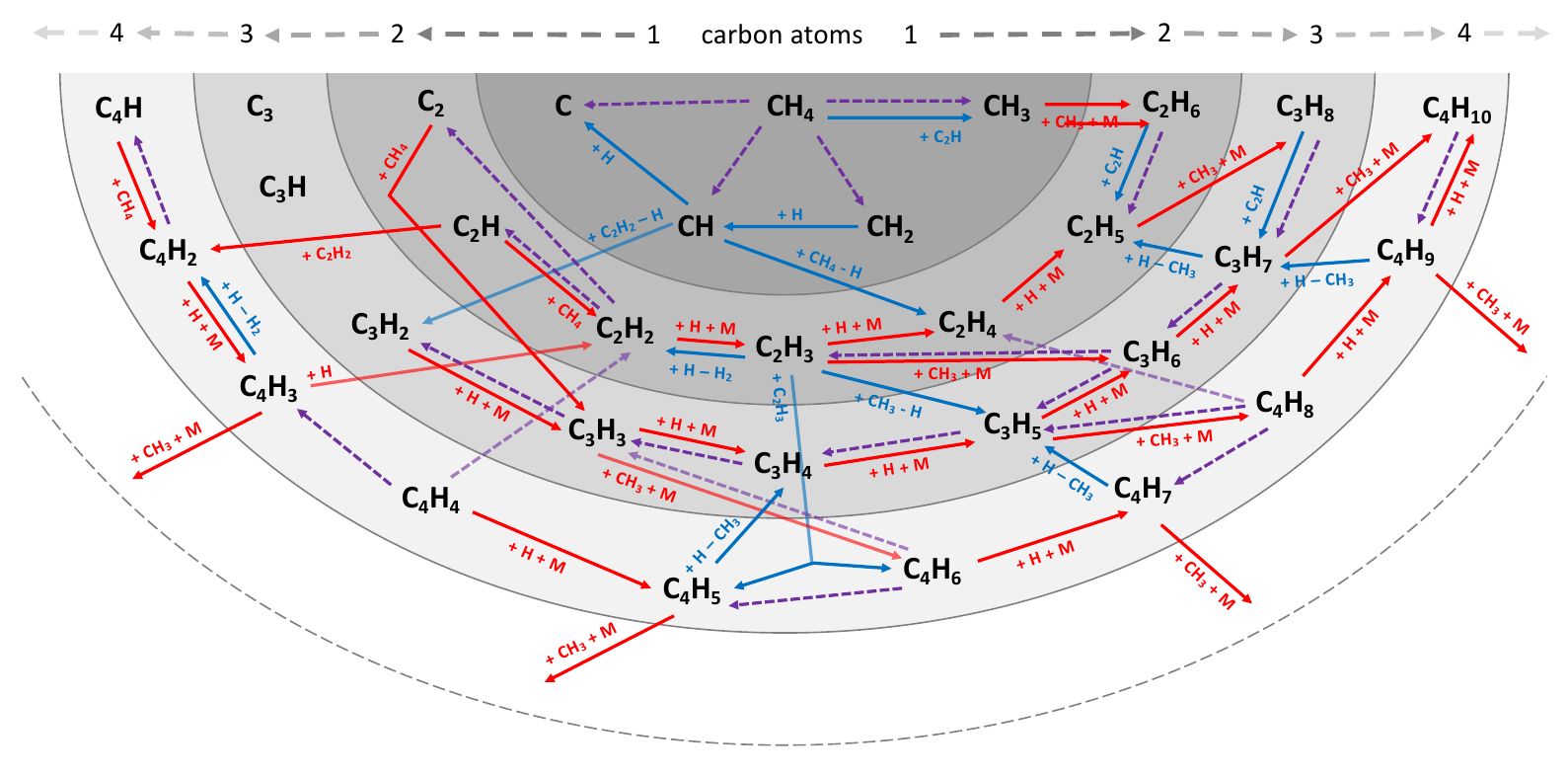}
  \caption{Hydrocarbon chemical reagent and reaction networks showing key pathways. Photolysis pathways are shown in purple dashed lines, three-body reactions in red and other reactions in blue. As the number of carbon atoms increases, the number of possible species multiplies. }
 \label{fig:hcnetwork}
\end{figure}

A network diagram showing the principal neutral pathways for hydrocarbon molecule formation is shown in Fig.~\ref{fig:hcnetwork}. Hereafter follows a high-level description of the key chemistry for each of the hydrocarbons and hydrogen leading to their relative abundances in the neutral atmosphere.

%%%%%%%%%%%%%%%%%%%%%%%
\subsubsection{	Hydrogen}

\begin{figure}
\includegraphics[scale=0.4]{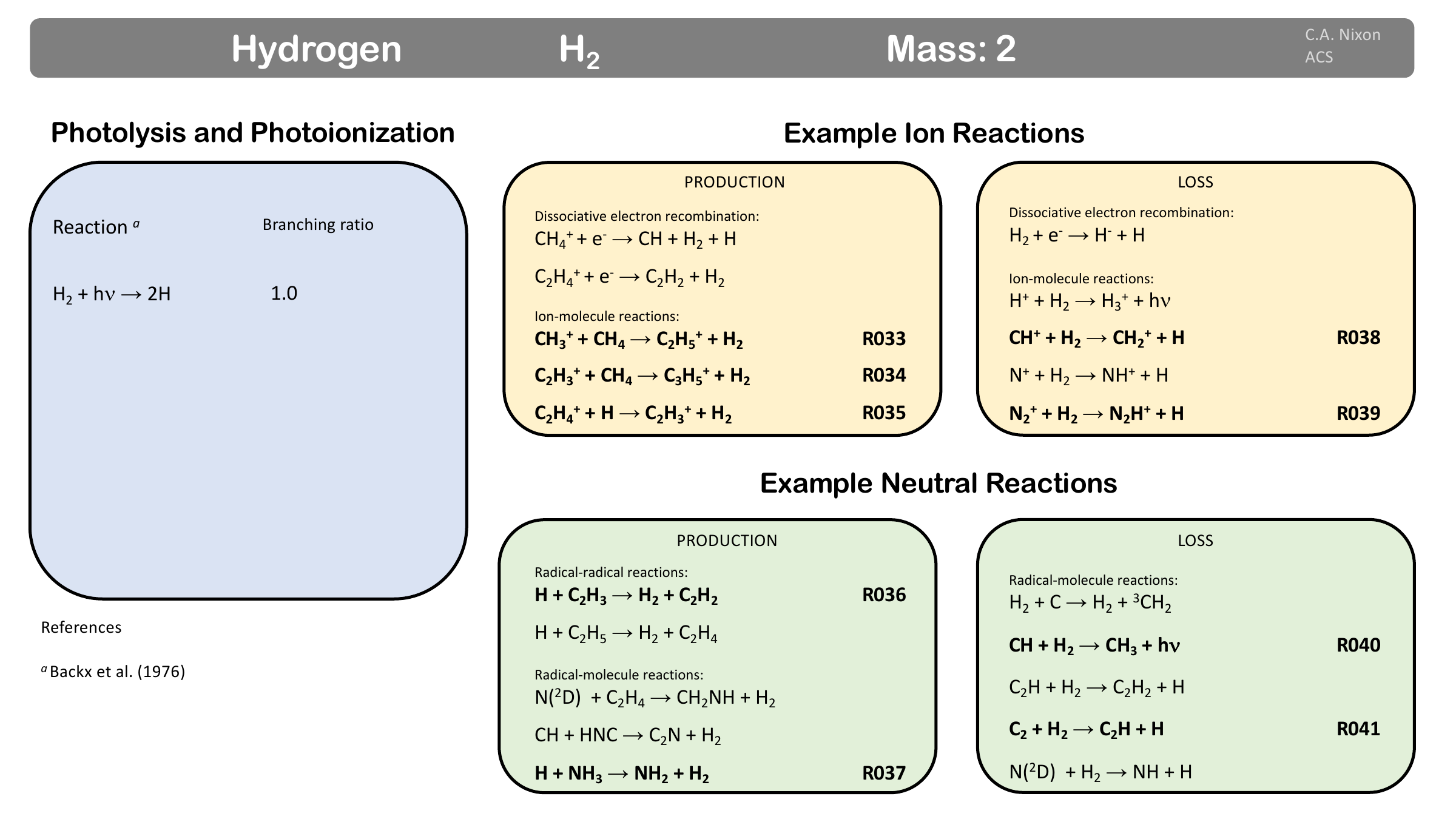}
  \caption{Hydrogen production and loss pathways. Reactions numbered and shown in bold correspond to discussion in the text.}
  \label{mol:hydrogen}
\end{figure}

{\em Detection:} Molecular hydrogen (\hydrogen ) was was tentatively detected by \citeauthor{trafton72a}\cite{trafton72a} using the 107 inch telescope at the McDonald Observatory,
via absorption in the S(0) and S(1) quadrupole lines of the 3-0 band at $\sim$0.82~\micron  . Hydrogen was clearly detected in the far-infrared by the {\em Voyager} 1 IRIS spectrometer \cite{courtin95} and confirmed by {\em Cassini} CIRS \cite{courtin12}. The \hydrogen\ Volume Mixing Ratio (VMR) was measured directly by the {\em Cassini} INMS instrument in the ionosphere as 0.4\% \cite{waite05} and in the lower stratosphere and troposphere by {\em Huygens} GCMS at 0.1\% \cite{niemann10}. The fact that hydrogen, presumed to be produced in the upper atmosphere by photolysis of methane \cite{strobel82,yung84}, was measured to have a decreasing abundance downwards, has proved difficult to replicate in models. Models have required a sink for \hydrogen\ at the surface \cite{strobel10a, strobel09, strobel14}, which has even been suggested as possibly biological in origin \cite{mckay05}. Important reactions for hydrogen are shown in Fig.~\ref{mol:hydrogen} \cite{backx76}.

{\em Production:} Molecular hydrogen can be produced in ion-phase reactions such as\cite{mcewan07}:

\begin{eqnarray}
{\rm CH_3^+ + CH_4 }  & {\longrightarrow} & {\rm C_2H_5^+ + H_2 } \\
{\rm C_2H_3^+ + CH_4 }  & {\longrightarrow} & {\rm C_3H_5^+ + H_2 } \\
{\rm C_2H_4^+ + H }  & {\longrightarrow} & {\rm C_2H_3^+ + H_2} 
\end{eqnarray}

and neutral phase radical reactions\cite{vuitton19,espinosa94}:

\begin{eqnarray}
{\rm H + {C_2H_3} }  & {\longrightarrow} & {\rm C_2H_2 + H_2 }  \\
{\rm H + NH_3 }  & {\longrightarrow} & {\rm NH_2 + H_2 } 
\end{eqnarray}

{\em Loss:} \hydrogen\ may be lost to photolysis, yielding 2H, although \citeauthor{vuitton19}\cite{vuitton19} has argued that this is a relatively small source of H in Titan's atmosphere, due to shielding of \hydrogen\ by \methane\ and \nitrogen , with most H-production coming from photolysis of methane.

\hydrogen\ may also be lost due to ion reactions, e.g.\cite{mcewan07,gerlich11,dutuit13}:

\begin{eqnarray}
{\rm CH^+ + H_2  }  & {\longrightarrow} & {\rm CH_2^+ + H} \\
{\rm N_2^+ + H_2  }  & {\longrightarrow} & {\rm N_2H^+ + H} 
\end{eqnarray}

and radical reactions \cite{brownsword97a, brownsword97b, klippenstein06}: 

\begin{eqnarray}
{\rm CH + H_2 }  & {\longrightarrow} & {\rm CH_3 }  + h\nu \\
{\rm C_2 + H_2 }  & {\longrightarrow} & {\rm C_2H+ H } 
\end{eqnarray}

{\em Future work:} Measurement and modeling work is still required to confirm our understanding of the vertical \hydrogen\ profile,\cite{strobel10a,strobel09,nixon18} and whether a solution to the vertical gradient lies in instrumental errors\cite{strobel22} or unknown processes in the atmosphere. 

%%%%%%%%%%%%%%%%%%%%%%%
\subsubsection{Methane}
\label{sect:methane}

\begin{figure}
\includegraphics[scale=0.4]{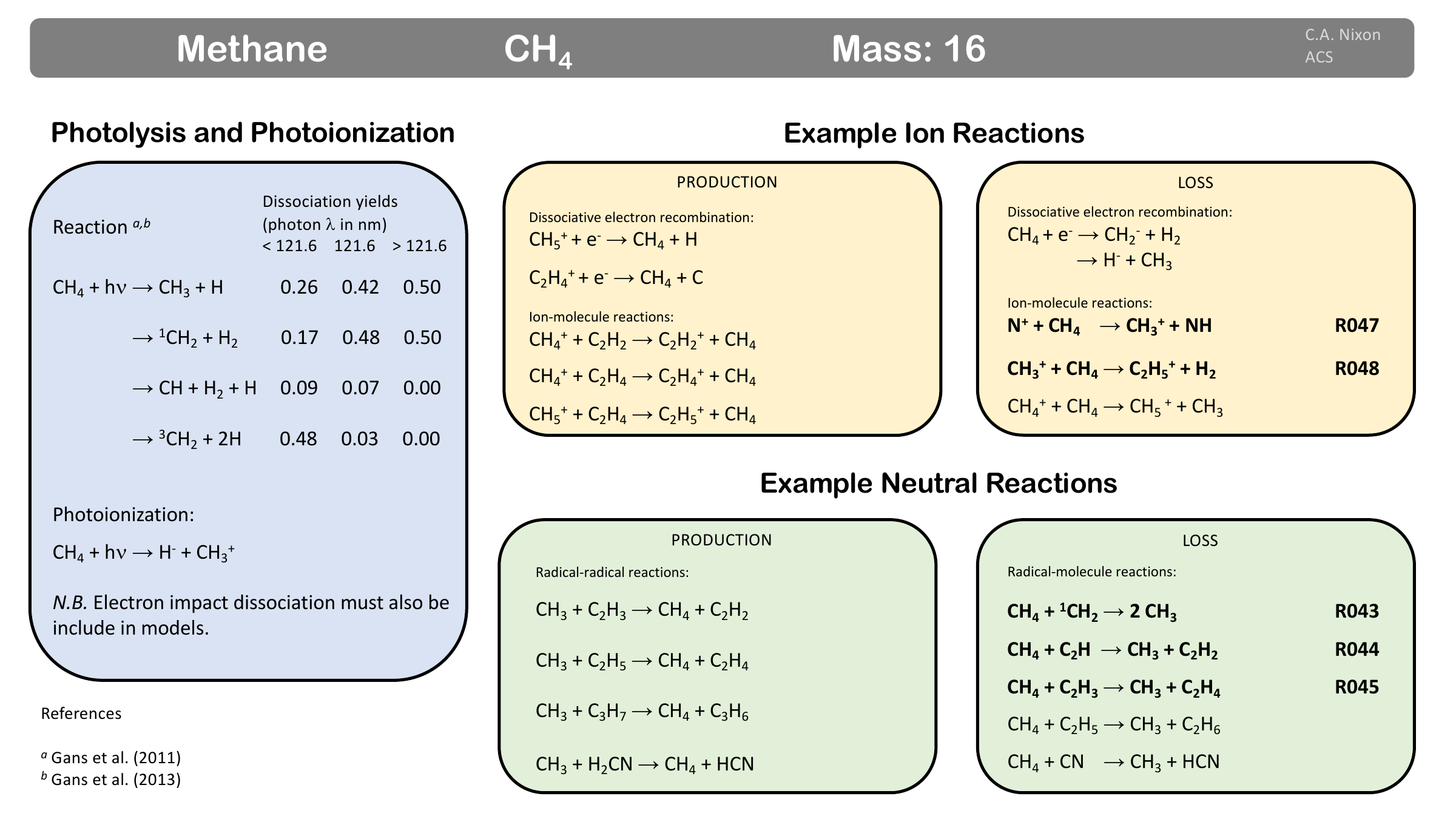}
  \caption{Methane production and loss pathways. Reactions numbered and shown in bold correspond to discussion in the text.}
  \label{mol:methane}
\end{figure}

{\em Profile:} Methane (CH$_4$) was the first molecule to be positively identified in Titan’s atmosphere, via visible and near-IR absorptions seen by Gerard Kuiper \citep{kuiper44}, using the 82 inch reflector at McDonald Observatory. We now know that methane is the basic ingredient enabling all of Titan's complex organic (i.e. carbon) chemistry, allowing reactions to proceed up to the creation of haze particles: some key reactions for methane are shown in Fig.~\ref{mol:methane}\cite{gans11, gans13}. 

Nevertheless, methane may be a gradually depleting resource since it is not permanently recycled, unless replenished by an as yet unidentified mechanism \cite{yung84, lorenz97a, nixon12b, mandt12, wong15}. Speculative mechanisms include: crustal destabilization leading to outgassing from methane clathrates,\cite{tobie06}
outgassing from cryovolcanism,\cite{fortes07, lopes13, sohl14}
and displacement from near-surface clathrate materials by condensed ethane\cite{choukroun12} - yet observational evidence for these processes remains inconclusive at best.

Methane’s vertical profile can be divided into three zones: (i) a tropospheric zone where the fractional abundance gradually decreases from 5.5\% at the surface to a minimum at the tropopause of around 1.4\%, due to reaching saturation at decreasing volume mixing ratios (VMRs) as the temperature decreases towards the tropopause (‘cold trap’); (ii) a relatively constant amount of ~1.4\% in the stratosphere, mesosphere and thermosphere; (iii) a gradually increasing mixing fraction above an altitude of ~800--850 km (the methane homopause) in the increasingly collisionless regime, due to the differing scale heights of different molecules \cite{strobel09}. It is presently uncertain whether methane has any variation with latitude on Titan, although a variation from $\sim$1.0\% to 1.5\% in the lower stratosphere has been reported \cite{lellouch14} based on infrared measurements by {\em Cassini} CIRS.\cite{lellouch14}

{\em Loss:} The photolysis of methane in the upper atmosphere leads to the formation of radicals including methyl (CH$_3$), methylidene (CH), and the carbene methylene ($\rm ^1CH_2$ or $\rm ^3CH_2$)\cite{vuitton19} which undergo a chain of reactions  to form all the hydrocarbons found in Titan's atmosphere. Note that as much as 75\% of methane photolysis above 700~km is due to the solar Lyman-$\alpha$ line at 121.6~nm.\cite{hebrard06} 

The fate of most methane is ultimately to form ethane via the addition of two methyl radicals (R \ref{reaction:ethane}) or via the creation of ethyl:

\begin{eqnarray}
{\rm ^1CH_2 + CH_4  }  & {\longrightarrow} & {\rm C_2H_5 + H  } 
\end{eqnarray}

\noindent
leading to permanent loss of methane. Methyl radicals are  produced either directly by primary photolysis, or by reaction of methane with radicals:\cite{berman82, fleurat02}

\begin{eqnarray}
{\rm ^1CH_2 + CH_4}  & {\longrightarrow} & {\rm 2 \:\: CH_3 } \\
{\rm C_2H + CH_4}  & {\longrightarrow} & {\rm CH_3 + C_2H_2 } \\
{\rm C_2H_3 + CH_4}  & {\longrightarrow} & {\rm CH_3 + C_2H_4 } 
\end{eqnarray}

Note that  the reaction of the ethynyl and vinyl radicals (amongst others) with methane is a catalytic destruction process, since the acetylene and ethylene generated are easily photolyzed back to the radical form where they can continue to destroy methane molecules. This process may be repeated hundreds of times before the radical catalysts are themselves lost to form higher hydrocarbons. This is the main source of methane depletion in Titan's atmosphere.\cite{vuitton19}

Methane fragments participate in ion chemistry, first being ionized by charge transfer:\cite{dutuit13}

\begin{eqnarray}
{\rm N_2^+ + CH_3 }  & {\longrightarrow} & {\rm CH_3^+ + N_2 } \\
{\rm N^+ + CH_4 }  & {\longrightarrow} & {\rm CH_3^+ + NH } 
\end{eqnarray}

\noindent
and then building up $\rm C_2H_x$ ions, e.g.:\cite{mcewan07}

\begin{eqnarray}
{\rm CH_3^+ + CH_4}  & {\longrightarrow} & {\rm C_2H_5^+ + H_2 }
\end{eqnarray}

{\em Future directions:} 
Although the chemistry of methane is perhaps one of the best-understood for any molecule on Titan, the most pressing questions remain the nature of its origin, and possible replenishment.\cite{nixon18}

%%%%%%%%%%%%%%%%%%%%%%%
\subsubsection{Acetylene}

\begin{figure}
\includegraphics[scale=0.4]{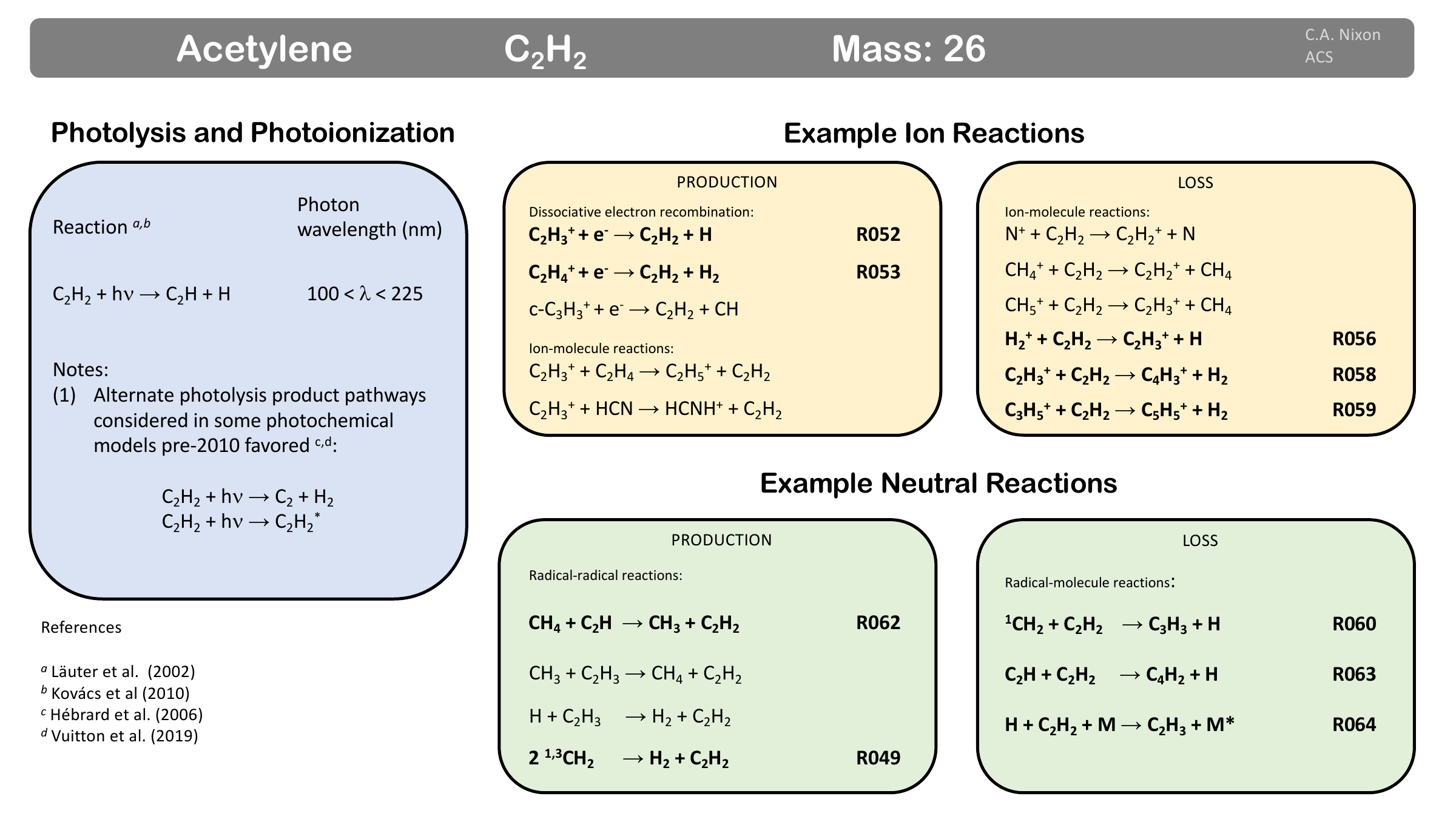}
  \caption{Acetylene production and loss pathways. Reactions numbered and shown in bold correspond to discussion in the text.}
  \label{mol:acetylene}
\end{figure}

Acetylene was the third molecule to be identified in Titan's atmosphere,\citep{gillett75} following the detection of methane and ethane, when Gillett observed its 13~\micron\ band in mid-infrared spectroscopy with 2~m and 4~m telescopes at Kitt Peak Observatory. Important reactions for acetylene are shown in Fig.~\ref{mol:acetylene}.\cite{lauter02,kovacs10,hebrard06,vuitton19}

{\em Production:} Acetylene may be produced either directly or indirectly from photolysis products of methane. In a direct process\cite{jasper07}:

\begin{eqnarray}
{\rm ^{3}CH_2 + \: ^{3}CH_2 }  & {\longrightarrow} & {\rm C_2H_2 + H_2}
\end{eqnarray}

\noindent or indirectly via \ethylene\ and other species from photolysis\cite{holland97} (Fig.~\ref{mol:ethylene}):

\begin{eqnarray}
{\rm C_2H_4} + h\nu   & {\longrightarrow} & {\rm C_2H_2 + H_2} \label{eq:ethylene-photo1} \\
& {\longrightarrow} & {\rm C_2H_2 + 2H} \label{eq:ethylene-photo2}
\end{eqnarray}

Dissociative electron recombination is another production pathway, e.g.:\cite{ehlerding04, kalhori02}

\begin{eqnarray}
{\rm C_2H_3^+} + {\rm e^-}  & {\longrightarrow} & {\rm C_2H_2 + H{^\cdotp}} \\
{\rm C_2H_4^+} + {\rm e^-}  & {\longrightarrow} & {\rm C_2H_2 + 2H / H_2}
\end{eqnarray}

{\em Loss:} In the ionosphere, photolysis of acetylene, and electron transfer to N$_2^+$ produces $\rm C_2H_2^+$\cite{dutuit13}, which reacts with neutrals to build heavier ions\cite{mcewan07}, e.g.:

\begin{eqnarray}
{\rm C_2H_2^+ + CH_4}  & {\longrightarrow} & {\rm C_3H_4^+ +H_2} \\
{\rm C_2H_2^+ + CH_4}  & {\longrightarrow} & {\rm C_3H_5^+ +H}  
\end{eqnarray}

\noindent which may (dissociatively) recombine with $e^-$ to form neutral $\rm C_3H_x$ species\cite{janev04,chabot13}. Likewise, proton transfer to neutral acetylene leads to $\rm C_2H_3^+$, which can also form $\rm C_3$ species\cite{mcewan07}:

\begin{eqnarray}
{\rm H_2^+ + C_2H_2}  & {\longrightarrow} & {\rm C_2H_3^+ +H} \\
{\rm C_2H_3^+ + CH_4}  & {\longrightarrow} & {\rm C_3H_5^+ +H_2}
\end{eqnarray}

\noindent
Finally, neutral \acet\ in the ionosphere may combine with other ions to build heavier species\cite{mcewan07}, e.g.:

\begin{eqnarray}
{\rm C_2H_3^+ + C_2H_2 }  & {\longrightarrow} & {\rm C_4H_3^+ + H_2} \\ 
{\rm C_3H_5^+ + C_2H_2 }  & {\longrightarrow} & {\rm C_5H_5^+ + H_2} 
\end{eqnarray}

In the neutral atmosphere, acetylene is lost by reaction with methylene forming propargyl\cite{gannon10a, gannon10b, gannon10c}:

\begin{eqnarray}
{\rm ^1CH_2 + C_2H_2}  & {\longrightarrow} & {\rm C_3H_3 + H}
\end{eqnarray}

Acetylene absorbs photons to longer wavelengths ($\sim$230 nm\cite{lavvas08a}) than methane, so its photolysis continues into the stratosphere (see Fig.~\ref{mol:acetylene}). This produces ethynyl (C$_2$H) and carbyne (C$_2$) which are potent means of methane depletion via hydrogen abstraction:\cite{baulch05}

\begin{eqnarray}
{\rm C_2 + CH_4} & {\longrightarrow} & {\rm C_2H + CH_3} \\
{\rm C_2H + CH_4}  & {\longrightarrow} & {\rm C_2H_2 + CH_3} 
\end{eqnarray}

\noindent
The acetylene produced by this reaction is recycled back to ethynyl by photolysis, and thereby each acetylene/ethynyl may cause the loss of hundreds of methane molecules before being lost itself to another reaction pathway such as:\cite{chastaing98}

\begin{eqnarray}
\label{eq:diacet-prod}
{\rm C_2H + C_2H_2}  & {\longrightarrow} & {\rm C_4H_2 + H}
\end{eqnarray}

\noindent Catalytic destruction of methane in this way is the principle means of methane depletion in Titan's atmosphere \cite{yung84, lavvas08a, vuitton19}.
In the lower atmosphere ($z < 500$~km), the dominant loss process for acetylene is conversion to ethylene by a two-step reaction with atomic hydrogen:\cite{vuitton12, vuitton19}

\begin{eqnarray}
{\rm C_2H_2 + H + M}  & {\longrightarrow} & {\rm C_2H_3 + M^*} \\
{\rm C_2H_3 + H + M}  & {\longrightarrow} & {\rm C_2H_4 + M^*} 
\end{eqnarray}

{\em Future directions:} Acetylene has been detected on Titan's surface \cite{niemann05, singh16}, and is likely to be present in the northern lakes and seas \cite{cordier09, cordier12}. Future investigation by the mass spectrometer (DrAMS) instrument of {\em Dragonfly}\cite{barnes21} will further refine the surface and near-surface abundance. 

\acet\ was one of the first molecules investigated to form a co-crystal \cite{cable18, cable19} at Titan surface temperatures, an organized co-condensate of two or more chemical species. The validity of multiple co-crystal types has since been established, but requires further laboratory work to determine the full parameter space of possible crystalline types. 

%%%%%%%%%%%%%%%%%%%%%%%
\subsubsection{	 Ethylene}

\begin{figure}
\includegraphics[scale=0.4]{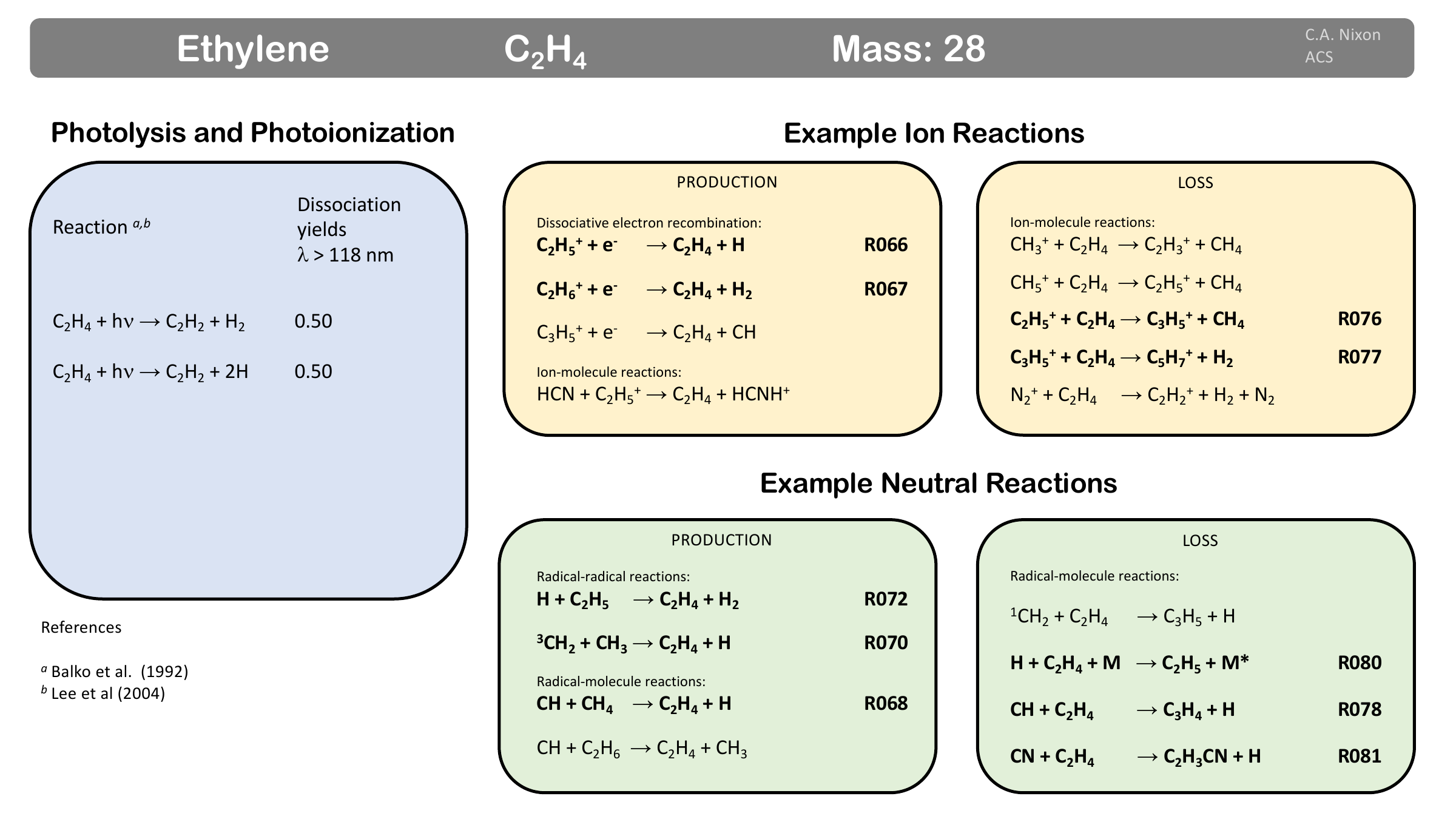}
  \caption{Ethylene production and loss pathways. Reactions numbered and shown in bold correspond to discussion in the text.}
  \label{mol:ethylene}
\end{figure}

Ethylene (C$_2$H$_4$) was discovered by infrared spectroscopy at the same time as acetylene \cite{gillett75}. The vertical profile of ethylene exhibited a surprising trend to decrease in abundance upwards in the lower stratosphere early in the {\em Cassini} mission\cite{vinatier07a,vinatier10a}, although this faded at later seasons \cite{teanby12, teanby19}. 

Ethylene is a crucial, two-carbon neutral molecule that provides a stepping stone from methane to higher hydrocarbons (Fig.~\ref{mol:ethylene})\cite{balko92,lee04}. Ethylene is remarkable in being one of the few molecules (along with \hydrogen\ and \nitrogen ) that does not condense at the tropopause, and therefore persists in significant quantities into the troposphere. 

{\em Production:} Ethylene may be produced in the ionosphere by dissociative recombination of heavier ions with an electron\cite{mclain04, geppert04, janev04}:

\begin{eqnarray}
{\rm C_2H_5^+ + e^- }  & {\longrightarrow} & {\rm C_2H_4+ H} \\ 
{\rm C_2H_6^+ + e^- }  & {\longrightarrow} & {\rm C_2H_4+ H_2}
\end{eqnarray}

In the neutral atmosphere, ethylene is largely formed through reactions between methane and its derived radicals, or between radicals\cite{canosa97,hebrard06,jasper07,lavvas08a, gannon10a, gannon10b}:

\begin{eqnarray}
{\rm CH + CH_4}  & {\longrightarrow} & {\rm C_2H_4 + H} \\
{\rm ^1CH_2 + CH_3}  & {\longrightarrow} & {\rm C_2H_4 + H} \\ 
{\rm ^3CH_2 + CH_3}  & {\longrightarrow} & {\rm C_2H_4 + H} 
\end{eqnarray}

At lower altitudes, production through the $\rm C_2H_5$ intermediary is also important\cite{baulch05}:

\begin{eqnarray}
{\rm ^1CH_2 + CH_4}  & {\longrightarrow} & {\rm C_2H_5 + H} \\ 
{\rm H + C_2H_5}  & {\longrightarrow} & {\rm C_2H_4 + H_2} 
\end{eqnarray}

{\em Loss:} In the ionosphere, the ethylenium ion - produced from ethylene photoionization - can be lost in various reactions with neutrals\cite{mcewan07}:

\begin{eqnarray}
{\rm C_2H_4^+ + H}  & {\longrightarrow} & {\rm C_2H_3^+ + H_2} \\
{\rm C_2H_4^+ + C_2H_2}  & {\longrightarrow} & {\rm c{\text -}C_3H_3^+ + CH_3} \\
{\rm C_2H_4^+ + C_2H_4}  & {\longrightarrow} & {\rm C_3H_5^+ + CH_3} 
\end{eqnarray}

\noindent
and ethylene can be lost during the formation of heavier ions\cite{mcewan07}:

\begin{eqnarray}
{\rm C_2H_5^+ + C_2H_4}  & {\longrightarrow} & {\rm C_3H_5^+ + CH_4} \\
{\rm C_3H_5^+ + C_2H_4}  & {\longrightarrow} & {\rm C_5H_7^+ + H_2} 
\end{eqnarray}

Photolysis of ethylene leads to acetylene ({R}\ref{eq:ethylene-photo1} and {R}\ref{eq:ethylene-photo2}). 
Insertion/addition reactions onto ethylene by CH, for example, can lead to higher hydrocarbons\cite{canosa97,mckee03,goulay09}:

\begin{eqnarray}
{\rm C_2H_4 + CH}  & {\longrightarrow} & {\rm CH_3CCH + H} \\
{\rm C_2H_4 + CH}  & {\longrightarrow} & {\rm CH_2CCH_2 + H} 
\label{eq:ethylene-ch}
\end{eqnarray}

Below 500 km, loss via H-addition becomes important\cite{vuitton12}:

\begin{eqnarray}
{\rm  H + C_2H_4 + M}  & {\longrightarrow} & {\rm C_2H_5 + M} 
\end{eqnarray}

The CN radical can also substitute onto ethylene to create vinyl cyanide (also known as acrylonitrile, or propenitrile):\cite{loison15,sims93,gannon07}

\begin{eqnarray}
{\rm C_2H_4 + CN}  & {\longrightarrow} & {\rm C_2H_3CN + H} 
\end{eqnarray}

{\em Future directions:} Ethylene is the simplest alkene, the family of hydrocarbons having a double C$=$C bond. Photolysis or other breaking of the alkene C$=$C bond leads to radicals which rapidly react, leading to formation of polymers. The role of polymer formation in Titan's atmosphere is incompletely understood, but is likely to be an important process in the formation of haze particles.

%%%%%%%%%%%%%%%%%%%%%%%
\subsubsection{ Ethane}

\begin{figure}
\includegraphics[scale=0.4]{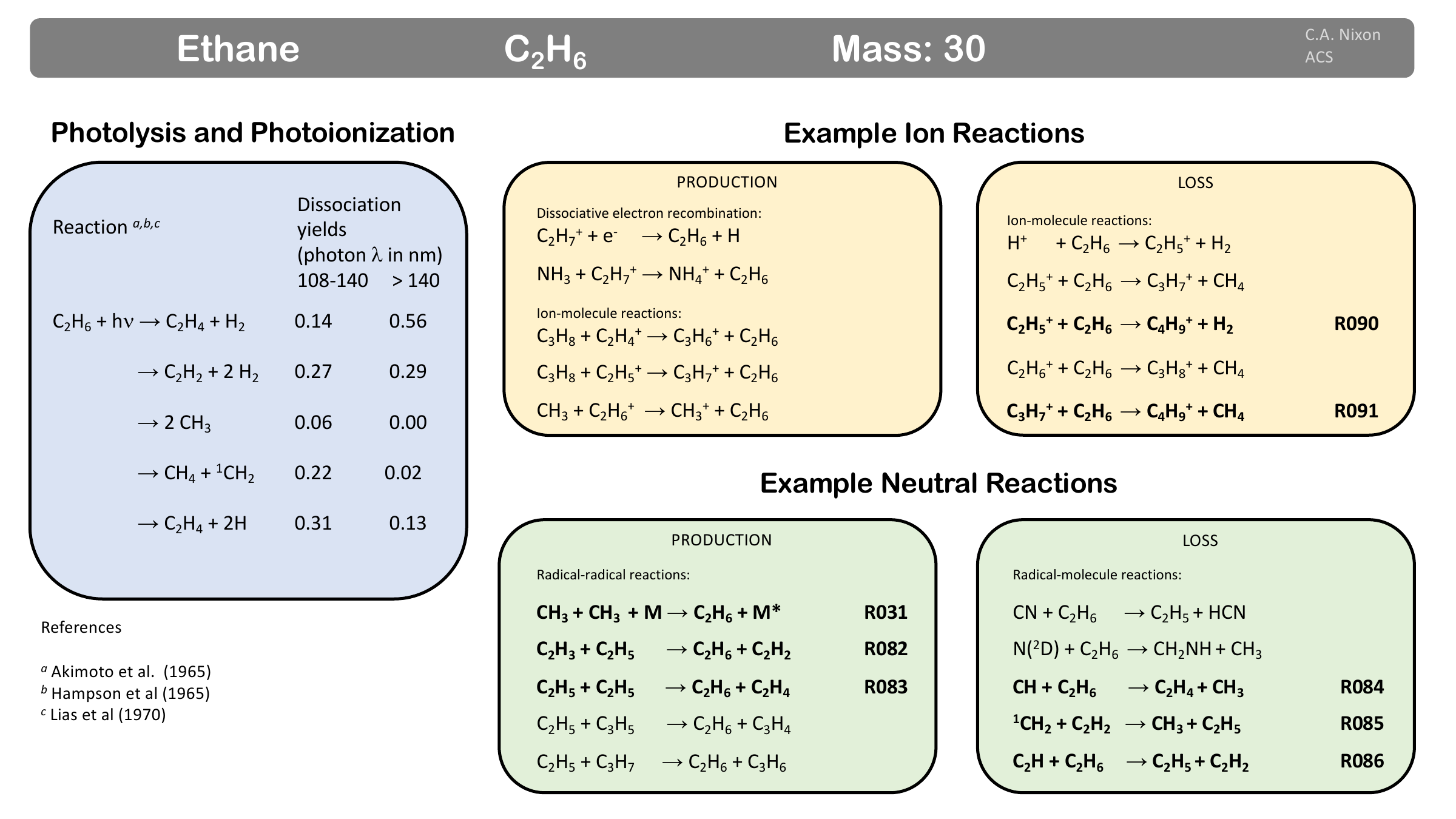}
  \caption{Ethane production and loss pathways. Reactions numbered and shown in bold correspond to discussion in the text.}
  \label{mol:ethane}
\end{figure}

Ethane was the second molecule to be discovered in Titan's atmosphere, via its strong $\nu_9$ band at 12 $\mu m$ \cite{gillett73} seen by Gillett with the 60-inch telescope on Mount Lemmon. Ethane forms one of the primary trace gases in Titan's atmosphere with concentrations greater than 1 ppm in the stratosphere \cite{coustenis89a, coustenis07, coustenis10}, and is the primary sink for methane loss \cite{yung84, wilson04}. Due to this observation, a global deep ethane ocean was originally predicted \cite{lunine83, flasar83} but later proved not to be the case \cite{muhleman90, smith96}. 

The liquid hydrocarbon bodies eventually detected on Titan's surface\cite{stofan07} have a measured ethane content that varies between different seas and is generally less than the methane fraction \cite{legall16, mastrogiuseppe18b}, except for the southern lake {\em Ontario Lacus} \cite{brown08,mastrogiuseppe18a}. Ethane may form co-crystals in Titan lakes with other organics, such as benzene \cite{cable14}. Ethane has also been implicated in displacing methane from clathrate hydrate, allowing for a partial resupply mechanism of methane to the atmosphere \cite{choukroun12}, which is otherwise continuously lost by chemistry \cite{yung84,wilson09}.

{\em Production:} Ethane is primarily produced by the addition of two methyl reactions ({R}\ref{reaction:ethane}) but also reforms from the ethyl radical\cite{laufer04}:

\begin{eqnarray}
{\rm C_2H_3 + C_2H_5}  & {\longrightarrow} & {\rm C_2H_6 + C_2H_2} \\
{\rm C_2H_5 + C_2H_5}  & {\longrightarrow} & {\rm C_2H_6 + C_2H_4 } 
\end{eqnarray}

Ethane also participates in ion reactions as shown in Fig.~\ref{mol:ethane}\cite{akimoto65, hampson65a, lias70}.

{\em Loss:} Ethane can be lost through photolysis back to $2\: {\rm CH_3}$, or to stable molecules such as ethylene and acetylene with loss of hydrogen (see Fig.~\ref{mol:ethane}). Ethane can also be attacked by reactive radicals such as methylene, methylidene and ethynyl (from acetylene photolysis)\cite{baulch05, lavvas08a, canosa97, mckee03}:

\begin{eqnarray}
{\rm  CH + C_2H_6}  & {\longrightarrow} & {\rm CH_3 + C_2H_4} \\
{\rm ^1CH_2 + C_2H_6}  & {\longrightarrow} & {\rm CH_3 + C_2H_5} \\
{\rm C_2H + C_2H_6}  & {\longrightarrow} & {\rm C_2H_2 + C_2H_5 } 
\end{eqnarray}

Ethane may form heavier ions through reactions such as\cite{mcewan07}:

\begin{eqnarray}
{\rm C_2H_2^+ + C_2H_6 }  & {\longrightarrow} & {\rm C_3H_5^+ + CH_3} \\
{\rm C_2H_2^+ + C_2H_6 }  & {\longrightarrow} & {\rm C_4H_7^+ + H} \\
{\rm C_2H_3^+ + C_2H_6 }  & {\longrightarrow} & {\rm C_4H_7^+ + H_2} \\
{\rm C_2H_5^+ + C_2H_6 }  & {\longrightarrow} & {\rm C_4H_9^+ + H_2} \\
{\rm C_3H_7^+ + C_2H_6 }  & {\longrightarrow} & {\rm C_4H_9^+ + CH_4} 
\end{eqnarray}

{\em Future directions: } Large amounts of ethane are thought to condense in Titan's lower stratosphere, and form a significant fraction of Titan's lakes and seas \cite{cordier09,cordier13}. Ethane was implicated in the formation of a vast north polar cloud seen during northern winter in 2005 by {\em Cassini} VIMS \cite{griffith06b}, although other interpretations have suggested that this cloud is condensed methane.\cite{anderson14} While the basic chemistry of ethane is well-understood, an improved understanding of its condensation - especially co-condensation with other gases - will be crucial to a more accurate interpretation of Titan's meteorology.

%%%%%%%%%%%%%%%%%%%%%%%
\subsubsection{	 Cyclopropenylidene}

\begin{figure}
\includegraphics[scale=0.4]{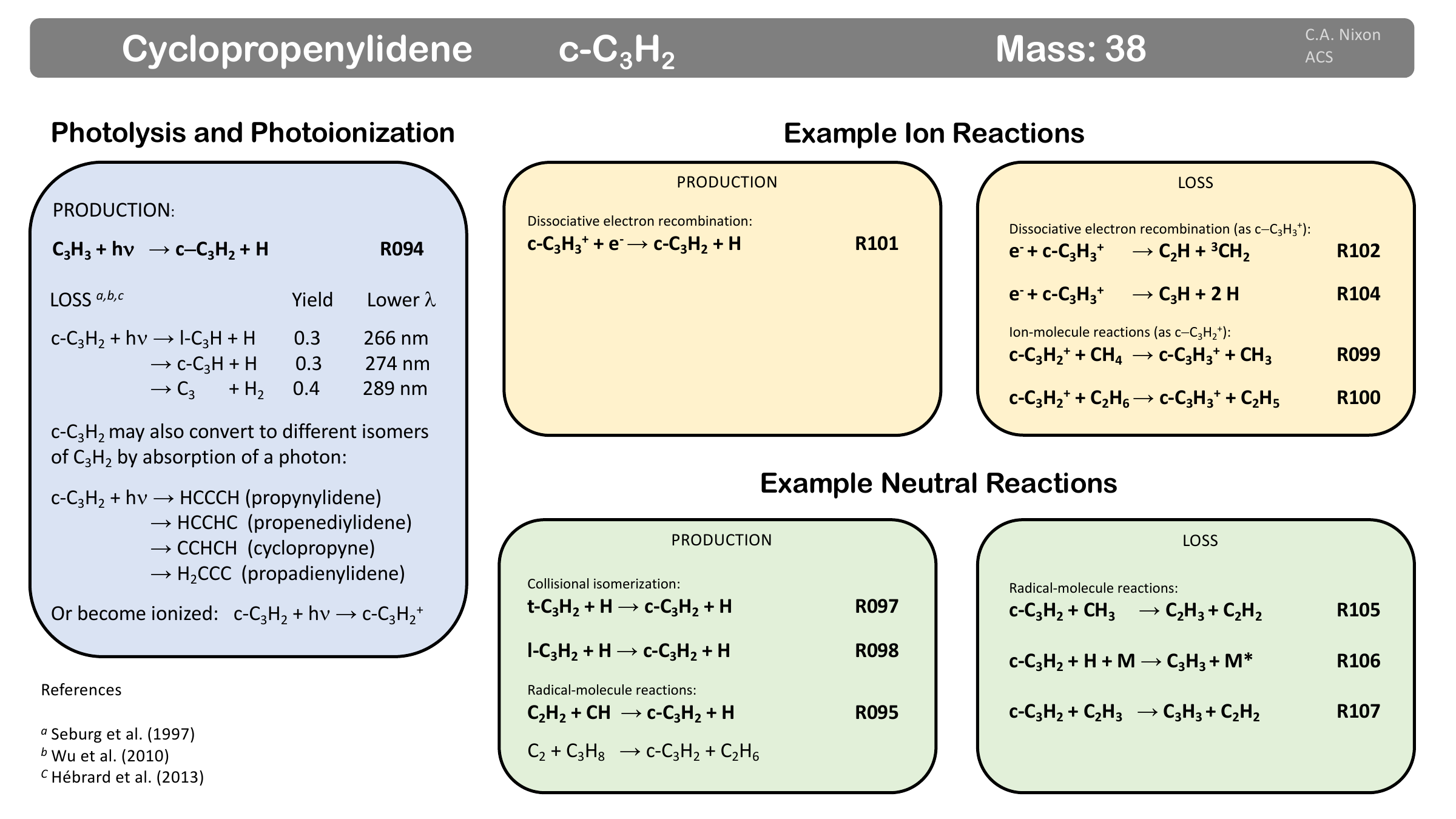}
  \caption{Cyclopropenylidene production and loss pathways. Reactions numbered and shown in bold correspond to discussion in the text.}
  \label{mol:cpld}
\end{figure}

Cyclopropenylidene (\cpld ) is the first carbene (a molecule having two unbonded, self-paired valence electrons from a carbon atom) and the second cyclic molecule to be found in Titan's atmosphere (after benzene). \cpld\ was detected using millimeter wavelength astronomy with ALMA \cite{nixon20}, the third molecule whose first detection on Titan was achieved with with this telescope. 

{\em Production:} In the upper atmosphere, production of a precursor, the cyclopropenyl cation ($\rm c{\text -}C_3H_3^+$) is thought to proceed by\cite{vuitton19}:

\begin{eqnarray}
{\rm C_2H_5^+ + C_2H_2}  & {\longrightarrow} & {\rm c{\text -}C_3H_3^+ + CH_4} \\
{\rm C_2H_4^+ + C_2H_2}  & {\longrightarrow} & {\rm c{\text -}C_3H_3^+ + CH_3 } 
\end{eqnarray}

\noindent which then recombines with $e^-$ to produce \cpld\ \cite{thaddeus85} and H. 

Other possible pathways include photolysis of propargyl:

\begin{eqnarray}
{\rm C_3H_3 + h\nu}  & {\longrightarrow} & {\rm c{\text -}C_3H_2 + H}
\end{eqnarray}

\noindent
In the neutral atmosphere the dominant production pathways may include CH addition to acetyene \cite{walch95, guadagnini98} (see Fig.~\ref{mol:cpld})\cite{seburg97,wu10}:

\begin{eqnarray}
{\rm CH + C_2H_2 }  & {\longrightarrow} & {\rm c{\text -}C_3H_2 + H}
\end{eqnarray}

\noindent and below 600 km:\cite{willacy22}

\begin{eqnarray}
{\rm H_2} + {\rm c{\text -}C_3H} & \longrightarrow &  {\rm c{\text -}C_3H_2} + {\rm H} 
\end{eqnarray}

\cpld\ can also result from collisional isomerization from its isomers propynylidene (t-C$_3$H$_2$) and propadienylidene (l-C$_3$H$_2$):\cite{willacy22}

\begin{eqnarray}
{\rm H + t{\text -}C_3H_2 }  & {\longrightarrow} & {\rm H + c{\text -}C_3H_2 } \\
{\rm H + l{\text -}C_3H_2 }  & {\longrightarrow} & {\rm H + c{\text -}C_3H_2 }
\end{eqnarray}

{\em Loss:} Once ionized, the $\rm c{\text -}C_3H_2^+$ ion may abstract hydrogen from neutrals to form the c-$\rm C_3H_3^+$  ion:\cite{prodnuk92,mcewan07}

\begin{eqnarray}
{\rm  c{\text -}C_3H_2^+  + CH_4 }      & {\longrightarrow} & {\rm C_3H_3^+ +  CH_3 }  \\
{\rm  c{\text -}C_3H_2^+  + C_2H_6 }  & {\longrightarrow} & {\rm  C_3H_3^+ + C_2H_5 }
\end{eqnarray}

\noindent
which in turn may dissociatively recombine with an electron, returning to \cpld , or splinter into smaller acyclic fragments:\cite{janev04, poterya05}

\begin{eqnarray}
{\rm  c{\text -}C_3H_3^+ } + e^-  & {\longrightarrow} & {\rm c{\text -}C_3H_2 + H } \\
{\rm  c{\text -}C_3H_3^+ } + e^-  & {\longrightarrow} & {\rm C_2H + {^3}CH_2 } \\
{\rm  c{\text -}C_3H_3^+ } + e^-  & {\longrightarrow} & {\rm C_2 +  CH_3 } \\
{\rm  c{\text -}C_3H_3^+ } + e^-  & {\longrightarrow} & {\rm C_3H + 2H  }
\end{eqnarray}

In the neutral atmosphere above 600 km, \cpld\ photodissociates to form c-C$_3$H, l-C$_3$H, and C$_3$. It is also lost by reaction with CH$_3$ to form acylic species:\cite{willacy22}

\begin{eqnarray}
{\rm c{\text -}C_3H_2} + {\rm CH_3} & {\longrightarrow} & {\rm C_2H_2} + {\rm C_2H_3} 
\end{eqnarray}

Below 600 km, these pathways continue to be significant, but there is additional \cpld\ loss via:\cite{willacy22}

\begin{eqnarray}
{\rm c{\text -}C_3H_2} + {\rm H + M } & {\longrightarrow} & {\rm C_3H_3 + M}  \\
{\rm c{\text -}C_3H_2} + {\rm C_2H_3} & {\longrightarrow} & {\rm C_3H_3} + {\rm C_2H_2}
\end{eqnarray}

{\em Future work:}  Chemical pathways leading to and from \cpld\ in Titan's atmosphere remain to be explored, especially given the multiple possible structures for $\rm C_3H_2$, including propynylidene ($\rm HC_3H, t{\text -}C_3H_2$ ), propadienylidene ($\rm H_2CCC, i{\text -}C_3H_2$), cyclopropyne ($\rm CCHCH$) and propenediylidene ($\rm HCCHC$)   \cite{seburg97, wu10}. Experimental and theoretical work on reaction pathways, rates and branching rates during photolysis will greatly help to clarify the production, loss and stability of cyclopropenylidene.

Several isomers of $\rm C_3H_2$ have been detected in space, including H$_2$CCC \cite{cernicharo91}, which should prompt further astronomical observations to determine if these isomers also exist in Titan's upper atmosphere. Intriguingly, the CHCCH form has been shown to dimerize to form {\em para}-benzyne \cite{walch95}, discussed in a later section.

%%%%%%%%%%%%%%%%%%%%%%%
\subsubsection{ Propyne}

\begin{figure}
\includegraphics[scale=0.4]{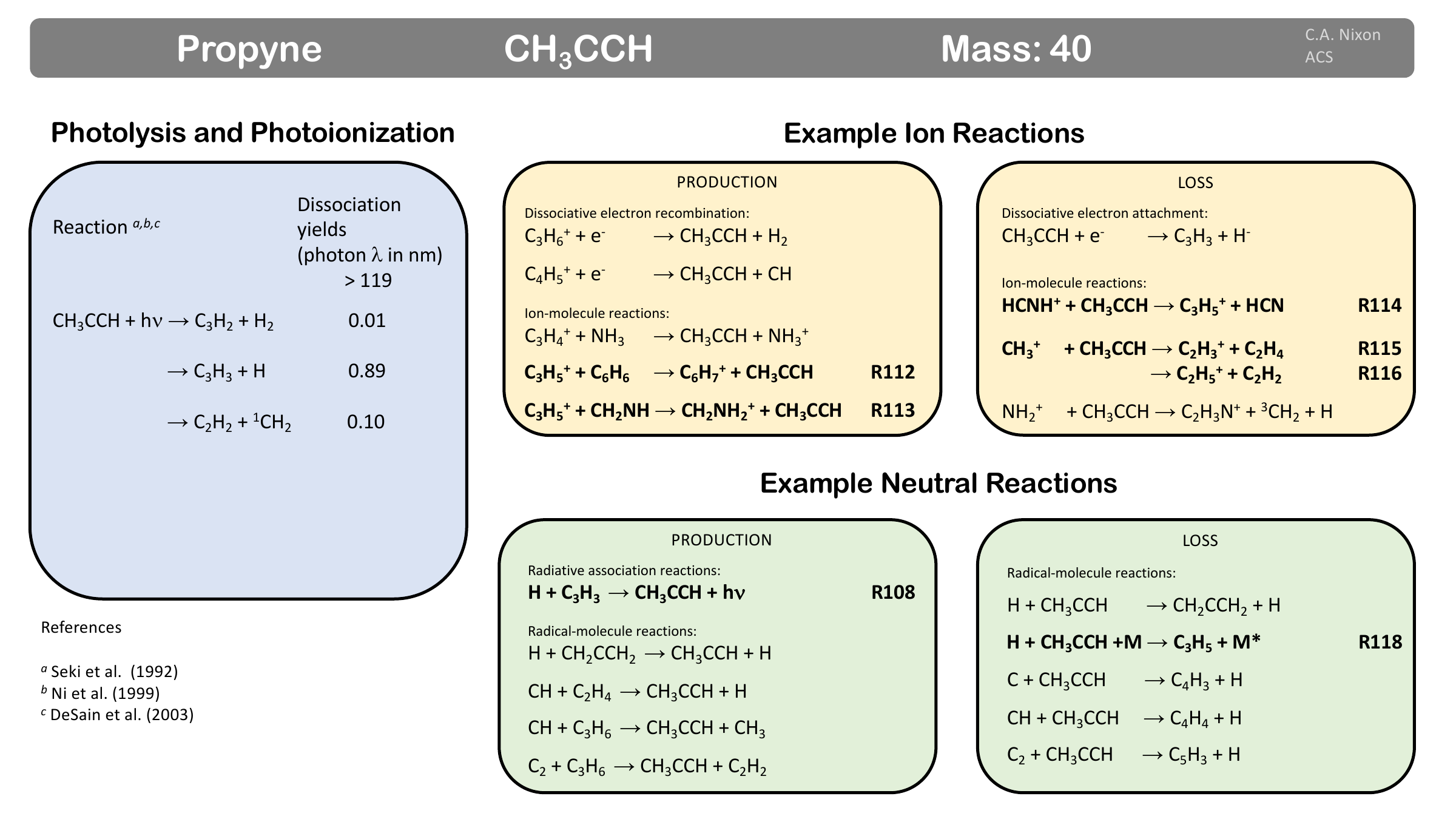}
  \caption{Propyne production and loss pathways. Reactions numbered and shown in bold correspond to discussion in the text.}
  \label{mol:propyne}
\end{figure}

The propyne (\propyne , methyl acetylene) isomer of  ($\rm C_3H_4$) was first detected in the infrared by the {\it Voyager} Infrared Interferometer Spectrometer (IRIS)\cite{hanel80} via long wavelength infrared emission bands at 328 and 633 \cm \cite{maguire81}.

{\em Production:} Propyne and its symmetric isomer propadiene (\propadiene , allene) are produced by addition of CH into ethylene (Eq.~\ref{eq:ethylene-ch}, Fig.~\ref{mol:propyne})\cite{seki92,ni99,desain03}, but also through H-addition to propargyl\cite{chastaing01}: 

\begin{eqnarray}
{\rm H + C_3H_3 }  & {\longrightarrow} & {\rm CH_3CCH + h\nu },
\end{eqnarray} 

\noindent
or through dissociation of \propene\ (see Fig.~\ref{mol:propene})\cite{collin88}:

\begin{eqnarray}
{\rm C_3H_6  + h\nu }  & {\longrightarrow} & {\rm CH_3CCH + 2H }
\end{eqnarray} 

In the ionosphere, the C$_3$H$_5^+$ ion is a precursor to C$_3$H$_4$, and produced via:\cite{vuitton07}

\begin{eqnarray}
{\rm C_2H_3^+ + CH_4 }  & {\longrightarrow} & {\rm C_3H_5^+ + H_2 } \\ 
{\rm C_2H_5^+ + C_2H_4 }  & {\longrightarrow} & {\rm C_3H_5^+ + CH_4 } 
\end{eqnarray}

\noindent
which then forms C$_3$H$_4$ by proton transfer:\cite{houriet78,edwards08}

\begin{eqnarray}
{\rm C_3H_5^+ + C_6H_6 }  & {\longrightarrow} & {\rm C_6H_7^+ + CH_3CCH } \\ 
{\rm C_3H_5^+ + CH_2NH }  & {\longrightarrow} & {\rm CH_2NH_2^+ + CH_3CCH } 
\end{eqnarray}

{\em Loss:} In the ionosphere, propyne is lost through ion reactions such as\cite{mcewan07}:

\begin{eqnarray}
{\rm HCNH^+ + CH_3CCH }  & {\longrightarrow} & {\rm C_3H_5^+ + HCN} \\ 
{\rm CH_3^+ + CH_3CCH  }  & {\longrightarrow} & {\rm C_2H_3^+ + C_2H_4} \\ 
				 & {\longrightarrow} & {\rm C_2H_5^+ + C_2H_2} 
\end{eqnarray} 

Propyne may also be lost by photolysis in the upper atmosphere\cite{fahr96}:

\begin{eqnarray}
{\rm CH_3CCH + h\nu }  & {\longrightarrow} & {\rm C_3H_3 + H},
\end{eqnarray} 

and to three-body reactions\cite{vuitton19}:

\begin{eqnarray}
{\rm CH_3CCH + H + M }  & {\longrightarrow} & {\rm C_3H_5 + M},
\end{eqnarray} 

{\em Future work:} Many reactions forming or depleting $\rm C_3H_4$ have uncertain branching ratios between \propyne\ and its isomer \propadiene . Further work is needed to improve knowledge of these quantities. Collisional interconversion between the two isomers may be mediated by atomic hydrogen\cite{lic15}, so accurate measurement of both isomers may be a way to provide constraint on the abundance of otherwise short-lived and difficult to measure H atom.

%%%%%%%%%%%%%%%%%%%%%%%
\subsubsection{ Propadiene}

\begin{figure}
\includegraphics[scale=0.4]{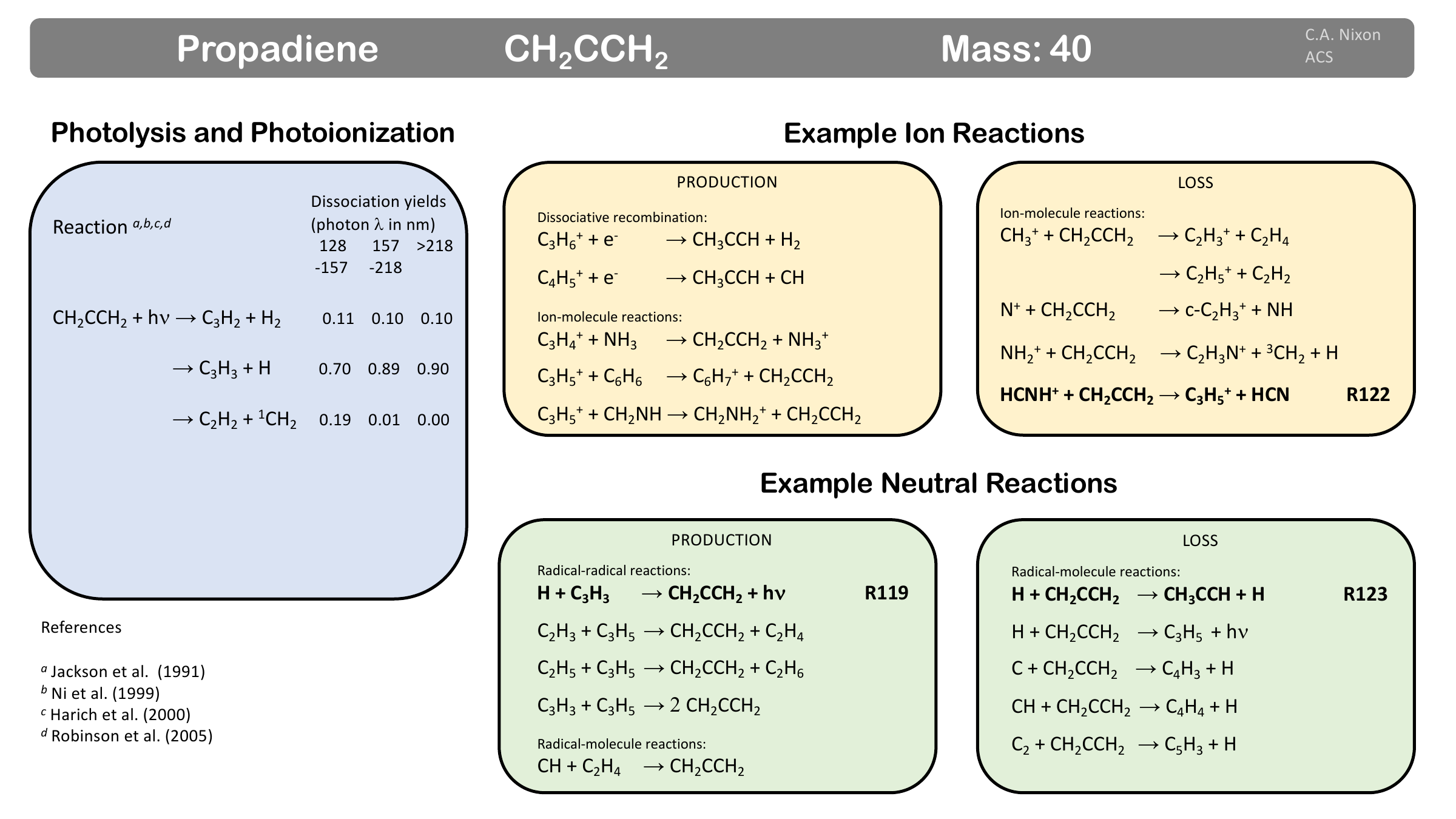}
  \caption{Propadiene production and loss pathways. Reactions numbered and shown in bold correspond to discussion in the text.}
  \label{mol:propadiene}
\end{figure}

Propadiene is a less abundant and less thermodynamically stable isomer of C$_3$H$_4$, which is more abundant in Titan's atmosphere in the form of propyne. Propadiene (\propadiene ) was detected in Titan's atmosphere using high-resolution ground-based spectroscopy at NASA's Infrared Telescope Facility (IRTF) with the Texas Echelon Cross Echelle Spectrograph (TEXES) instrument via its $\nu_{10}$ band at $\sim$845~\cm\ \cite{lombardo19c}.

{\em Production:} Like propyne (\propyne ), propadiene is produced in the upper atmosphere by CH addition to ethylene (Eq.~\ref{eq:ethylene-ch},
Fig.~\ref{mol:propadiene})\cite{jackson91,ni99,harich00,robinson05}, 
and by H addition to \propargyl :\cite{chastaing01}

\begin{eqnarray}
{\rm H + C_3H_3 }  & {\longrightarrow} & {\rm CH_2CCH_2 + h\nu } 
\end{eqnarray} 

Lower in the atmosphere, where propene is more plentiful, it can be photodissociated to produce propadiene (see Fig.~\ref{mol:propene})\cite{gierczak88}:

\begin{eqnarray}
{\rm C_3H_6 + h\nu  }  & {\longrightarrow} & {\rm CH_2CCH_2 + H_2 } 
\end{eqnarray} 

Ion formation pathways of \propadiene\ are less certain, but may follow similar channels as propyne, with branching ratios that are currently uncertain.

{\em Loss:} Propadiene is lost to direct photolysis\cite{ni99}:

\begin{eqnarray}
{\rm CH_2CCH_2 + h\nu }  & {\longrightarrow} & {\rm C_3H_3 + H} 
\end{eqnarray} 

\noindent
to ion reactions such as\cite{vuitton19}:

\begin{eqnarray} 
{\rm HCNH^+ + CH_2CCH_2  }  & {\longrightarrow} & {\rm C_3H_5^+ + HCN} 
\end{eqnarray} 

\noindent or by collision with H to form propyne (R\ref{eq:ccchhhh})\cite{vuitton19}.

\begin{eqnarray}
{\rm CH_2CCH_2 + H}  & {\longrightarrow} & {\rm CH_3C_2H + H} 
\label{eq:ccchhhh}
\end{eqnarray} 

{\em Future directions: } As with propyne, the branching ratios of reactions implicated in the formation of $\rm C_3H_4$ isomers remain uncertain, so  further experimental and theoretical  work is required. Accurate measurement of both propyne and propadiene is a possible means to indirectly inferring the abundance of atomic hydrogen\cite{lic15} in the lower atmosphere, in the absence of direct, in situ measurements.

%%%%%%%%%%%%%%%%%%%%%%%
\subsubsection{	 Propene}

\begin{figure}
\includegraphics[scale=0.4]{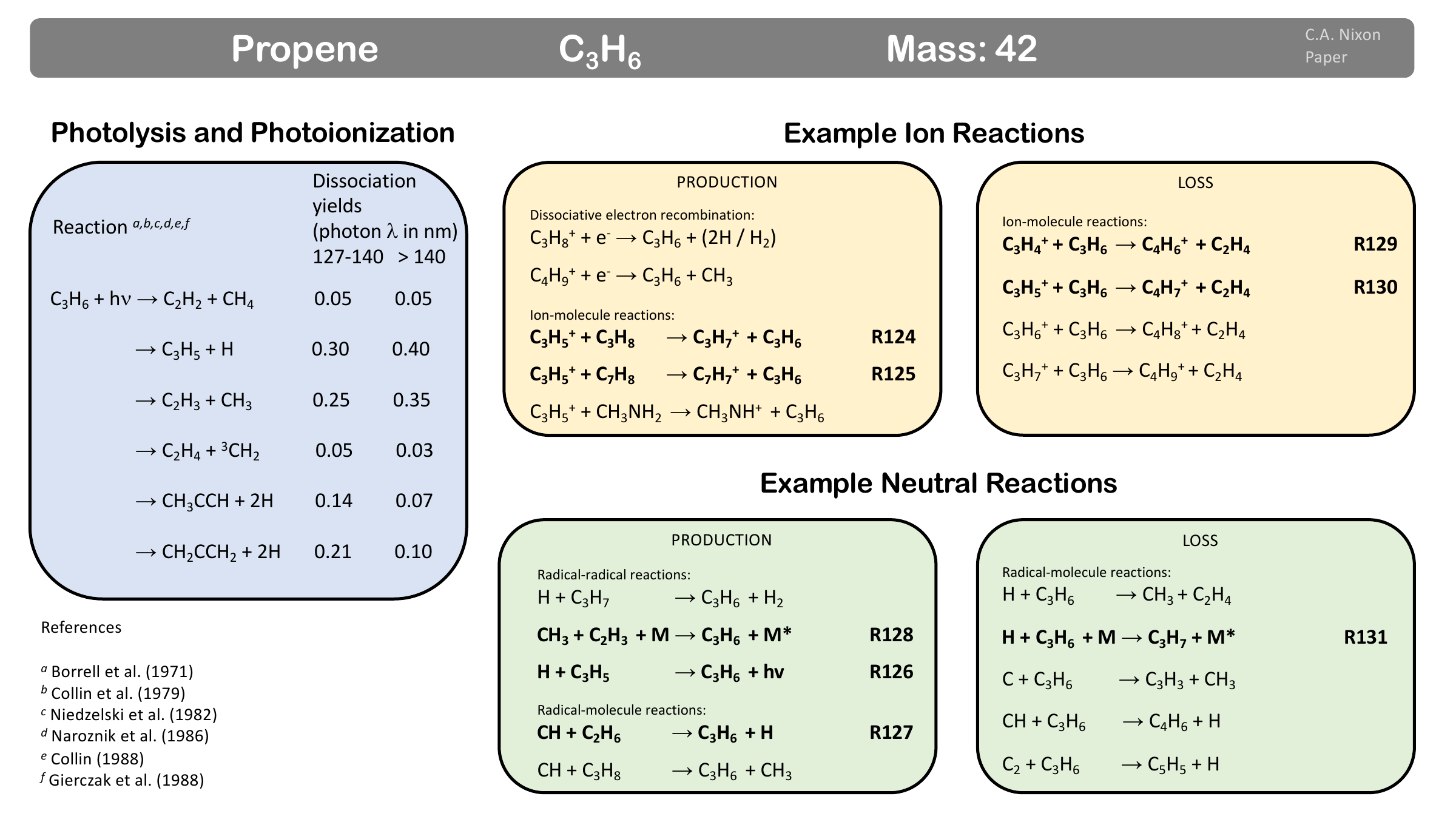}
  \caption{Propene production and loss pathways. Reactions numbered and shown in bold correspond to discussion in the text.}
  \label{mol:propene}
\end{figure}

Propene (\propene , propylene) was first detected using data from the {\it Cassini} CIRS infrared spectrometer via its $\nu_{19}$ band emission near 11 \micron\ \cite{nixon13a, lombardo19b}.

{\em Production:} Ion reactions lead to creation of propene from $\rm C_3H_5^+$, \eg :\cite{mcewan07}

\begin{eqnarray}
{\rm C_3H_5^+ + C_3H_8 }  & {\longrightarrow} & {\rm C_3H_7^+ + C_3H_6 }  \\
{\rm C_3H_5^+ + C_7H_8 }  & {\longrightarrow} & {\rm C_7H_7^+ + C_3H_6 } 
\end{eqnarray}

Propene is predicted to be produced in the upper atmosphere by both H addition to $\rm C_3H_5$ (62\%), and CH insertion into ethane (38\%)\cite{hebrard13}:

\begin{eqnarray}
{\rm H + C_3H_5 }  & {\longrightarrow} & {\rm C_3H_6 }  + h\nu \\
{\rm CH + C_2H_6 }  & {\longrightarrow} & {\rm C_3H_6 + H } 
\label{eq:propene-prod}
\end{eqnarray}

\noindent and also lower in the atmosphere by the termolecular reaction\cite{vuitton19}:

\begin{eqnarray}
{\rm C_2H_3 + CH_3 +M }  & {\longrightarrow} & {\rm C_3H_6 +M } 
\end{eqnarray}

{\em Loss:} Propene is lost both through photodissociation (Fig.~\ref{mol:propene}\cite{borrell71,collin79,niedzielski82,naroznik86,collin88,gierczak88}) and ion reactions e.g.:\cite{mcewan07}

\begin{eqnarray}
{\rm C_3H_4^+ + C_3H_6 }  & {\longrightarrow} & {\rm C_4H_6^+ + C_2H_4 } \\
{\rm C_3H_5^+ + C_3H_6 }  & {\longrightarrow} & {\rm C_4H_7^+ + C_2H_4 } 
\end{eqnarray}

In the lower atmosphere, propene is predicted to be lost through a termolecular reaction with H atom addition:\cite{vuitton19}

\begin{eqnarray}
{\rm H + C_3H_6 +M}  & {\longrightarrow} & {\rm C_3H_7 + M } 
\end{eqnarray}

{\em Future directions:} Propene, as an alkene, may also undergo polymerization to form polypropylene, a notable and widespread plastic used on Earth. Most likely polyynes in Titan's atmosphere are not pure polymers of a single repeated monomer type (ethylene, propylene, {\em etc}) but an assorted mixture of many types, with the lighter, more abundant alkenes more heavily represented than larger, heavier units. Further research into polymerization of mixed monomers will yield insights into the formation of Titan's haze.

%%%%%%%%%%%%%%%%%%%%%%%
\subsubsection{	Propane}

\begin{figure}
\includegraphics[scale=0.4]{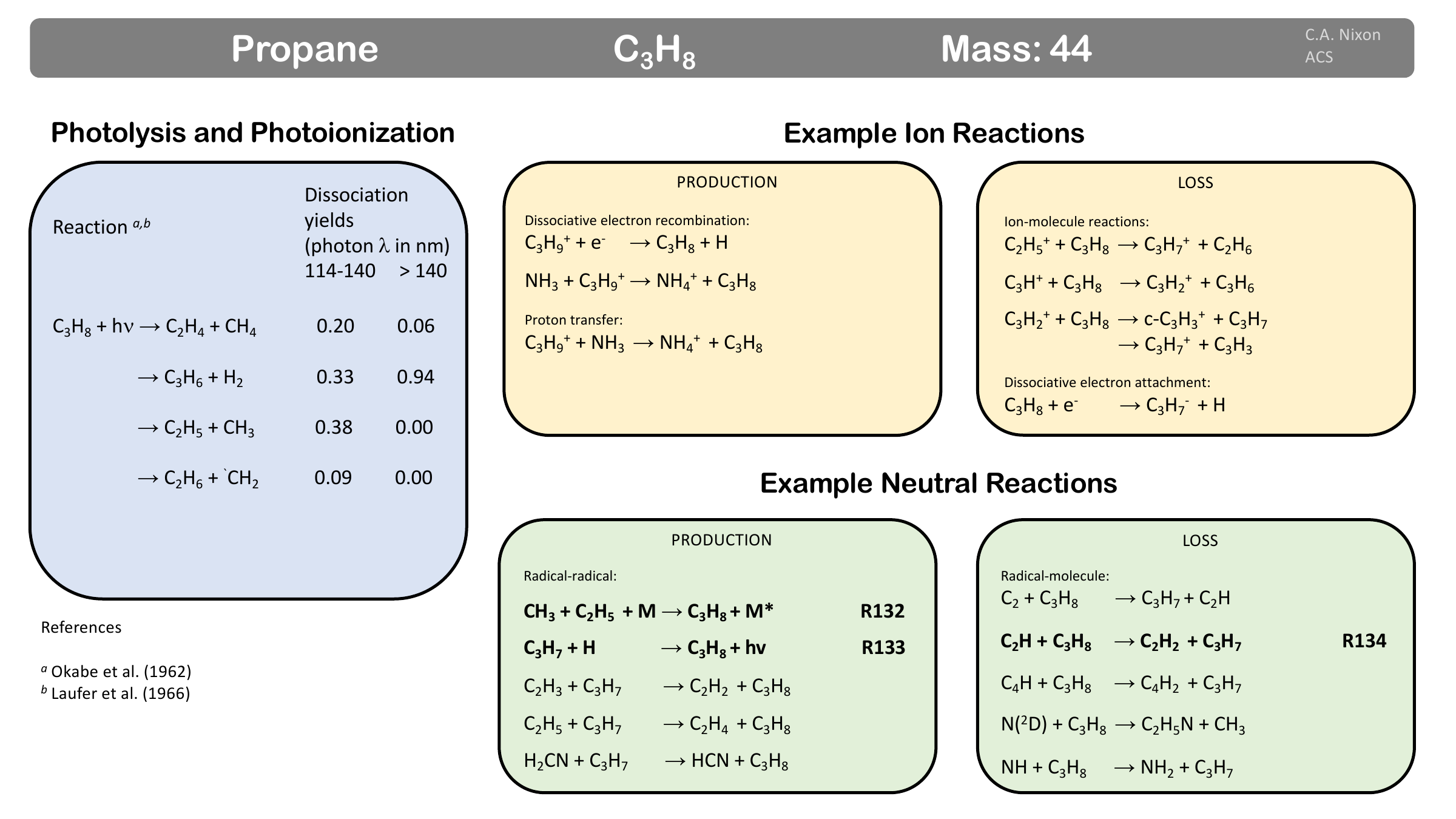}
  \caption{Propane production and loss pathways. Reactions numbered and shown in bold correspond to discussion in the text.}
  \label{mol:propane}
\end{figure}

Propane (\propane ) was detected contemporaneously with propyne (\propyne ) by {\it Voyager}'s IRIS instrument \cite{maguire81} via an infrared band at 748 \cm\ and subsequently confirmed by ground-based observations\cite{roe03} and with {\it Cassini} CIRS \cite{coustenis07, vinatier07a, nixon09b}.

{\em Production:} A significant pathway for the production of propane is by addition of $\rm CH_3$ to $\rm C_2H_5$ (Fig.~\ref{mol:propane})\cite{okabe62a,laufer66}: 

\begin{eqnarray}
{\rm C_2H_5 + CH_3 + M}  & {\longrightarrow} & {\rm C_3H_8 + M} 
\end{eqnarray}

but also to a lesser extent by the association reaction:\cite{harding07, vuitton19}

\begin{eqnarray}
{\rm C_3H_7 + H }  & {\longrightarrow} & {\rm C_3H_8 + h\nu} 
\end{eqnarray}

{\em Loss:} Propane is primarily lost in the upper atmosphere by photolysis to propene, but participates in other reactions as shown in Fig.~\ref{mol:propane}.
Propane also undergoes H-abstraction by ethynyl to recycle acetylene:\cite{murphy03}

\begin{eqnarray}
{\rm C_3H_8 + C_2H }  & {\longrightarrow} & {\rm C_3H_7 + C_2H_2} 
\end{eqnarray}

\noindent
but the fate of $\rm C_3H_7$ is largely to react with H to reform propane.\cite{vuitton19}

{\em Future directions:} While the chemistry of propane remains relatively well known, its role in cloud formation and lake composition on Titan remains to be fully explored.
Quantum mechanical analysis of propane's 23 infrared active bands\cite{nixon09b} remains incomplete, preventing accurate modeling at high resolution for these bands. However in recent years the pseudo-linelist technique has proved useful for providing practical absorption coefficients across a wide bandwidth for calculation at medium resolution \cite{sung13}. 

%%%%%%%%%%%%%%%%%%%%%%%
\subsubsection{	Diacetylene}

\begin{figure}
\includegraphics[scale=0.4]{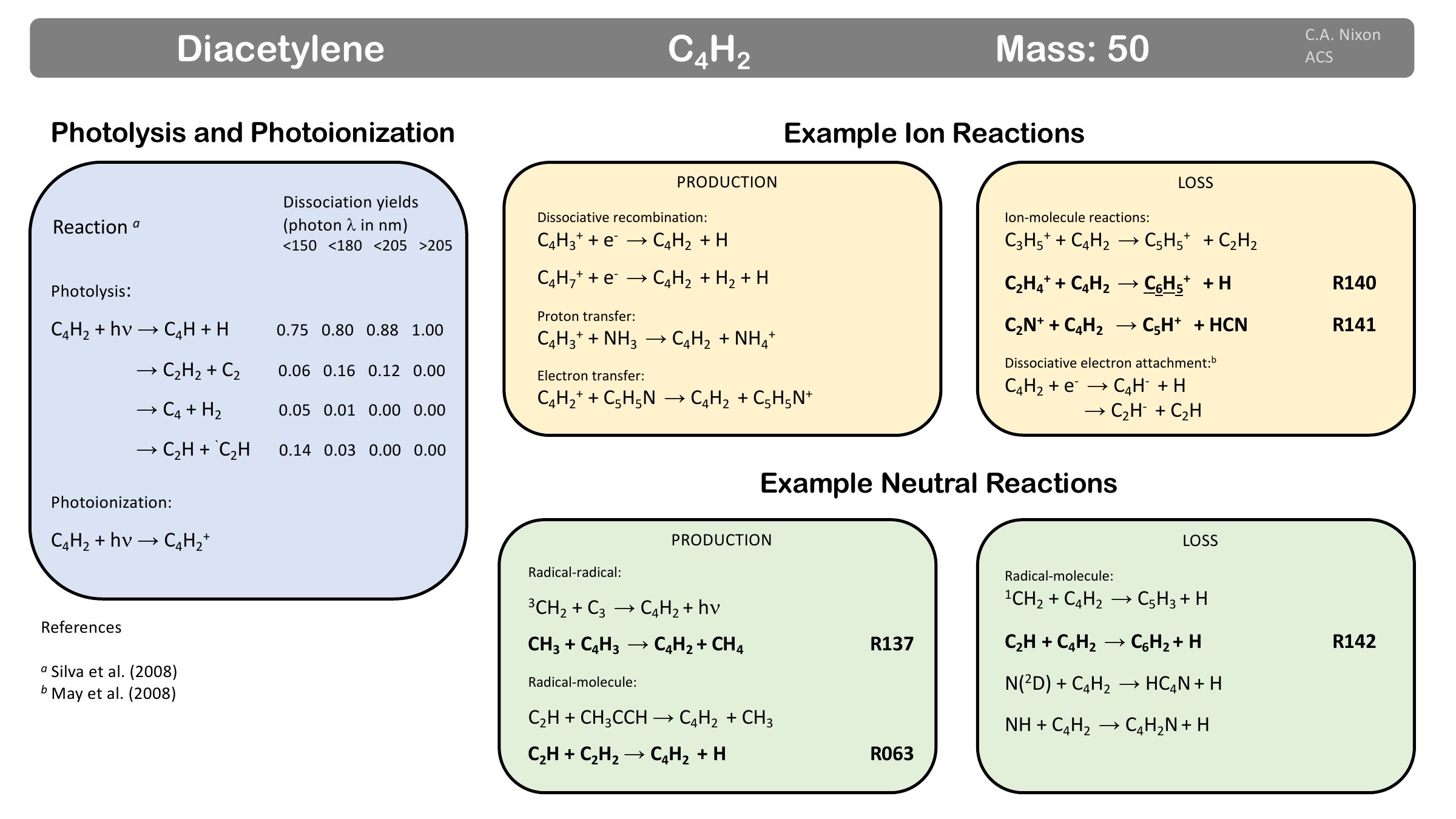}
  \caption{Diacetylene production and loss pathways. Reactions numbered and shown in bold correspond to discussion in the text.}
  \label{mol:diacet}
\end{figure}

The presence of diacetylene (\diacet , butadiyne) was inferred from infrared spectroscopy of Titan's atmosphere with {\it Voyager}'s IRIS instrument \cite{kunde81} via emission bands at 220 and 628 \cm . At present, it remains the only C$_4$ hydrocarbon species confirmed in Titan's atmosphere (although note that the nitrile CH$_3$C$_3$N, detected with ALMA\cite{thelen20}, also has four carbon atoms.)

{\em Production:} Diacetylene can be produced by the aforementioned reaction of the ethynyl radical with acetylene (R \ref{eq:diacet-prod}), or by stepwise addition to acetylene \cite{canosa97, vereecken99, vuitton19}:

\begin{eqnarray}
{\rm CH + C_2H_2 }  & {\longrightarrow} & {\rm C_3H_2 + H} \\
{\rm ^3CH_2 + C_3H_2 }  & {\longrightarrow} & {\rm C_4H_3 + H} \\
{\rm CH_3 + C_4H_3 }  & {\longrightarrow} & {\rm C_4H_2 + CH_4} 
\end{eqnarray}

{\em Loss:} Diacetylene may undergo ionization to $\rm C_4H_2^+$ (Fig.~\ref{mol:diacet})\cite{silva08,may08}, and subsequent loss to processes such as:\cite{mcewan07}

\begin{eqnarray}
{\rm C_4H_2^+ + C_2H_4 }  & {\longrightarrow} & {\rm C_6H_5^+ + H} \\
{\rm C_4H_2^+ + C_2H_4 }  & {\longrightarrow} & {\rm C_4H_4^+ + C_2H_2} 
\end{eqnarray}

\noindent
while neutral \diacet\ may be lost through ion reactions, including:\cite{nahar97,dutuit13}

\begin{eqnarray}
{\rm C_2H_4^+ + C_4H_2 }  & {\longrightarrow} & {\rm C_5H_3^+ + CH_3} \\
{\rm C_2N^+ + C_4H_2 }  & {\longrightarrow} & {\rm C_5H^+ + HCN} 
\end{eqnarray}

Insertion of ethynyl is a way to lengthen the polyyne chain from diacetylene to triacetylene:\cite{gu09b}

\begin{eqnarray}
{\rm C_2H + C_4H_2 }  & {\longrightarrow} & {\rm C_6H_2 + H} 
\end{eqnarray}

{\em Future directions:} To date, the triacetylene molecule ($\rm C_6H_2$) has remained elusive on Titan, despite its detection in space at high relative abundances compared to \diacet\ \cite{cernicharo01}. Detection of triacetylene would help to clarify the importance of the $\rm C_6H$ radical, which contributes to the depletion of methane in photochemical models along with the smaller related radicals $\rm C_2H$ and $\rm C_4H$, as well as the efficacy of polyyne formation in general.

%%%%%%%%%%%%%%%%%%%%%%%
\subsubsection{	Benzene}

\begin{figure}
\includegraphics[scale=0.4]{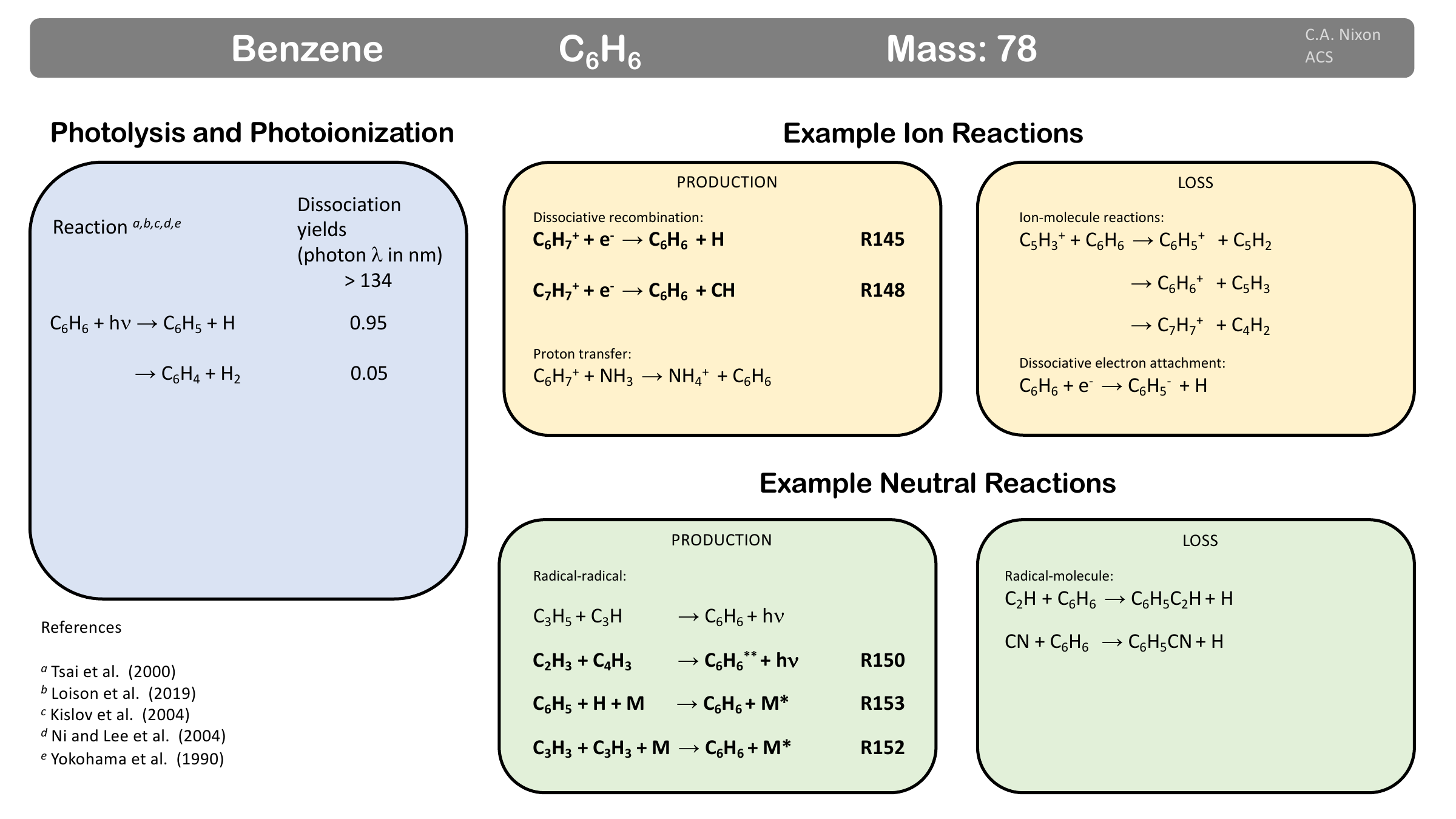}
  \caption{Benzene production and loss pathways. Reactions numbered and shown in bold correspond to discussion in the text.}
  \label{mol:benzene}
\end{figure}

Benzene (\benzene , Fig.~\ref{mol:benzene})\cite{tsai00, loison19, kislov04, ni04, yokoyama90} was the second new species detection on Titan made by the {\it Infrared Space Observatory} (ISO) in 2003 \cite{coustenis03}, via its strong hydrogen bending mode at 674 \cm . Benzene was the first cyclic (closed ring) molecule to be detected on Titan, and remains the only confirmed aromatic molecule (molecules with a de-localized $\pi$ electron orbital). The detection of benzene is highly significant, since it provides a measurement of the basic six-membered ring from which larger, multi-ring molecules can be formed,\cite{kaiser21, kaiser21b} building towards macromolecular haze particles - see discussion later in this article.

{\em Production:} 
In the upper atmosphere (800-950~km), a significant pathway to creation of benzene is dissociative recombination (DR) of the phenylium ion. Phenylium is created through:\cite{vuitton08,loison19}

\begin{eqnarray}
{\rm C_6H_5^+ + H_2 }  & {\longrightarrow} & {\rm C_6H_7^+ + h\nu } \\
{\rm C_4H_5^+ + C_2H_4 }  & {\longrightarrow} & {\rm C_6H_7^+ + H_2} 
\end{eqnarray}

\noindent  which then forms benzene through:\cite{hamberg11}

\begin{eqnarray}
{\rm C_6H_7^+ + e^- }  & {\longrightarrow} & {\rm C_6H_6 + H } 
\end{eqnarray}

\begin{figure}
\includegraphics[scale=0.5]{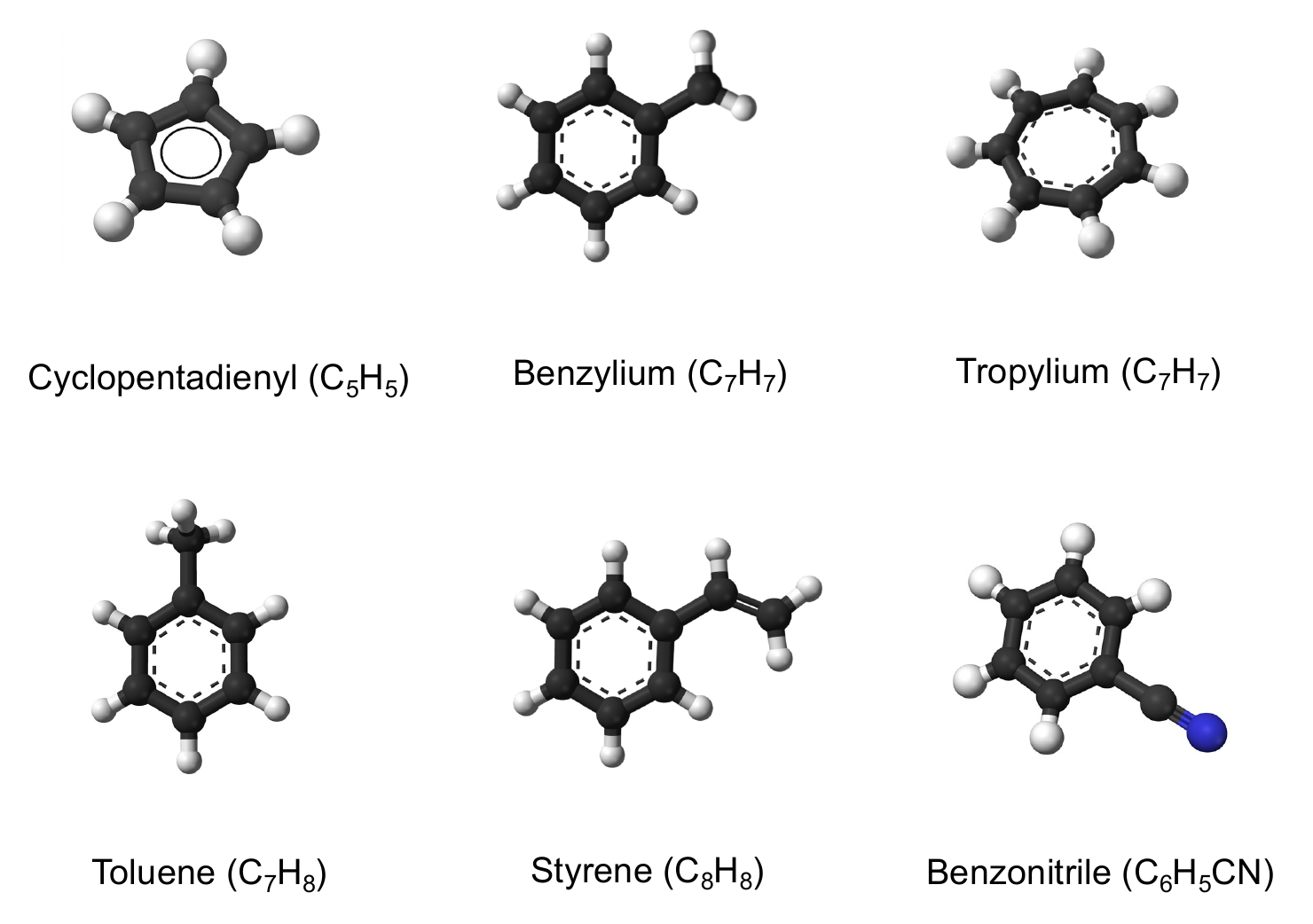}
  \caption{Potential benzene precursor molecules (upper) and products (lower). Molecule images: wikimedia commons.}
  \label{benzene-derivs}
\end{figure}

A second source roughly equal in importance is thought to be dissociative recombination of the C$_7$H$_7^+$ ion (benzylium or tropylium, Fig.~\ref{benzene-derivs}):\cite{loison19}

\begin{eqnarray}
{\rm C_5H_5^+ + C_2H_2 }  & {\longrightarrow} & {\rm C_7H_7^+ + h\nu } \\
{\rm C_6H_5^+ + CH_4 }  & {\longrightarrow} & {\rm C_7H_7^+ + H_2 }
\end{eqnarray}

\noindent followed by:

\begin{eqnarray}
{\rm C_7H_7^+ + e^- }  & {\longrightarrow} & {\rm C_6H_6 + CH } .
\end{eqnarray}

Yet a third ion channel is the DR of C$_8$H$_{11}^+$ with $e^-$ leading to benzene plus other hydrocarbon fragments.\cite{loison19} 

Radical chemistry also leads to to benzene, such as C$_2$ addition to 1,3 butadiene proposed to occur in the ISM:\cite{jones11}

\begin{eqnarray}
{\rm C_2H + C_4H_6  }  & {\longrightarrow} & {\rm C_6H_6 + H} 
\end{eqnarray}

An alternate pathway:

\begin{eqnarray}
{\rm C_2H_3 + C_4H_3  }  & {\longrightarrow} & {\rm C_6H_6^{**} + h\nu} 
\end{eqnarray}

\noindent
also leads to benzene, but in a highly excited state where it will mostly dissociate to C$_6$H$_5$ + H.\cite{loison19} Recently, a new pathway via a smaller, five-membered ring radical (cyclopentadienyl  Fig.~\ref{benzene-derivs}) molecule has been proposed by Kaiser et al.\cite{kaiser21b}:

\begin{equation}
{\rm c{\text -}C_5H_5 + CH_3  }  \:\: {\longrightarrow}  \:\: {\rm c{\text -}C_5H_5CH_3 } \:\:  {\longrightarrow}  \:\: {\rm c{\text -}C_6H_7 + H }  \:\: {\longrightarrow}  \:\: {\rm c{\text -}C_6H_6 + 2H } 
\end{equation}

\noindent but this has yet to be added to photochemical models to assess its relative importance. 

At higher pressures lower in the atmosphere, the three-body reaction combining two propargyl radicals becomes the dominant pathway for creation of benzene: \cite{lavvas08a,lavvas08b,vuitton08,loison19}

\begin{eqnarray}
{\rm C_3H_3 + C_3H_3 + M }  & {\longrightarrow} & {\rm C_6H_6 + M^{\ast} } 
\end{eqnarray}

{\em Loss:} Benzene is lost through ionization to the phenylium ion (C$_6$H$_5^+$) and through photolysis to form phenyl ($\rm C_6H_5$)\cite{vuitton08, capalbo16}. The phenyl radical then either reforms benzene:\cite{vuitton12}

\begin{eqnarray}
{\rm C_6H_5 + H + M }  & {\longrightarrow} & {\rm C_6H_6 + M^{\ast} } 
\end{eqnarray}

\noindent or reacts with other radicals and neutral species, leading to molecules such as toluene ($\rm C_6H_5CH_3$), styrene ($\rm C_6H_5C_2H$) and benzonitrile ($\rm C_6H_5CN$).\cite{gu09a}

{\em Future directions: } Benzene is a highly significant molecule, as the precursor to larger, multi-ring molecules (Fig.~\ref{benzene-derivs}). Further study of its creation and loss mechanisms, especially pathways to larger molecules\cite{kaiser21}, are important future directions.

%%%%%%%%%%%%%%%%%%%%%%%%%%%%%%%%%%%%%%%%%%%%%%%%%%%%%%%%%%%%%%%%%%%%%
\subsection{Nitrogen Compounds}

\begin{figure}
\includegraphics[scale=0.60]{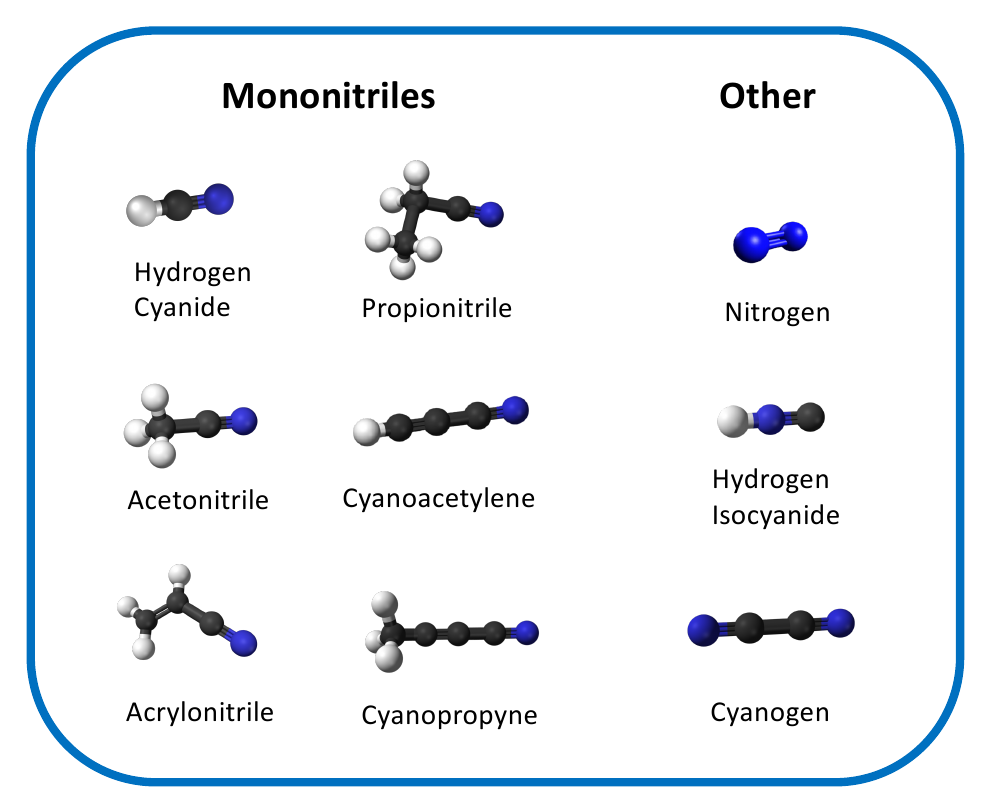}
  \caption{Nitrogen-bearing molecules detected on Titan.}
  \label{fig:nitrogen-mols}
\end{figure}

Nitrogen compounds are formed by chemical combination of dissociation products from initial \nitrogen\ and \methane , and have formulas $\rm C_xH_yN_z $. All of the eight known heteroatomic nitrogen compounds are cyanides, wherein nitrogen is bonded to carbon by a triple bond ($\text -$C$\equiv$N) and therefore have a formula: $\rm C_xH_y$-$\rm (CN)_z$. These are HCN, HNC, \acetonitrile , \acrylonitrile , \propionitrile , \cyanoacet , \butynenitrile\ and \cyanogen . Other than the light molecules HCN and HNC, the remaining molecules are nitriles (organic cyanides). See Fig.~\ref{fig:nitrogen-mols}.

Hydrogen isocyanide (HNC) is less stable than  hydrogen cyanide (HCN) and is converted exothermically to HCN as it descends in the atmosphere. This leads to a predicted steep decrease in abundance with increasing pressure \cite{cordiner14} and its present non-detection at lower altitudes. 

In the lower atmosphere, nitrogen has always been found to date to be triple-bonded in the terminal position of a molecule: other types of species (amines, imines etc) have not yet been detected. We will return to the topic of what additional nitrogen compounds may be waiting to be discovered in a later section. The chemistry of known N-bearing molecules in the neutral atmosphere is now summarized.

% (network diagram)

% (vertical reactions diagram)

%%%%%%%%%%%%%%%%%%%%%%%
\subsubsection{	Nitrogen}

\begin{figure}
\includegraphics[scale=0.4]{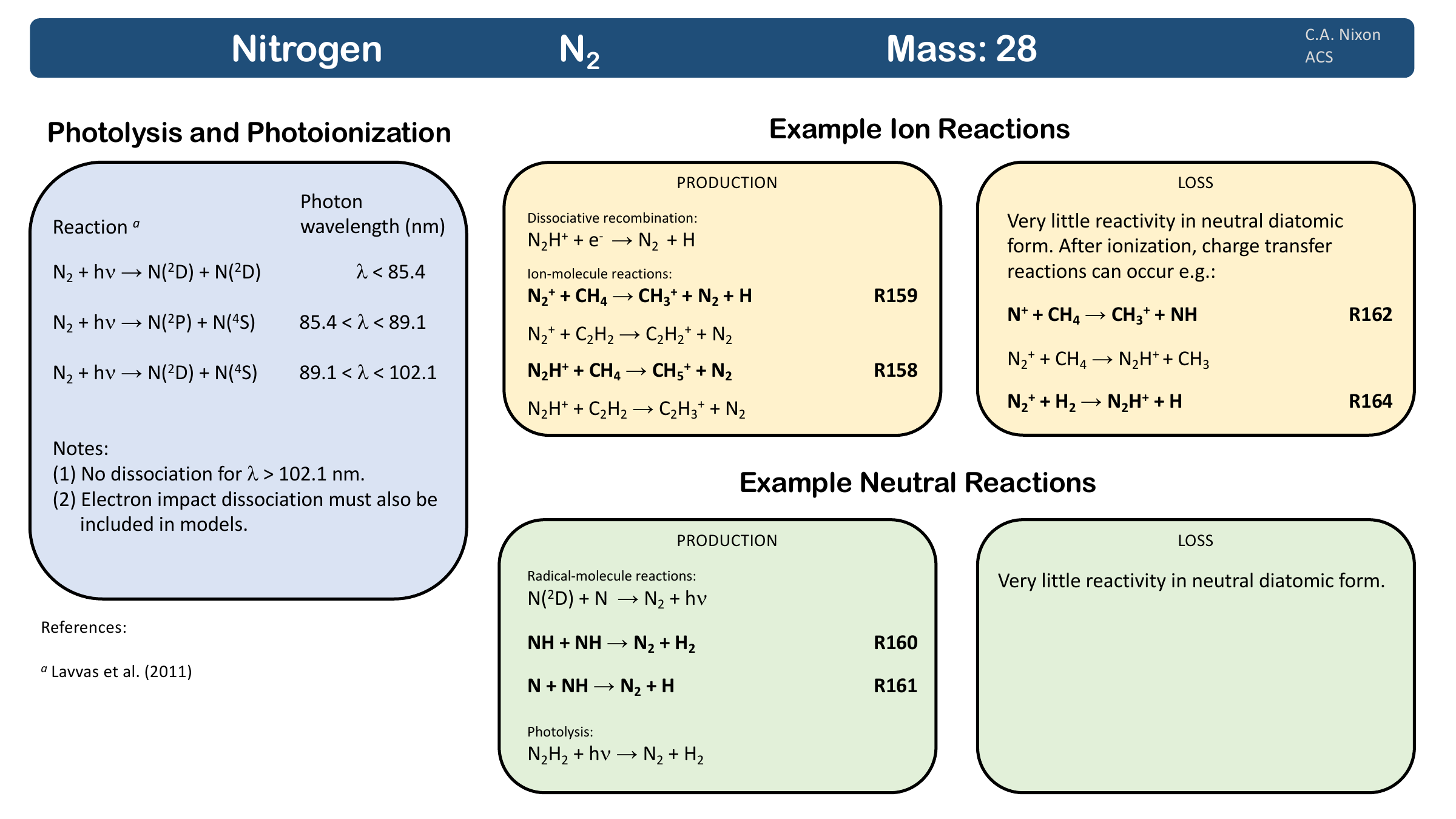}
  \caption{Production and loss pathways for molecular nitrogen. Reactions numbered and shown in bold correspond to discussion in the text.}
  \label{mol:nitrogen}
\end{figure}

A major but unobserved constituent in Titan's atmosphere was necessitated by the observed collisional broadening of methane spectral lines \cite{trafton72b}: this was hypothesized to be molecular nitrogen \cite{hunten73} which would be invisible at visible and longer wavelengths. The first conclusive observations of nitrogen were by {\em Voyager 1}'s UVS instrument, which detected dayside airglow at 96 and 98 nm, and longer wavelength absorptions with occultation measurements \cite{broadfoot81,vervack04}. Measurements of nitrogen were greatly extended by {\em Cassini}'s UVIS instrument \cite{esposito04, ajello07, ajello08, stevens11, capalbo13}. 

{\em Production:}  The origin of nitrogen in Titan's atmosphere has been long debated, and is not the subject of this paper. In brief, two major theories exist: enclathratization of \nitrogen\ gas in the protosolar nebula\cite{owen82}, or accretion in the form of \ammonia\ ice followed by later photodissociation to eventually form \nitrogen\ through a reaction cascade \cite{atreya78}:

\begin{eqnarray}
{\rm NH_3 + h\nu }  & {\longrightarrow} & {\rm NH_2 + H }  \\
{\rm 2 \:\: NH_2 + M }  & {\longrightarrow} & {\rm N_2H_4 + M^{\ast}  } \\
{\rm N_2H_4 + H }  & {\longrightarrow} & {\rm N_2H_3 + H_2 }  \\
{\rm 2 \:\: N_2H_3 }  & {\longrightarrow} & {\rm 2 \:\: NH_3 + N_2 } 
\end{eqnarray}

The latter scenario is currently favored due to the low temperatures in the sub-nebula required to capture molecular nitrogen directly. Variations on the theory include impact conversion of either \ammonia\ or  ammonium sulfate ($\rm (NH_4)_2SO_4$) to \nitrogen\ \cite{mckay88,fukuzaki10}. 

\nitrogen\ can also be recycled by recombination or proton transfer of one its ions\cite{mcewan07, dutuit13, xu13}:

\begin{eqnarray}
{\rm N_2H^+ + CH_4 }  & {\longrightarrow} & {\rm CH_5^+ + N_2  }  \\
{\rm N_2^+ + CH_4 }  & {\longrightarrow} & {\rm CH_3^+ + N_2 + H } 
\end{eqnarray}

\noindent or through recycling of one of its radicals\cite{wakelam15, vuitton19}:

\begin{eqnarray}
{\rm NH + NH }  & {\longrightarrow} & {\rm N_2 + H_2  }  \\
{\rm N + NH }  & {\longrightarrow} & {\rm N_2 + H  }  
\end{eqnarray}

{\em Loss:} Molecular nitrogen is dissociated and/or ionized by short-wavelength solar radiation at $\lambda < 127$~nm \cite{lavvas11a}, Saturn magnetosphere electrons \cite{lavvas15} and Galactic Cosmic Rays (GCRs)\cite{capone76, capone83, molina-cuberos99a, molina-cuberos99b} (Fig.~\ref{mol:nitrogen})\cite{lavvas11b}. 

Nitrogen ions react with abundant neutrals including \methane\ and \hydrogen \cite{marquette88, dutuit13}:

\begin{eqnarray}
{\rm N^+ + CH_4 }  & {\longrightarrow} & {\rm CH_3^+ + NH \: (and \: others) }  \\
{\rm N^+ + H_2 }  & {\longrightarrow} & {\rm NH^+ + H }  \\
{\rm N_2^+ + H_2 }  & {\longrightarrow} & {\rm N_2H^+ + H  } 
\end{eqnarray}

\noindent and can recycle to $\rm N_2$ through reaction with hydrogen, methane and other hydrocarbons, e.g. \cite{dutuit13}:

\begin{eqnarray}
{\rm N_2^+ + H_2 }  & {\longrightarrow} & {\rm H_2^+ + N_2}  \\
{\rm N_2^+ + CH_4 }  & {\longrightarrow} & {\rm CH_3^+ + N_2 + H \: (and \: others) }  \\
{\rm N_2^+ + C_2H_2 }  & {\longrightarrow} & {\rm C_2H_2^+ + N_2}   \\
{\rm N_2^+ + C_2H_6 }  & {\longrightarrow} & {\rm CH_3^+ + N_2 + CH_3 \: (and \: others) }  
\end{eqnarray}

However, molecular nitrogen in the un-ionized state has very low reactivity,  which in part contributes to its great abundance and significant longevity in the atmosphere.

{\em Isotopes:} Since $\rm ^{14}N_2$ and $\rm ^{14}N^{15}N$ have significantly different UV cross-sections \cite{liang07}, it is important to correctly account for both isotopes and the wavelength variation of the solar spectrum to arrive at correct dissociation rates. Self-shielding by the more abundant $\rm ^{14}N^{14}N$ is thought to reduce photolysis rates relative to the less abundant, less shielded $\rm ^{14}N^{15}N$, causing a lower $\rm ^{14}N/^{15}N$ ratio in nitrogen atoms than in the original molecules. Since significant amounts of atomic nitrogen go on to form nitriles, this skew towards increased production of $\rm ^{15}N$ may explain the lower $\rm ^{14}N/^{15}N$ in nitriles than in \nitrogen\ itself \cite{liang07, vinatier07b, jennings08}.

{\em Future directions: } The dissociation and reaction pathways for \nitrogen\ and its daughter ions and radicals remain one of the better known areas of Titan chemistry. However gaps remain, in particular whether nitrogen exists in chemicals such as amines and imines, or if it is incorporated in to heterocyclic ring molecules. This is further discussed in a later section.

%%%%%%%%%%%%%%%%%%%%%%%
\subsubsection{Hydrogen Cyanide}

\begin{figure}
\includegraphics[scale=0.4]{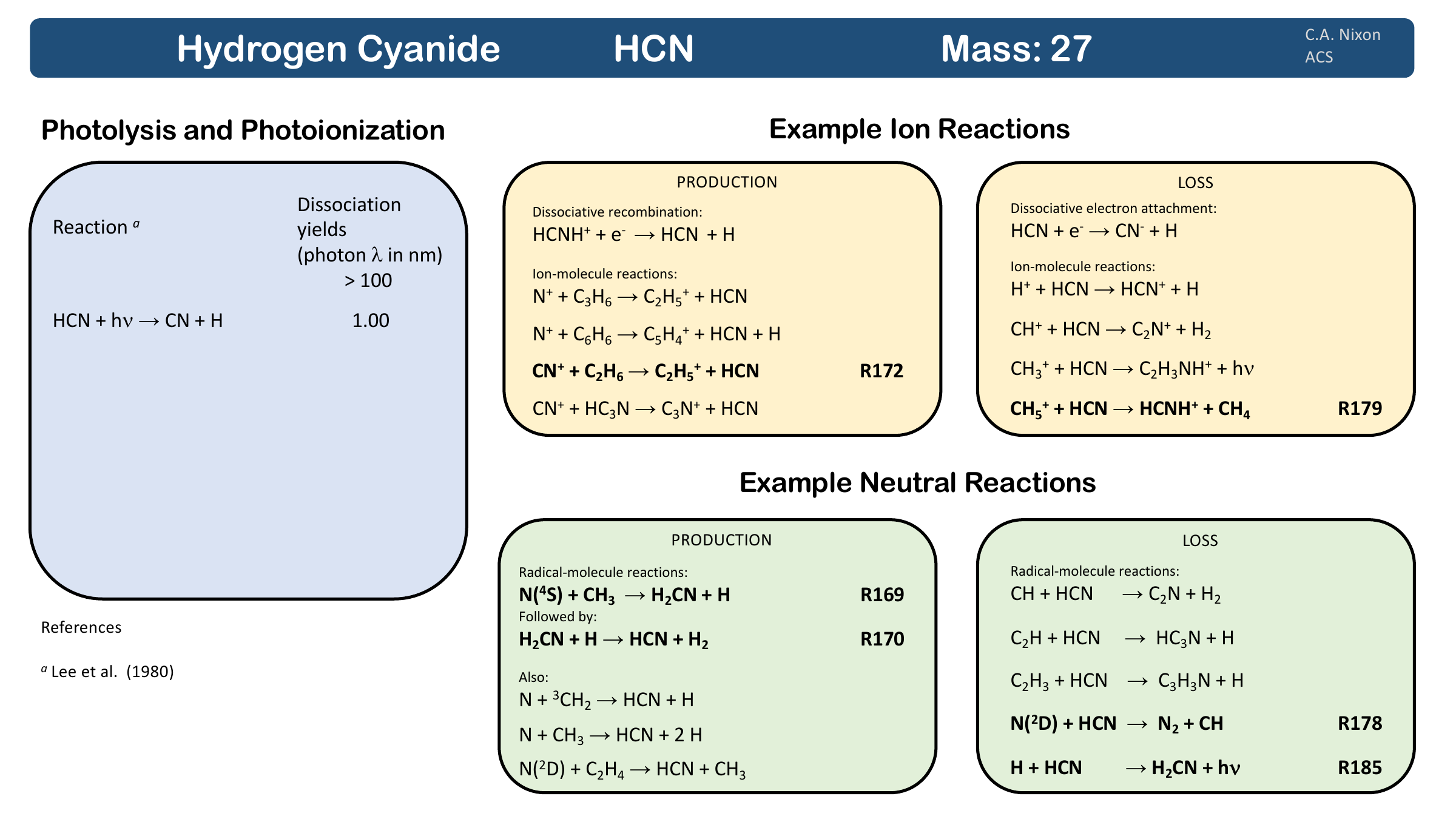}
  \caption{Hydrogen cyanide production and loss pathways. Reactions numbered and shown in bold correspond to discussion in the text.}
  \label{mol:hcn}
\end{figure}

Hydrogen cyanide was first detected by the {\em Voyager} 1 IRIS spectrometer through its strong infrared emission at 712 \cm\ \cite{hanel81}, and later at sub-millimeter wavelengths from ground-based observatories \cite{marten88, tanguy90, hidayat97, marten02}. Although a relatively simple molecule that has been included in photochemical models for more than four decades, gaps in our knowledge of HCN formation may still exist, and new pathways have been identified recently \cite{pearce19}.

{\em Production:} HCN is primarily produced in the upper atmosphere by the reaction of methane and nitrogen dissociation products (Fig.~\ref{mol:hcn}\cite{lee80}):

\begin{eqnarray}
{\rm N(^4S) + CH_3} & \longrightarrow & {\rm H_2CN + H} \\
{\rm H_2CN + H} & \longrightarrow & {\rm HCN + H_2 , or} \\
{\rm H_2CN + h\nu } & \longrightarrow & {\rm HCN + H  } 
\end{eqnarray}

\noindent and may be reformed from its ion by ion-molecule reactions, e.g.\cite{mcewan07}:

\begin{eqnarray}
{\rm CN^+ + C_2H_6} & \longrightarrow & {\rm C_2H_5^+ + HCN }
\end{eqnarray}

In the lower atmosphere, reactions with CN radicals become important:\cite{sims93, hebrard13, gannon07, fukuzawa98}

\begin{eqnarray}
{\rm CN + CH_4H} & \longrightarrow & {\rm HCN + CH_3 } \\
{\rm CN + C_2H_6} & \longrightarrow & {\rm HCN + C_2H_5 } \\
{\rm CN + C_3H_8 } & \longrightarrow & {\rm HCN + C_3H_7  } 
\end{eqnarray}

\noindent
along with photodissociation of \acrylonitrile\ (Fig. N)\cite{wilhelm09} and reaction of other nitriles with H:\cite{hebrard13, vuitton19}

\begin{eqnarray}
{\rm H + C_2N_2} & \longrightarrow & {\rm HCN + CN } \\
{\rm H + H_2CN} & \longrightarrow & {\rm HCN + H_2 }
\end{eqnarray}

{\em Loss:}  
At high altitudes ($z \geq 1000$~km) HCN is primarily destroyed by reaction with $\rm N(^2D)$: \cite{hebrard13, vuitton19}

\begin{eqnarray}
{\rm HCN + N(^2D)} & \longrightarrow & {\rm CH + N_2 }
\end{eqnarray}

HCN may also be lost in a two-step process, beginning with proton-transfer from a lower proton affinity molecule, e.g.:\cite{vuitton19}

\begin{eqnarray}
{\rm CH_5^+ + HCN } & \longrightarrow & {\rm HCNH^+ + CH_4 } 
\end{eqnarray}

\noindent followed by dissociative recombination:\cite{wakelam15}

\begin{eqnarray}
{\rm HCNH^+ + e^- } & \longrightarrow & {\rm CN + 2 H } 
\end{eqnarray}

Lower in the atmosphere, radical reactions and photolysis become important:\cite{hebrard13}

\begin{eqnarray}
{\rm HCN + C_2 } & \longrightarrow & {\rm C_3N + H } \\
{\rm HCN + C_2N} & \longrightarrow & {\rm C_4N_4 + H }
\end{eqnarray}

As noted by previous authors, the C${\equiv}$N triple bond is extremely stable and therefore the CN unit tends to persist when HCN is photolyzed, being incorporated into heavier nitriles, e.g.:

\begin{eqnarray}
{\rm HCN + h\nu } & \longrightarrow & {\rm H + CN } \\
{\rm C_2H_2 + CN} & \longrightarrow & {\rm HC_3N + H } 
\end{eqnarray}

Also at low altitudes ($z < 650 $ km \cite{vuitton19}) H-addition can lead to formation of the methylene-amidogen radical:

\begin{eqnarray}
{\rm HCN + H } & \longrightarrow & {\rm H_2CN + h\nu } 
\end{eqnarray}

{\em Future directions: } Although well-studied for decades, recent work\cite{pearce19, pearce20} has identified new pathways to the formation of HCN in planetary atmospheres for which reaction rates are currently unknown. Theoretical predictions now exist, but experimental confirmation is needed.

HCN has been shown to form co-crystals with hydrocarbons at Titan-relevant temperatures \cite{ennis20}, the study of which will be important for understanding the solids and liquids on the surface. HCN, along with \cyanoacet , has also been implicated in the formation of \dicyanoacet\ in grain-surface chemical reactions \cite{anderson16}, which requires further study to elucidate reaction rates and whether this process is sufficient to explain observed ice spectral properties \cite{anderson10}. 

Finally, HCN has been implicated in processes of astrobiological importance. A well known example is its proposed ability to directly form the amino acid adenine (\adenine )\cite{oro60,oro61a,oro61b,oro61c} from the rearrangement (oligomerization) of five HCN molecules. Although the importance of this reaction for the seeding of life on the early Earth has been disputed\cite{shapiro84, shapiro95}, it may be more prevalent on Titan where HCN occurs in greater abundance.\cite{jung13, he14}
HCN may also have the potential to polymerize into polyimines, structures that may catalyze astrobiologically important reactions \cite{rahm16}. The astrobiological potential of HCN therefore remains under continued investigation\cite{jeilani16, pearce19, sandstrom21, sharma22}.

%%%%%%%%%%%%%%%%%%%%%%%
\subsubsection{Hydrogen Isocyanide}

\begin{figure}
\includegraphics[scale=0.4]{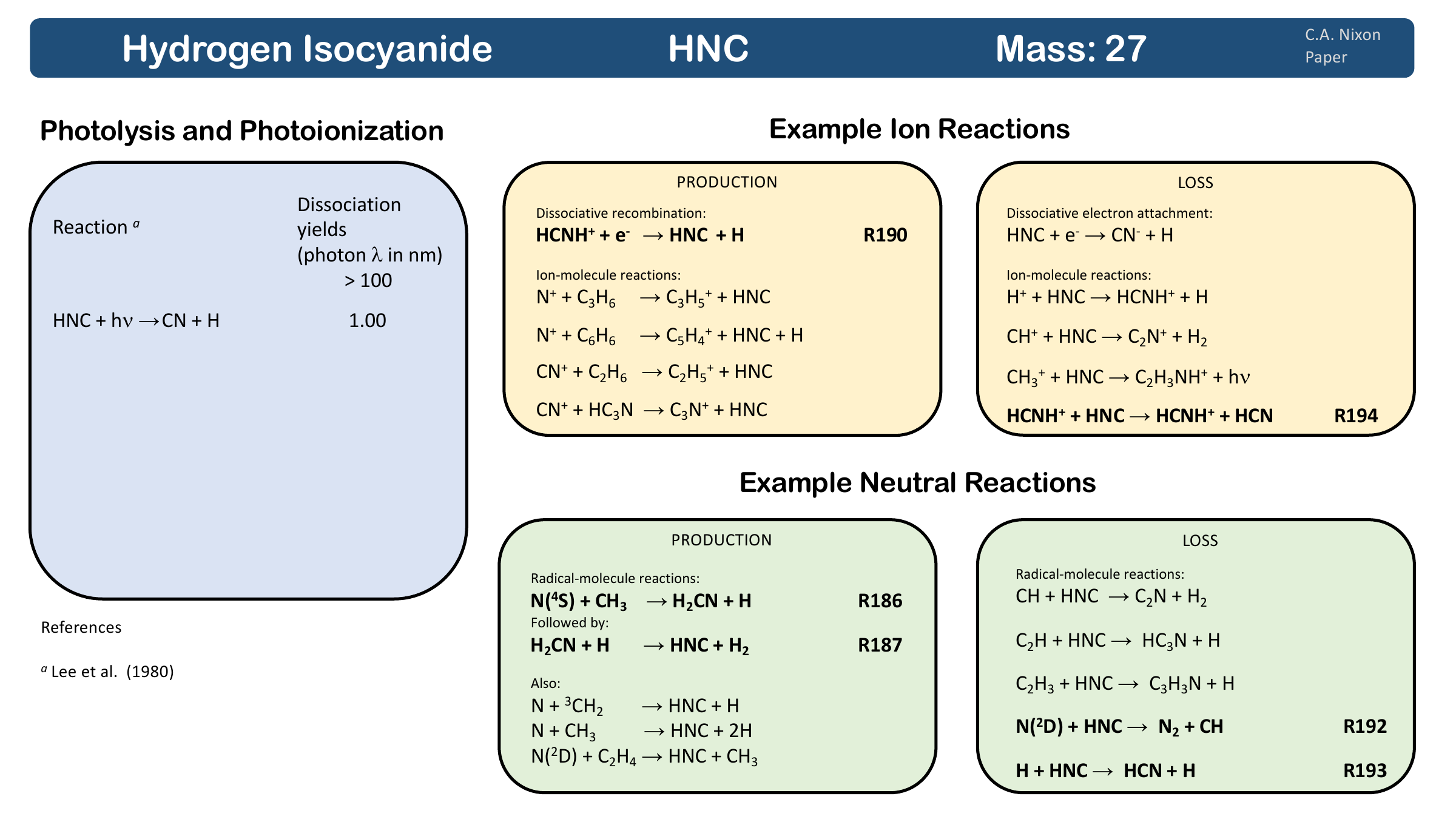}
  \caption{Hydrogen isocyanide production and loss pathways. Reactions numbered and shown in bold correspond to discussion in the text.}
  \label{mol:hnc}
\end{figure}

Hydrogen isocyanide, a higher energy isomer of hydrogen cyanide,\cite{nguyen15} was discovered on Titan using the {\em Herschel} space observatory by its sub-millimeter transition at 544 Ghz \cite{moreno11}, and subsequently measured by ALMA as well \cite{cordiner14, lellouch19}. HNC is readily interconverted to the more stable HCN (releasing $14.4 \pm 1.0$ kcal/mol),\cite{lee91} and therefore is predicted to have a steeply diminishing mixing ratio profile with altitude.\cite{loison15, vuitton19}

{\em Production:} HNC is produced by the same neutral reactions as HCN: 

\begin{eqnarray}
{\rm N(^4S) + ^3CH_2} & \longrightarrow & {\rm HNC + H} \:\:\:\: (34\%) \label{reaction:hcn1} \\
{\rm H + H_2CN} & \longrightarrow & {\rm HNC + H_2 } \:\:\:\: (64\%) 
\end{eqnarray}

\noindent 
where the relative productions are estimated at 1300~km.\cite{hebrard12} At 1000~km  R{\ref{reaction:hcn1} becomes dominant. Note that there are two important production pathways for ${\rm H_2CN}$: }

\begin{eqnarray}
{\rm N(^4S) + CH_3} & \longrightarrow & {\rm H_2CN + H} \label{reaction:h2cn1} \\
{\rm CH_2NH + h\nu } & \longrightarrow & {\rm H_2CN+ H} \label{reaction:h2cn2}
\end{eqnarray}

\noindent 
with R{\ref{reaction:h2cn1}} dominating in the thermosphere and R{\ref{reaction:h2cn2}} becoming important in the mesosphere and below.\cite{hebrard12}
Ion pathways may also be similar (see Fig.~\ref{mol:hnc})\cite{lee80}, although branching ratios are in most cases more uncertain than for HCN, e.g. through dissociative recombination of HCNH$^+$:\cite{mendes12}

\begin{eqnarray}
{\rm HCNH^+ + e^-} & \longrightarrow & {\rm HNC + H} 
\end{eqnarray}

HNC may also be produced as photodissociation product of \acrylonitrile \cite{wilhelm09} in the upper atmosphere, and a further production peak may occur due to cosmic ray chemistry at 100-150~km \cite{loison15}.

{\em Loss:} 
At high altitudes ($\sim$1300~km) the principal loss channels for HNC are:\cite{hebrard12}

\begin{eqnarray}
{\rm N(^2D) + HNC } & \longrightarrow & {\rm CN_2 + H} \\
{\rm N(^2D) + HNC } & \longrightarrow & {\rm CH+ N_2} 
\end{eqnarray}

\noindent
While at lower altitudes collisional isomerization to the lower energy HCN becomes important, and dominant by 600~km:\cite{hebrard12}

\begin{eqnarray}
{\rm HNC + H } & \longrightarrow & {\rm HCN + H}  \:\: (z < 1000 \: {\rm km}) \\
{\rm HNC + HCNH^+ } & \longrightarrow & {\rm HCN + HCNH^+ } \: \: (z > 1000 \: {\rm km})
\end{eqnarray}

{\em Future directions: } HNC/HCN is now one of two isomer pairs known in Titan's atmosphere (the other being $\rm C_3H_4$). Study of the branching ratios and reaction rates leading to and from isomer pairs/triples etc is of importance because the less stable isomer(s) may follow different reaction pathways compared to the more abundant molecule(s). Therefore for a complete understanding of Titan's atmospheric chemistry, all isomers must be included in models. Study of the vertical ratio between HCN/HNC and \propyne /\propadiene\ may also provide useful information on the abundance of atomic H, collisions which can cause conversion between the isomers.

%%%%%%%%%%%%%%%%%%%%%%%
\subsubsection{Acetonitrile}

\begin{figure}
\includegraphics[scale=0.4]{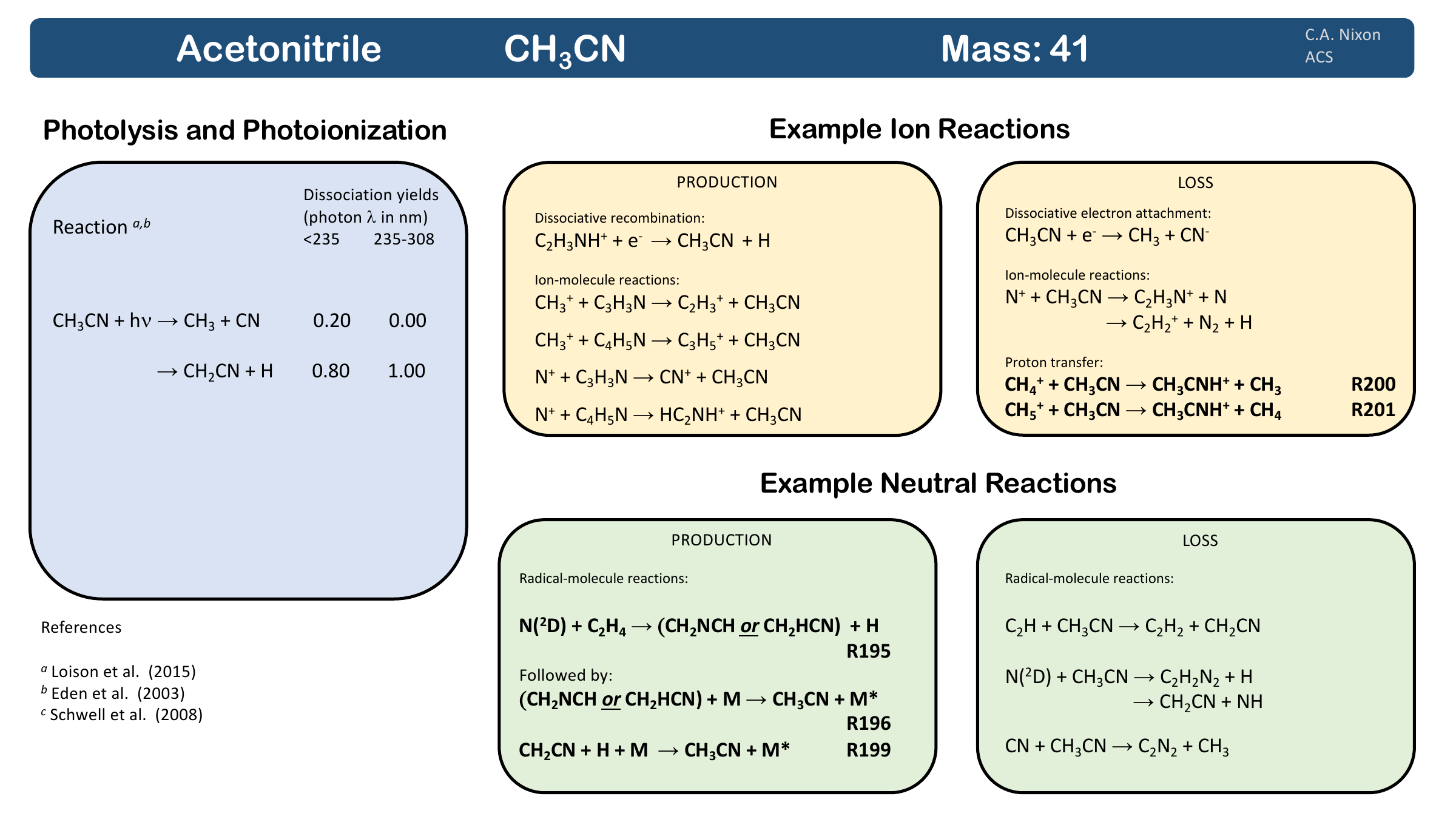}
  \caption{Acetonitrile production and loss pathways. Reactions numbered and shown in bold correspond to discussion in the text.}
  \label{mol:acetonitrile}
\end{figure}

Acetonitrile was first detected on Titan in the early 1990s by millimeter wavelength astronomy \cite{bezard92}, followed ten years later by the first measurement of its vertical profile \cite{marten02} using the 30~m telescope at IRAM. \acetonitrile\ was the first Titan molecule to be first detected at millimeter wavelengths, an astronomical technique that was to yield many other discoveries later with ALMA.

{\em Production:} Acetonitrile is produced in the upper atmosphere by the reaction of N-radicals with ethylene\cite{balucani12}:

\begin{eqnarray}
{\rm N(^2D) + C_2H_4  } & \longrightarrow & ( {\rm c{\text -}CH_2HCN} { \: or \:} {\rm CH_2(N)CH} )  + {\rm H } \\
( {\rm c{\text -}CH_2HCN} { \: or \: } {\rm CH_2(N)CH}  )  + {\rm M } & \longrightarrow & {\rm CH_3CN + M^{\ast} }
\end{eqnarray}

\noindent
and by the termolecular reaction of H with cyanomethyl (CH$_2$CN), in a chain that begins with acrylonitrile (C$_2$H$_5$CN):\cite{loison15,vuitton19}

\begin{eqnarray}
{\rm H + C_2H_3CN + M } & \longrightarrow & {\rm C_2H_4CN + M^{\ast} } \\
{\rm H + C_2H_4CN  } & \longrightarrow & {\rm CH_2CN + CH_3 } \\
{\rm H + CH_2CN + M } & \longrightarrow & {\rm CH_3CN + M^{\ast} } 
\end{eqnarray}

{\em Loss:} The major loss mechanism for acetonitrile is proton transfer from another ion to form $\rm CH_3CNH^+$:\cite{blair73}

\begin{eqnarray}
{\rm CH_4^+ + CH_3CN } & \longrightarrow & {\rm CH_3CNH^+ + CH_3 } \\
{\rm CH_5^+ + CH_3CN } & \longrightarrow & {\rm CH_3CNH^+ + CH_4 } 
\end{eqnarray}

\noindent
followed by dissociative recombination:\cite{vigren08, vigren12}

\begin{eqnarray}
{\rm CH_3CNH^+ + e^- } & \longrightarrow & {\rm HNC + CH_3 } \\
  & \longrightarrow & {\rm HCN + {^3}CH_2 + H } 
\end{eqnarray}

\noindent
and in the lower atmosphere by photolysis\cite{halpern85, suto85} (Fig.~\ref{mol:acetonitrile})\cite{loison15,eden03,schwell08}.

{\em Future directions: } Acetonitrile, as many other simple molecules, has  been implicated in formation of a co-crystal with acetylene \cite{cable20}, providing an interesting avenue for further investigation of its solid phase properties, with possible implications for cloud particle growth.

%%%%%%%%%%%%%%%%%%%%%%%
\subsubsection{Cyanoacetylene}

\begin{figure}
\includegraphics[scale=0.4]{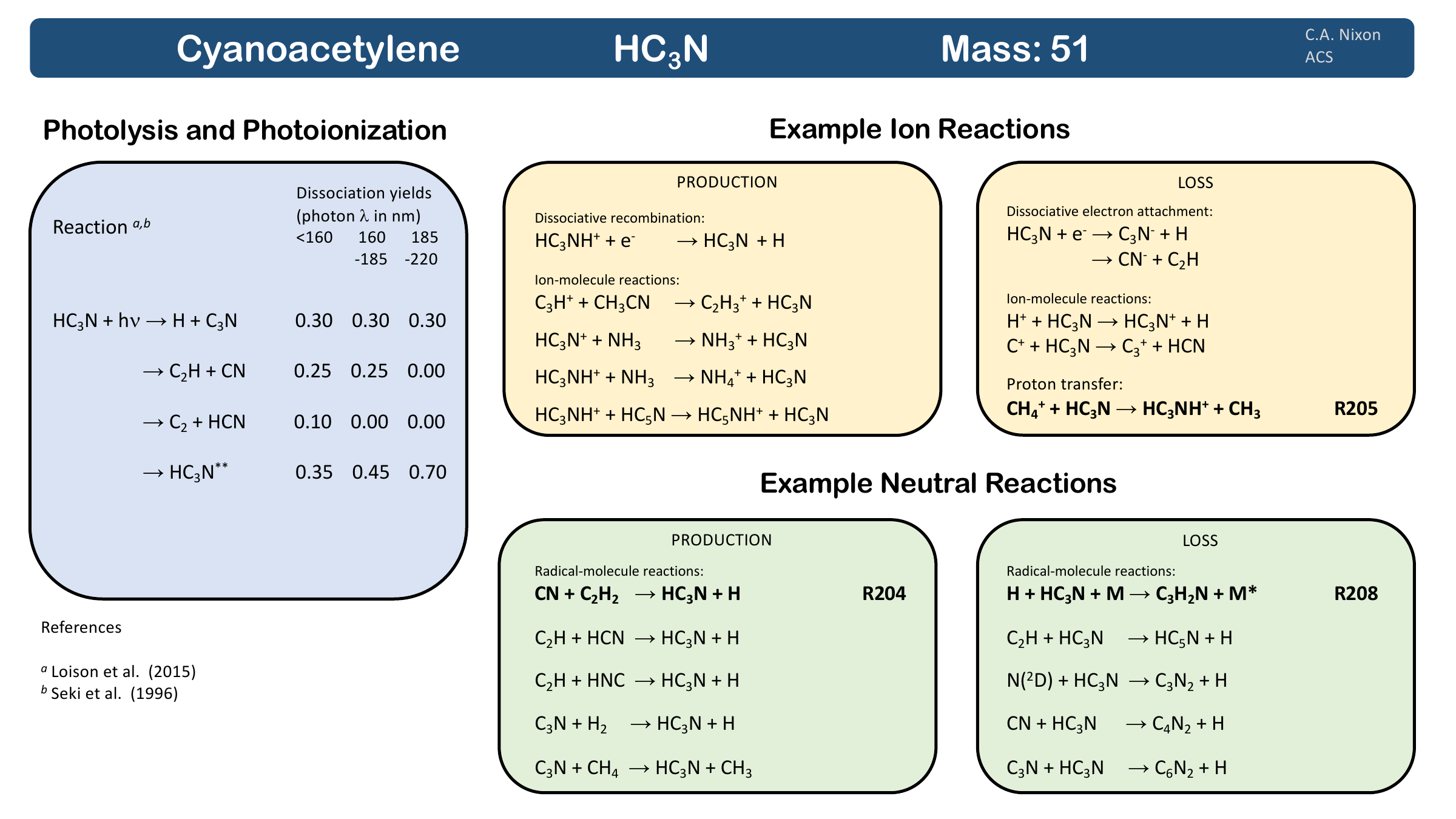}
  \caption{Cyanoacetylene production and loss pathways. Reactions numbered and shown in bold correspond to discussion in the text.}
  \label{mol:cyanoacet}
\end{figure}

Cyanoacetylene (\cyanoacet , propynenitrile) was first detected in Titan's atmosphere by the {\em Voyager} IRIS spectrometer in the infrared \cite{kunde81} at 500 and 663 \cm , following a prediction by \citeauthor{capone81}\cite{capone81} Cyanoacetylene, like diacetylene and cyanogen, was found in 1980 to be greatly enhanced over Titan's northern (winter) pole - interpreted as evidence of a global stratospheric circulation cell. Gases such as \cyanoacet\ with relatively short photochemical lifetimes (compared to a Titan year) have volume mixing profiles with steep vertical gradients at most latitudes, decreasing in a downwards direction as the gases become depleted and diluted. However, the presence of a strong downward motion from the mesosphere ($\sim$500~km) causes enrichment in trace species to show up much lower down in the lower stratosphere ($\sim$100~km).

{\em Production:} Cyanoacetylene is produced above 1000 km by reaction of the CN radical from photolysis of HCN with acetylene (see Fig.~\ref{mol:cyanoacet})\cite{sims93,loison15,seki96,gannon07}:

\begin{eqnarray}
{\rm C_2H_2 + CN}  & {\longrightarrow} & {\rm HC_3N + H} 
\end{eqnarray}

\noindent
and to a lesser extent by photodissociation of acrylonitrile (see Fig.~\ref{mol:acrylonitrile}).

{\em Loss:} As with other nitriles, the principal loss pathway for cyanoacetylene in the upper atmosphere is proton transfer, forming $\rm HC_3NH^+$, e.g.:\cite{mcewan07}

\begin{eqnarray}
{\rm CH_4^+ + HC_3N }  & {\longrightarrow} & {\rm HC_3NH^+ + CH_3 }  
\end{eqnarray}

\noindent
followed by dissociative recombination breaking up the molecule \cite{vigren12}:

\begin{eqnarray}
{\rm HC_3NH^+}  + e^-  & {\longrightarrow} & {\rm C_2H_2 + CN }  
\end{eqnarray}

\noindent
On the other hand, photolysis is not a significant loss channel, since $\rm C_3N$ is thought to rapidly recycle back to $\rm HC_3N$ through reaction with methane:\cite{loison15}

\begin{eqnarray}
{\rm C_3N}  + {\rm CH_4}  & {\longrightarrow} & {\rm HC_3N + CH_3 }  
\end{eqnarray}

While HC$_3$N does react without a barrier with radicals such as CN and C$_2$H, the main loss channel in the neutral atmosphere is thought to be successive hydrogen addition:\cite{loison15}

\begin{eqnarray}
{\rm HC_3N}  + {\rm H + M}  & {\longrightarrow} & {\rm H_2C_3N + M^{\ast} } \\  
{\rm H_2C_3N}  + {\rm H}  & {\longrightarrow} & {\rm HCN + C_2H_2  } \\  
{\rm H_2C_3N}  + {\rm H + M} & {\longrightarrow} & {\rm C_2H_3CN  + M^{\ast}} 
\end{eqnarray}

{\em Future directions: } In interstellar space (\eg\ molecular clouds such as TMC-1), cyanopolyynes of the form $\rm HC_xN$ have been detected with $x = 1, 3, 5, 7, 9, 11$ \cite{winnewisser78, broten78, bell82}. $\rm HC_5N$ has been sought, but not yet detected, in Titan's neutral atmosphere. Detection of this molecule may provide some clues as to the relative abundances of cyanopolyynes vs N-heterocycles.

%%%%%%%%%%%%%%%%%%%%%%%
\subsubsection{Cyanogen}

\begin{figure}
\includegraphics[scale=0.4]{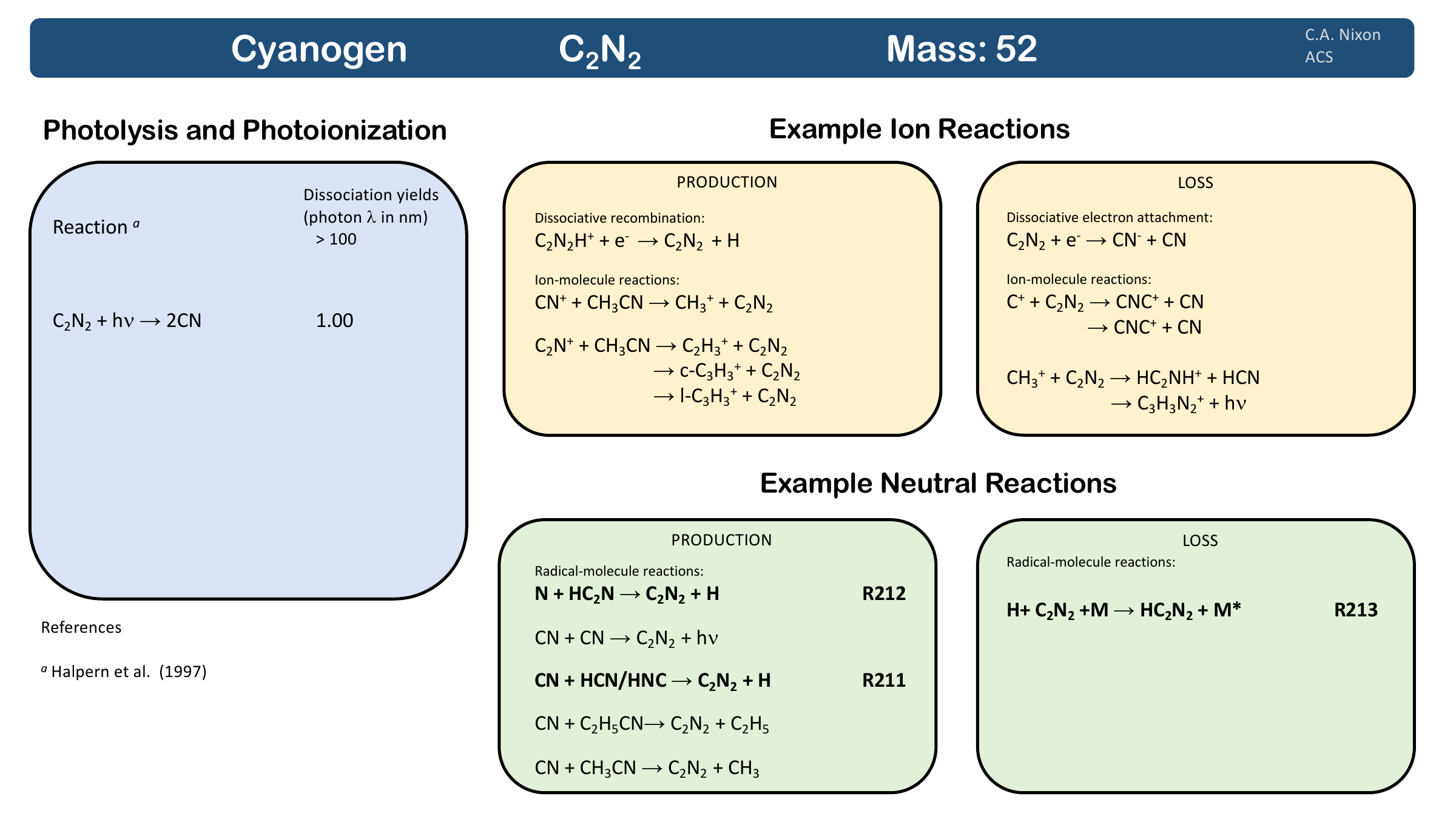}
  \caption{Cyanogen production and loss pathways. Reactions numbered and shown in bold correspond to discussion in the text.}
  \label{mol:cyanogen}
\end{figure}

Cyanogen (\cyanogen ), like cyanoacetylene was first detected in Titan's atmosphere by the {\em Voyager} IRIS spectrometer in the infrared \cite{kunde81} at 233 \cm .

{\em Production:} Cyanogen is thought to be produced mainly by addition of CN to HNC:

\begin{eqnarray}
{\rm CN + (HCN / HNC)}   & {\longrightarrow} & {\rm C_2N_2 + H }  
\end{eqnarray}

\noindent
and through the radical-radical reaction:\cite{loison15}

\begin{eqnarray}
{\rm N + HCCN}   & {\longrightarrow} & {\rm C_2N_2 + H }  
\end{eqnarray}

\noindent 
via the intermediate adduct NCHCN. Neither of these reactions are expected to have an entrance barrier\cite{loison15, petrie04, osamura04}, while the reaction of CN with HCN is inefficient due to the low rate constant.\cite{zabarnick89, yang92a, yang92b}

{\em Loss:} Cyanogen is lost by photodissociation (Fig.~\ref{mol:cyanogen}\cite{halpern97}) and by H-addition (with an entrance barrier of $\sim$14-30~kJ/mol):\cite{loison15}

\begin{eqnarray}
{\rm H + C_2N_2 + M}   & {\longrightarrow} & {\rm HC_2N_2 + M^{\ast} }  \\
{\rm H + HC_2N_2 }   & {\longrightarrow} & {\rm 2 \: HCN  }
\end{eqnarray}

{\em Future directions: } A larger cousin to cyanogen, dicyanoacetylene ($\rm C_4N_2$) is likely to exist in Titan's atmosphere, and detection of its ice has been proposed to explain a feature seen in {\em Voyager} IRIS and {\em Cassini} CIRS spectra at 478~\cm , \cite{khanna87, samuelson97, anderson10} although a lack of detection of the corresponding gas emission at 471 \cm\ has remained puzzling. \citeauthor{anderson16}\cite{anderson16} have proposed a possible explanation by way of ice grain surface chemistry combining HCN and \cyanoacet , but further laboratory and perhaps in situ experimental measurement is required to verify this hypothesis. For the time being, \cyanogen\ remains the only dicyanide molecule known in Titan's atmosphere.

%%%%%%%%%%%%%%%%%%%%%%%
\subsubsection{Acrylonitrile}

\begin{figure}
\includegraphics[scale=0.4]{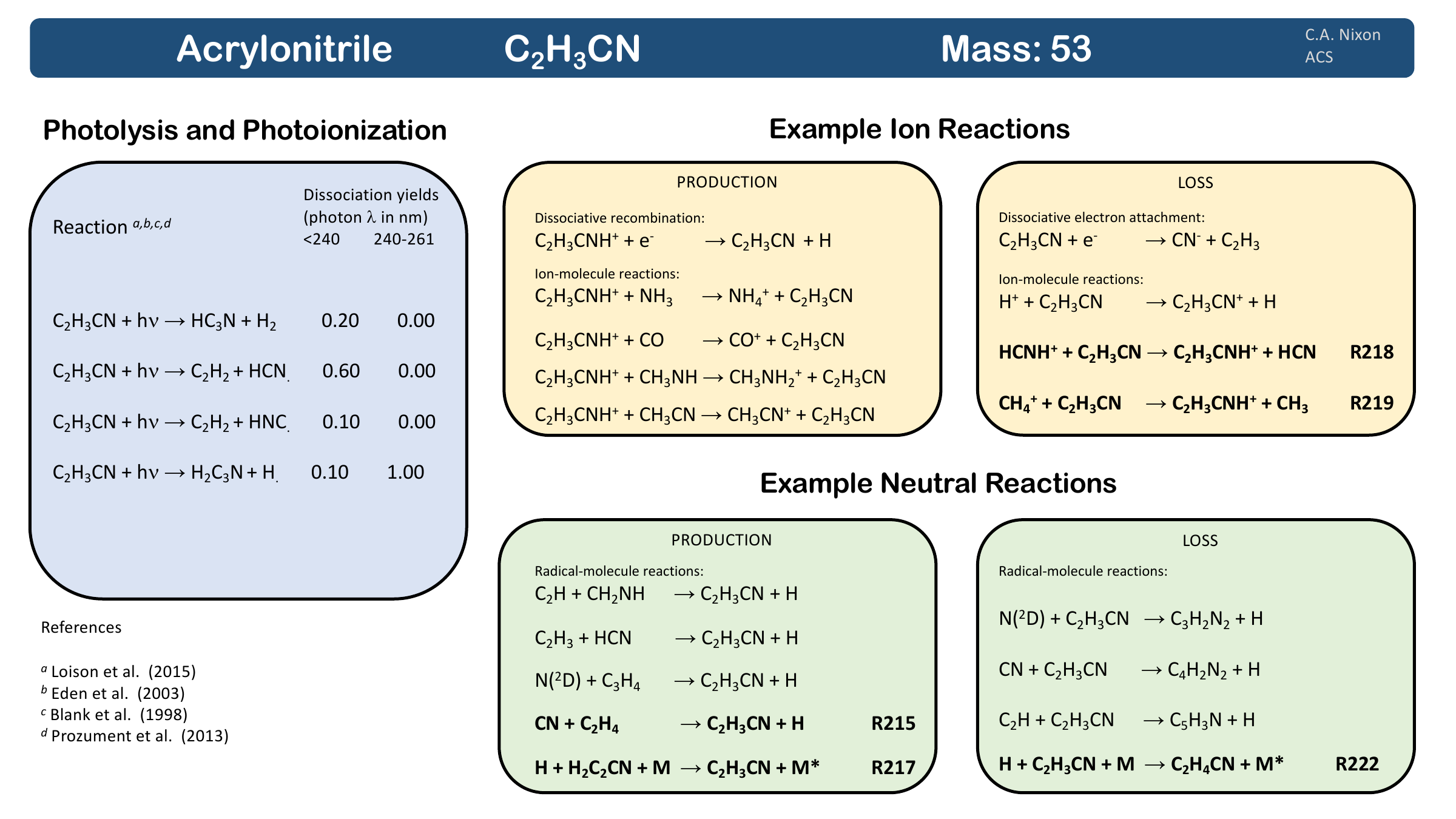}
  \caption{Acrylonitrile production and loss pathways. Reactions numbered and shown in bold correspond to discussion in the text.}
  \label{mol:acrylonitrile}
\end{figure}

Acrylonitrile (\acrylonitrile ) was the second molecule to be discovered on Titan at millimeter wavelengths using ALMA \cite{palmer17}, following the detection of propionitrile,\cite{cordiner15} discussed in the next section.

{\em Production:} Acrylonitrile (see Fig.~\ref{mol:acrylonitrile}\cite{loison15,eden03, blank98, prozument13}) is produced above 800 km by substitution of the CN radical onto ethylene\cite{sims93, gannon07}:

\begin{eqnarray}
{\rm C_2H_4 + CN}  & {\longrightarrow} & {\rm C_2H_3CN + H} 
\end{eqnarray}

Below 800 km acrylonitrile may be produced by the termolecular reaction chain:\cite{loison15,vuitton19}

\begin{eqnarray}
{\rm H + HC_3N + M}    & {\longrightarrow} & {\rm H_2C_2CN + M^{\ast}}  \\
{\rm H + H_2C_2CN + M}    & {\longrightarrow} & {\rm C_2H_3CN + M^{\ast}}
\end{eqnarray}

{\em Loss:} In a similar manner to HCN, $\rm HC_3N$ and other nitriles, \acrylonitrile\ is lost in the ionosphere by the two-step process of proton transfer:\cite{petrie91,petrie92}

\begin{eqnarray}
{\rm HCNH^+}  + {\rm C_2H_3CN}  & {\longrightarrow} & {\rm C_2H_3CNH^+ + HCN }  \\
{\rm CH_4^+}  + {\rm C_2H_3CN}  & {\longrightarrow} & {\rm C_2H_3CNH^+ + CH_3 }  
\end{eqnarray}

\noindent followed by dissociative electron recombination:\cite{vigren09}

\begin{eqnarray}
{\rm C_2H_3CNH^+}  + e^-  & {\longrightarrow} & {\rm C_2H_3 + HNC }  \\
& {\longrightarrow} & {\rm C_2H_2 + HCN + H }
\end{eqnarray}

In the lower atmosphere it may be lost to photodissociation (which tends to recycle acrylonitrile) or by H-addition:\cite{vuitton12}

\begin{eqnarray}
{\rm C_2H_3CN + H + M } & \longrightarrow & {\rm C_2H_4CN + M^{\ast} } 
\end{eqnarray}

{\em Future directions: }
Several small nitrile molecules, which tend to exhibit polar properties, have been investigated in a theoretical study for the potential to self-organize into spherical vesicles or membranes in non-polar liquids (such as, for example, Titan lakes and seas of methane-ethane-nitrogen). These calculations showed that acrylonitrile was the best candidate for forming so-called `azotosomes' \cite{stevenson15}, which, if experimentally confirmed, could be significant for astrobiology, as vesicles (containers) for self-replicating organisms. However at this time experimental verification of azotosomes is still lacking, while a later study has questioned the ability of these structures to form \cite{sandstrom20}.

%%%%%%%%%%%%%%%%%%%%%%%
\subsubsection{Propionitrile} 

\begin{figure}
\includegraphics[scale=0.4]{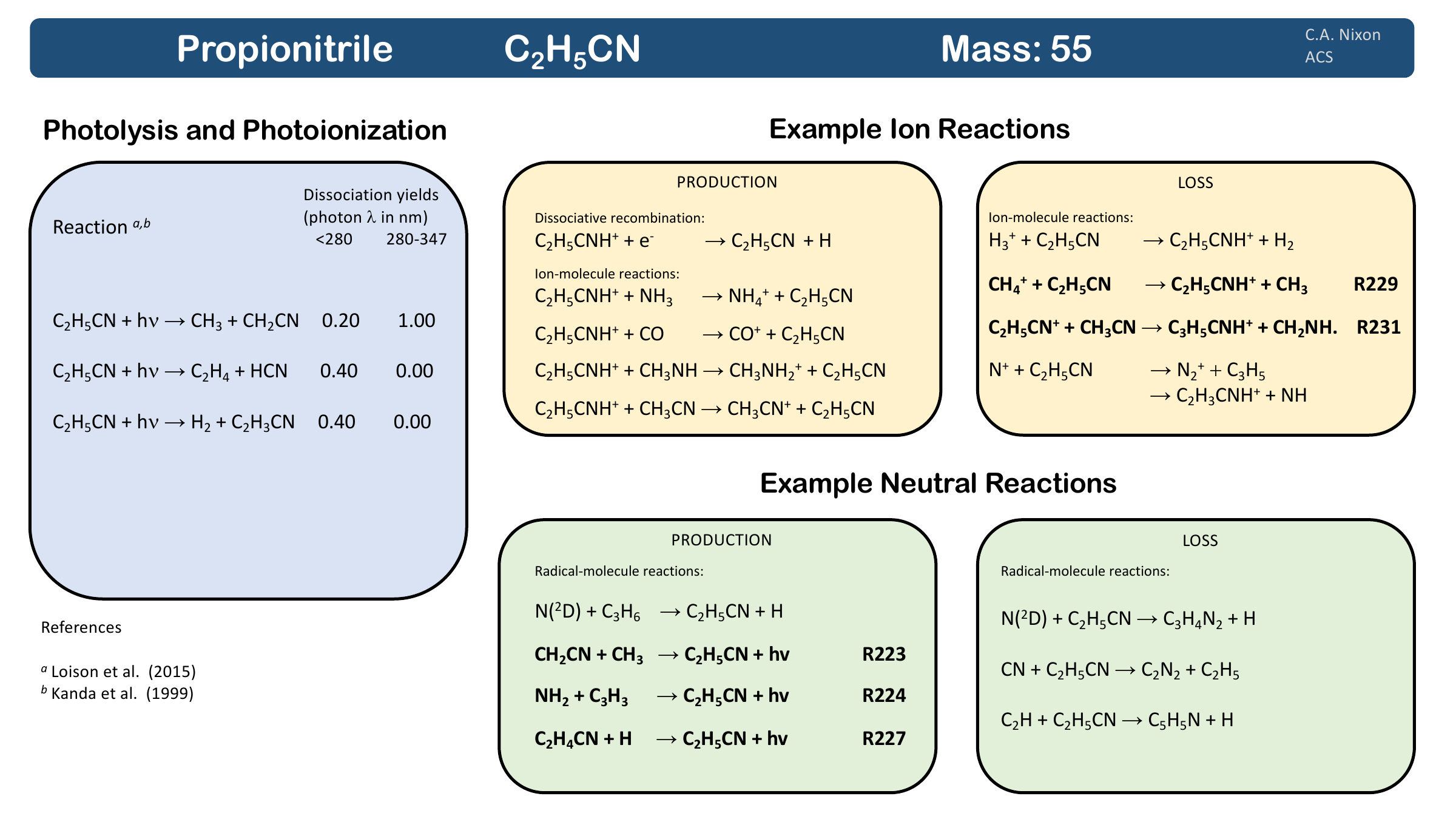}
  \caption{Propionitrile production and loss pathways. Reactions numbered and shown in bold correspond to discussion in the text.}
  \label{mol:propionitrile}
\end{figure}

Propionitrile (\ethylcyn ) was the second molecule to be originally detected using sub-millimeter astronomy and the first molecule with ALMA \cite{cordiner15}.

{\em Production:} Propionitrile has been posited to be produced above 900 km \cite{vuitton19} by the association reactions (see Fig.~\ref{mol:propionitrile})\cite{loison15,kanda99}:

\begin{eqnarray}
{\rm CH_3 + CH_2CN }  & {\longrightarrow} & {\rm C_2H_5CN + h\nu  } \\
{\rm C_3H_3 + NH_2 }  & {\longrightarrow} & {\rm C_2H_5CN + h\nu } 
\end{eqnarray}

Another proposed route to propionoitrile formation is:\cite{ krasnopolsky09, umemoto85}

 \begin{eqnarray}
{\rm C_3H_6 + N^{\ast}}  & {\longrightarrow} & {\rm C_2H_5CN + H  }  
\end{eqnarray}

In the middle atmosphere (400--900 km) \citep{vuitton19}, successive rounds of hydrogen addition to acrylonitrile via termolecular reactions can produce propionitrile:

\begin{eqnarray}
{\rm H + C_2H_3CN + M}    & {\longrightarrow} & {\rm C_2H_4CN + M^{\ast}}  \\
{\rm H + C_2H_4CN + M}    & {\longrightarrow} & {\rm C_2H_5CN + M^{\ast}}
\end{eqnarray}

\noindent or the termolecular reaction:\cite{loison15}

\begin{eqnarray}
{\rm CH_3 + CH_2CN + M}  & {\longrightarrow} & {\rm C_2H_5CNH + M^{\ast}  }  
\end{eqnarray}

{\em Loss:} As with other nitriles, the first step in loss of this nitrile in the ionosphere is proton transfer, forming $\rm C_2H_5CNH^+$:

\begin{eqnarray}
{\rm CH_4^+ + C_2H_5CN}  & {\longrightarrow} & {\rm C_2H_5CNH^+ + CH_3  }  
\end{eqnarray}

\noindent
This is followed by either dissociative electron recombination:\cite{vigren10}

\begin{eqnarray}
{\rm C_2H_5CNH^+}  + e^-  & {\longrightarrow} & {\rm CH_2CN + CH_3 + H }  
\end{eqnarray}

\noindent
or ion-neutral reactions such as:\cite{edwards08}

\begin{eqnarray}
{\rm C_2H_5CNH^+  + CH_3CN}  & {\longrightarrow} & {\rm C_3H_5CNH^+ + CH_2NH }  
\end{eqnarray}

{\em Future directions: } Propionitrile has been asserted to condense in pure crystalline form in Titan's atmosphere \cite{khanna05}, on the basis of an unexplained feature in Titan's far-infrared spectrum. This has been questioned based on vapor pressure of the gaseous form \cite{samuelson07, dekok08}, although it is possible that a co-condensed ice containing \propionitrile\ along with other gases may replicate the unexplained `haystack' emission \cite{nna-mvondo19}. Further work on spectroscopy will be required to determine if this is a unique solution, or if other possibilities exist.

%%%%%%%%%%%%%%%%%%%%%%%
\subsubsection{Cyanopropyne}

\begin{figure}
\includegraphics[scale=0.4]{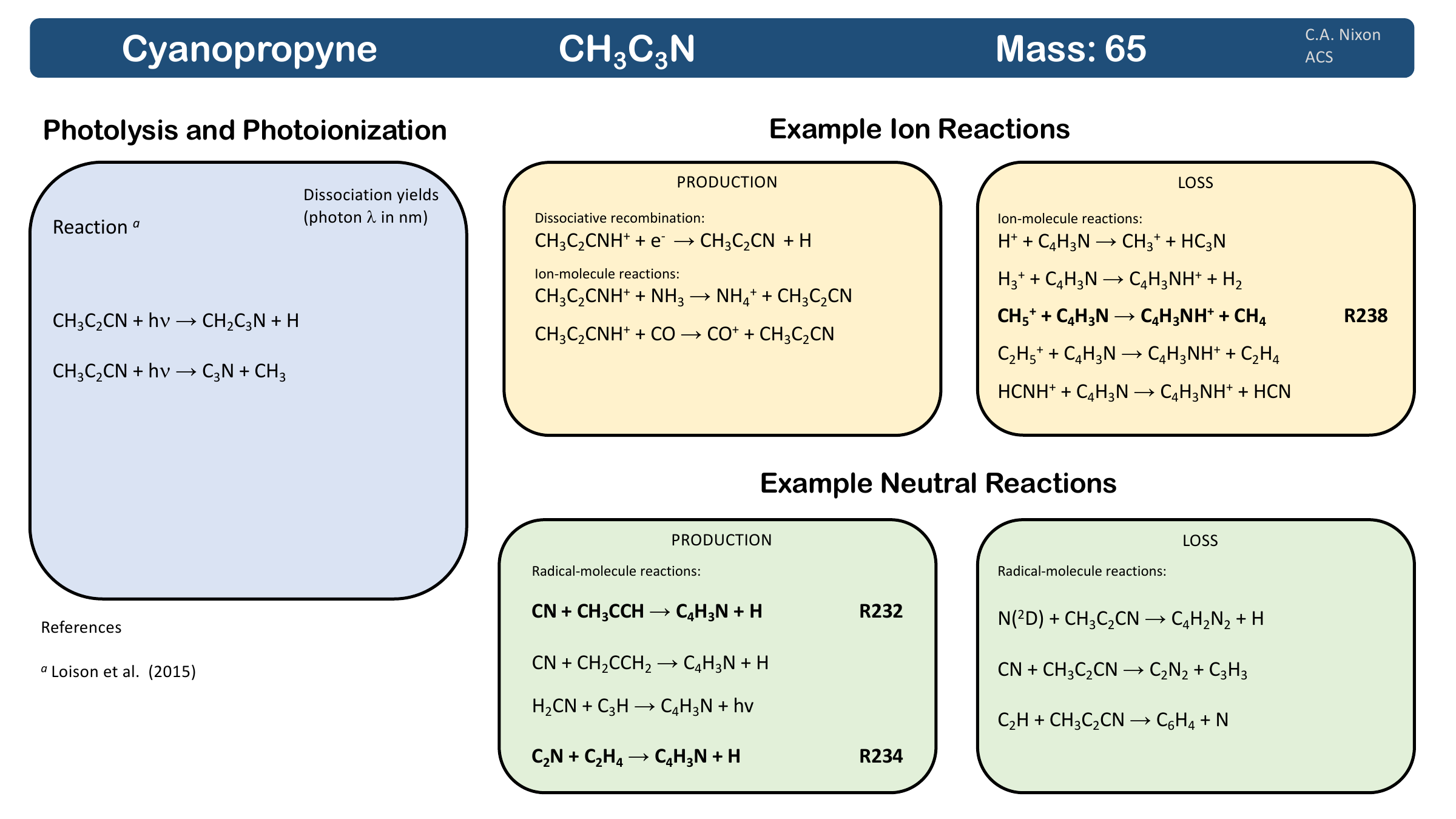}
  \caption{Cyanopropyne production and loss pathways. Reactions numbered and shown in bold correspond to discussion in the text.}
  \label{mol:cyanopropyne}
\end{figure}

Cyanopropyne {\cyanopropyne } was the fourth molecule to be discovered by ALMA spectroscopy of Titan at mm wavelengths \cite{thelen20}, following previous detection in the ISM \cite{broten84}.

{\em Production:} Production pathways for cyanopropyne (see Fig.~\ref{mol:cyanopropyne}) are more uncertain than for many other molecules due to the size and complexity of the molecule, allowing for more numerous reaction pathways, and multiple isomers of $\rm C_4H_3N$. Pathways involving radicals include CN substitution onto propyne or butadiene \cite{carty01, balucani02}:

\begin{eqnarray}
{\rm CN + CH_3CCH }    & {\longrightarrow} & {\rm CH_3C_3N + H}  \\
{\rm CN + CH_3CCCH_3 }    & {\longrightarrow} & {\rm CH_3C_3N + CH_3}
\end{eqnarray}

or $\rm C_2N$ attack on ethylene:\cite{zhu03}
\begin{eqnarray}
{\rm C_2N + C_2H_4 }    & {\longrightarrow} & {\rm CH_3C_3N + H}  
\end{eqnarray}

or acetylene via a three-step process with three-body reactions:\cite{zhu03, wang06, loison15}

\begin{eqnarray}
{\rm C_2N + C_2H_2}    & {\longrightarrow} & {\rm HC_4N + H } \\
{\rm HC_4N + H + M }    & {\longrightarrow} & {\rm CH_2C_3N + M^{\ast} } \\
{\rm CH_2C_3N + H + M}    & {\longrightarrow} & {\rm CH_3C_3N + M^{\ast} } 
\end{eqnarray}

{\em Loss:} Cyanopropyne is thought to be lost through either photolysis, or through protonation, e.g.:\cite{vuitton19}

\begin{eqnarray}
{\rm CH_5^+ + CH_3C_2CN}   & {\longrightarrow} & {\rm CH_3C_2CNH^+ + CH_4 }  
\end{eqnarray}

\noindent
followed by dissociative electron recombination:\cite{vuitton19}

\begin{eqnarray}
{\rm CH_3C_2CNH^+}  + e^-  & {\longrightarrow} & {\rm C_3H_3 + HNC }  \\
& {\longrightarrow} & {\rm HC_3N + CH_3 }
\end{eqnarray}

{\em Future directions: }
$\rm C_4H_3N$, has at least three stable isomers that have been detected in space \cite{mcguire18b, marcelino21}. Besides the currently detected isomer (cyanopropyne, \cyanopropyne , butynenitrile or methylcyanoacetylene), there is also cyanoallene ($\rm CH_2C_2HCN$)\cite{lovas06} and propargyl cyanide ($\rm HC_3H_2CN$)\cite{mcguire20}, both being first detected in the Taurus Molecular Cloud (TMC-1) at radio wavelengths. These provide good targets for detection on Titan, and their measurement would help to constrain photochemical pathways and models. Further more exotic arrangements of the same atoms may also exist and remain to be detected.

%%%%%%%%%%%%%%%%%%%%%%%%%%%%%%%%%%%%%%%%%%%%%%%%%%%%%%%%%%%%%%%%%%%%%
\subsection{Oxygen Compounds}

\begin{figure}
\includegraphics[scale=0.60]{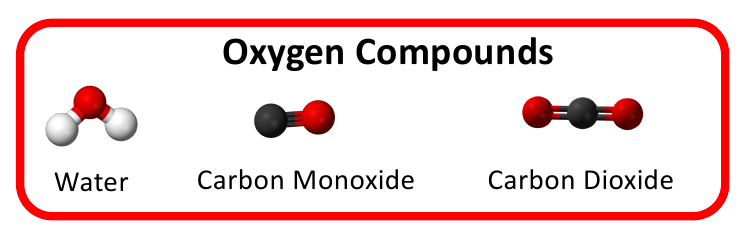}
  \caption{Oxygen-bearing molecules detected on Titan.}
  \label{fig:oxygen-mols}
\end{figure}

The oxygen chemistry of Titan's atmosphere is apparently straightforward, with few species involved - only CO, \coo\ and \water\ are presently observed (see Fig.~\ref{fig:oxygen-mols}) - but has proven remarkably difficult to replicate in models. Early work showed difficulty in producing sufficient CO from an external flux of water (OH) \cite{english96, lara96}, which was originally presumed to come from meteoritic and cometary materials. The discovery of the Enceladus plumes \cite{porco06,dougherty06,spahn06}, the connection to Saturn's E-ring (or Enceladus torus) and subsequent finding of both OH and O$^+$ entering Titan's upper atmosphere \cite{hartle06} apparently from Enceladus, provided an abundant and unambiguous source of oxygen. Subsequent work by \citeauthor{horst08}\cite{horst08} showed that the combination flux of O$^+$ could finally explain the abundance of CO. In the most recent work, \citeauthor{vuitton19}\cite{vuitton19} have shown that OH alone is sufficient to produce the CO via previously unrecognized reaction intermediaries.

%(network diagram)

%(vertical reactions diagram)

%%%%%%%%%%%%%%%%%%%%%%%
\subsubsection{Water}

Water was first detected in Titan's atmosphere through infrared spectroscopy with ISO \cite{coustenis98} through detection of emission lines at 39.4 and 43.9 \micron , and subsequently confirmed with {\em Cassini} CIRS \cite{cottini12b, bauduin18}.

\begin{figure}
\includegraphics[scale=0.4]{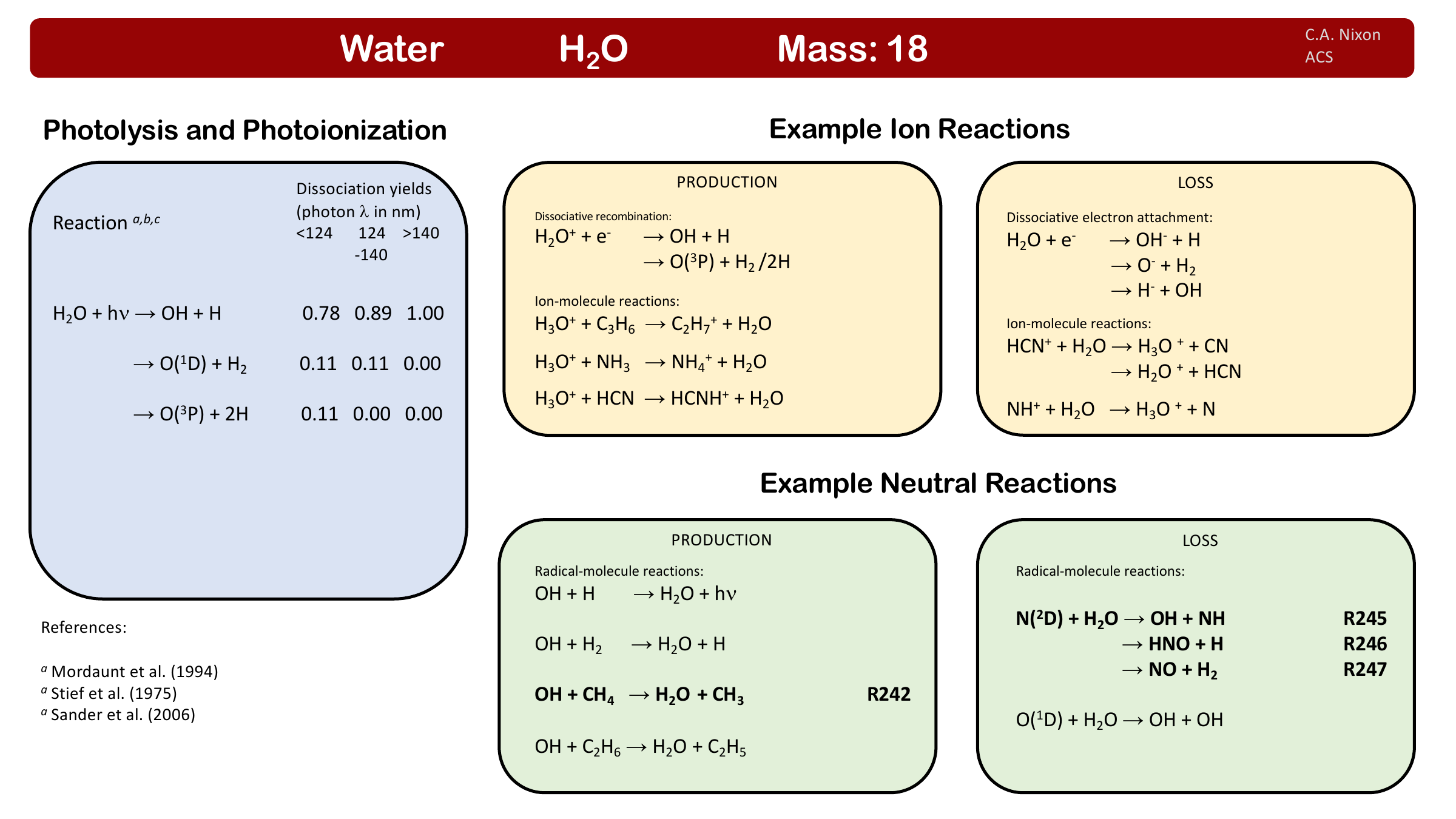}
  \caption{Water production and loss pathways. Reactions numbered and shown in bold correspond to discussion in the text.}
  \label{mol:water}
\end{figure}

{\em Production:} Water (Fig.~\ref{mol:water}\cite{mordaunt94, stief75, sander06a}) is thought to mainly be derived by the re-combination of OH infalling at the top of the atmosphere, primarily sourced from dissociated Enceladus water, with methane and its dissociation products :

\begin{eqnarray}
{\rm OH + CH_3 } & \longrightarrow & {\rm H_2O + {^1}CH_2 }  \\
{\rm OH + CH_4 } & \longrightarrow & {\rm H_2O + CH_3 } 
\label{eq:water}
\end{eqnarray}

{\em Loss:} Water is lost to photolysis throughout the atmosphere, reforming hydroxyl (OH). A large fraction of this OH reacts with $\rm CH_3$ to reform water (see previous equation). However, OH participates in several other reactions. Above 1000~km, it reacts with $\rm N(^4S)$ to form NO:\cite{dobrijevic14, wakelam15}

\begin{eqnarray}
{\rm OH + N(^4S) } & \longrightarrow & {\rm NO + H} 
\end{eqnarray}

\noindent
while in the middle atmosphere it reacts with CO to form \coo : \cite{dobrijevic14}

\begin{eqnarray}
{\rm OH + CO } & \longrightarrow & {\rm CO_2 + H} 
\end{eqnarray}

Water also reacts directly with excited state nitrogen atoms above 900~km:\cite{herron99, dobrijevic14}

\begin{eqnarray}
{\rm H_2O + N(^2D) } & \longrightarrow & {\rm NH + OH } \\
& \longrightarrow & {\rm HNO + H } \\
& \longrightarrow & {\rm NO + H_2 }
\end{eqnarray}

\noindent
and any remaining unreacted water is ultimately lost by condensation in the lower stratosphere.

{\em Future directions: } Due to its low vapor pressure, water remains difficult to measure in Titan's atmosphere. Currently there are large uncertainties in its vertical profile \cite{cottini12b, bauduin18}, and its latitudinal distribution remains unknown. Further work to better constrain these distributions may help to elucidate the relative importance of meteoritic versus Enceladus sources \cite{dobrijevic14}.

%%%%%%%%%%%%%%%%%%%%%%%
\subsubsection{Carbon Monoxide}

\begin{figure}
\includegraphics[scale=0.4]{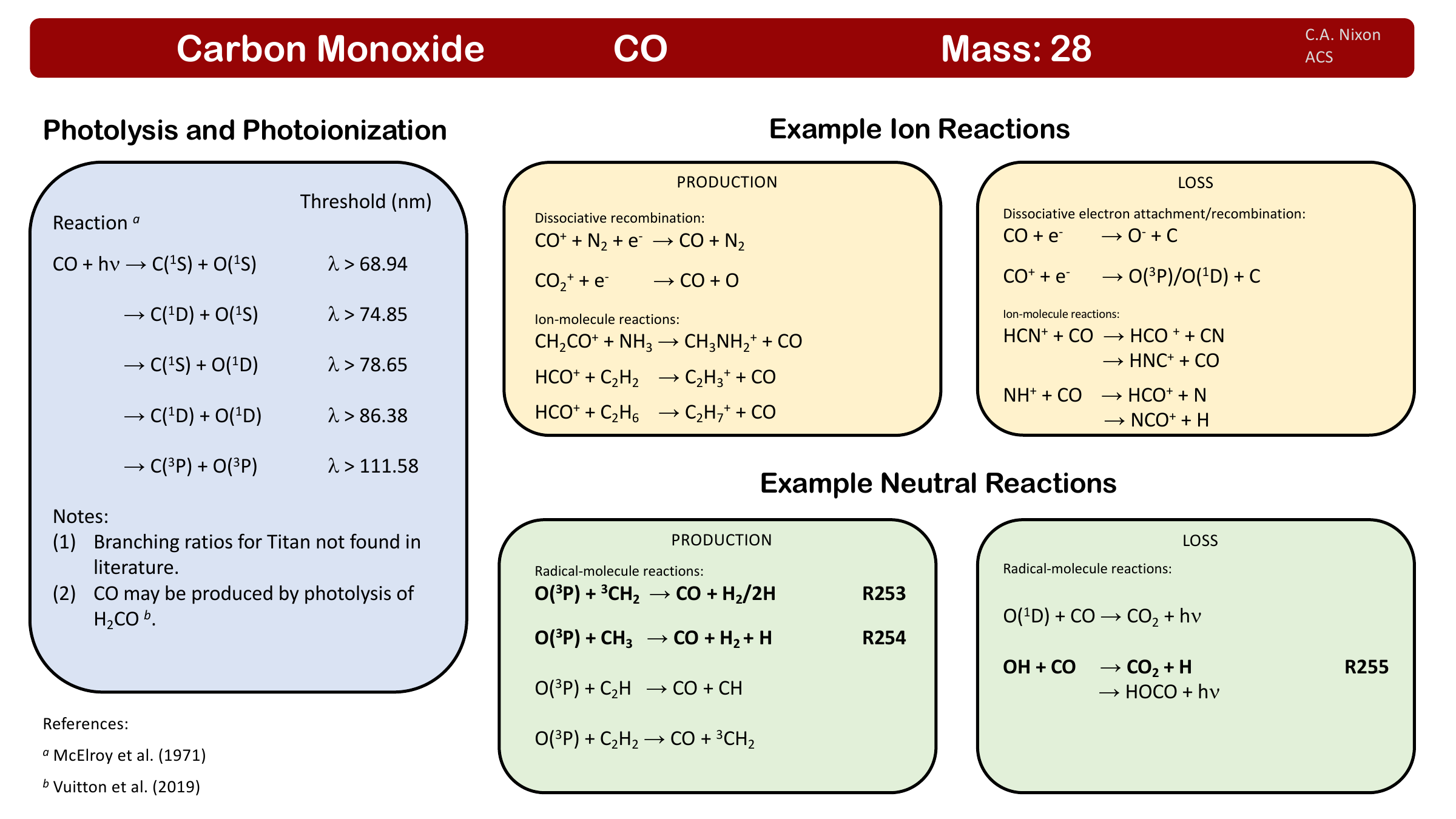}
  \caption{Carbon monoxide production and loss pathways. Reactions numbered and shown in bold correspond to discussion in the text.}
  \label{mol:co}
\end{figure}

Carbon monoxide (CO, Fig.~\ref{mol:co}\cite{mcelroy71}) was first detected on Titan by near-IR spectroscopy, showing an absorbance of CO at 1.6 \micron\ \cite{lutz83}, and the detection was soon confirmed at radio wavelengths \cite{marten88}. Estimates of its abundance fluctuated throughout the years following its discovery \cite{noll96}, and because these measurements were often sensitive to different altitudes, this led to the suggestion that the vertical profile was non-uniform \cite{marten88, hidayat98}. Subsequent measurements with high sensitivity telescope arrays at Owens Valley and Mauna Kea however showed evidence for a uniform profile, converging on a mixing ratio of $\sim 50$~ppb \cite{gurwell00, gurwell04}.

Recently, high-sensitivity observations with ALMA have narrowed the experimental error to a range of $50 \pm 2$~ppm \cite{serigano16}, making it the fourth most abundance species in Titan's atmosphere after \nitrogen , \methane\ and \hydrogen . As a triple-bonded molecule, CO once produced is both resistant to photolysis and chemical reaction, as well having no loss through condensation, and hence all evidence available at present points to a uniform vertical profile.

{\em Production:} In the model of \citeauthor{vuitton19}\cite{vuitton19} CO is mostly produced via formation of formaldehyde, and subsequent photolysis:\cite{glicker71, cooper96, meller00}

\begin{eqnarray}
{\rm OH + CH_3} & \longrightarrow & {\rm CHOH + H_2 }  \\
{\rm CHOH + H } & \longrightarrow & {\rm H_2CO + H } \\
{\rm H_2CO + h\nu } & \longrightarrow & {\rm CO + H_2 } 
\end{eqnarray}

(Formaldehyde may also be created by reaction of OH with $^3$CH$_2$, C$_2$H$_4$ etc). 

Alternatively, \formaldehyde\ may undergo a two-step process to form CO:\cite{baulch05, tsang86}

\begin{eqnarray}
{\rm H_2CO + h\nu } & \longrightarrow & {\rm HCO + H } \\ 
{\rm HCO + (H/CH_3) } & \longrightarrow & {\rm CO + (H_2/CH_4) }
\end{eqnarray}

\noindent
(also OH + $^3$CH$_2$, OH + C$_2$H$_4$ etc). Note that formaldehyde has yet to be detected in Titan's atmosphere.

Lesser routes to CO production may be through reaction of atomic oxygen (deposited to the top of the atmosphere from Enceladus) with methane fragments, e.g.:\cite{horst08}

\begin{eqnarray}
{\rm O({^3}P) + {^3}CH_2 } & \longrightarrow & {\rm CO + 2H / H_2} \\ 
{\rm O({^3}P) + CH_3 } & \longrightarrow & {\rm CO + H_2 + H }
\end{eqnarray}

{\em Loss:} CO is primarily lost slowly through reaction with OH forming \coo :\cite{sander06b}

\begin{eqnarray}
{\rm CO + OH } & \longrightarrow & {\rm CO_2 + H } 
\label{eq:coo}
\end{eqnarray}

\noindent
which is in turn lost through condensation.

{\em Future directions:} Laboratory chemistry simulations of Titan's atmosphere have shown that CO may react with methane and nitrogen when sufficiently stimulated, forming amino acids and even nucleobases \cite{horst12}. This provides an exciting possibility of astrobiology that now requires remote and in situ measurements to confirm.

%%%%%%%%%%%%%%%%%%%%%%%
\subsubsection{Carbon Dioxide}

\begin{figure}
\includegraphics[scale=0.4]{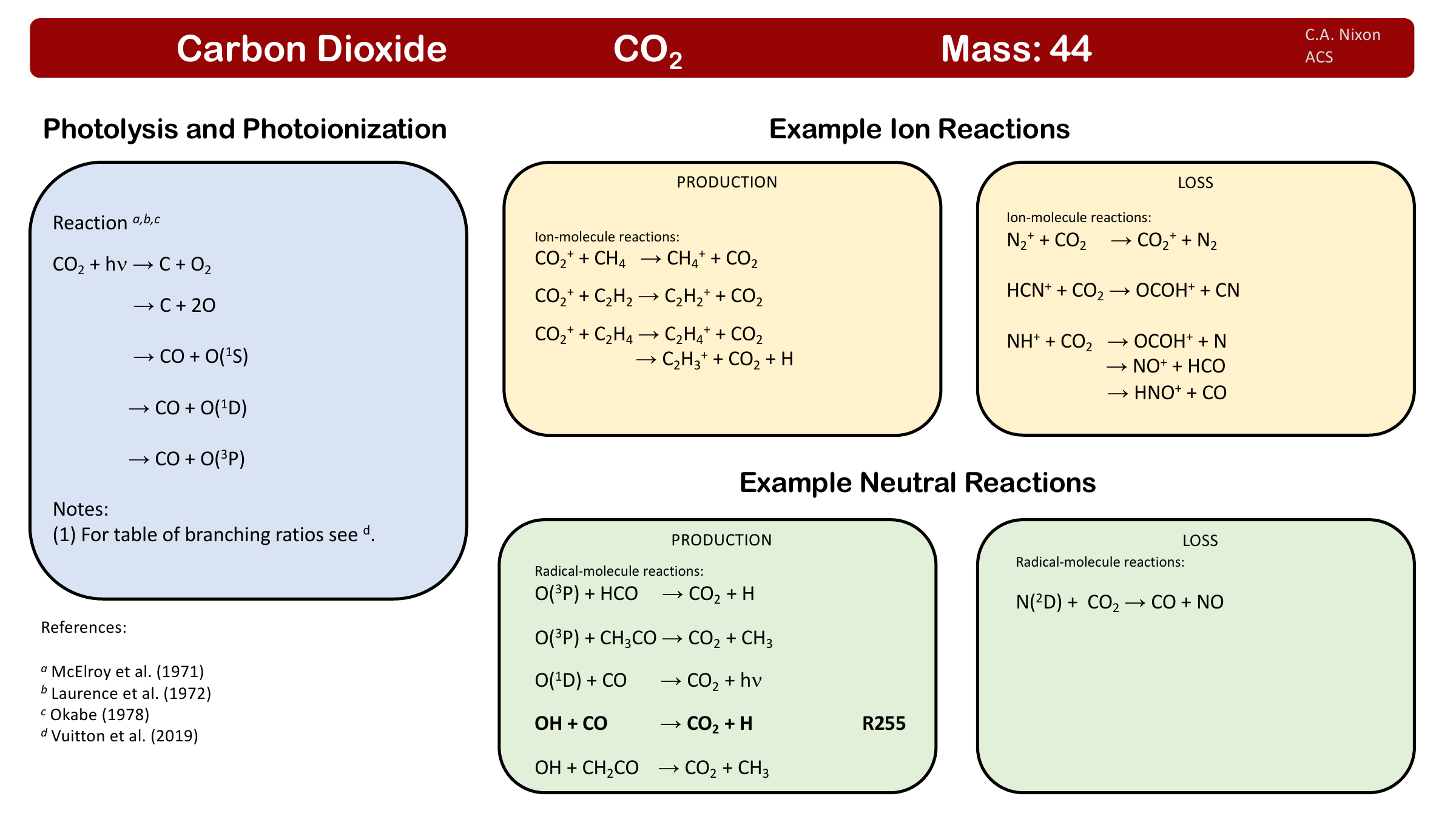}
  \caption{Carbon dioxide production and loss pathways.}
  \label{mol:coo}
\end{figure}

Carbon dioxide (\coo ) was one of seven new gas species detected by {\em Voyager}'s IRIS infrared spectrometer \cite{samuelson83, coustenis95b} and subsequently by {\em Cassini} CIRS \cite{dekok07a, vinatier10a}. Unlike shorter-lived chemical chemical species (\eg\ \cyanoacet , \diacet ), \coo\ exhibits little variation with latitude in the lower stratosphere, lacking a polar enhancement.

{\em Production:} \coo\ is thought to be mostly produced from CO + OH as shown in the previous section (R \ref{eq:coo}). See also Fig.~\ref{mol:coo}\cite{mcelroy71,okabe78,vuitton19}.

{\em Loss:} 
The principle loss pathway of \coo\ is through photolysis:\cite{dobrijevic14, chan93, yoshino96, parkinson03, stark07, shemansky72}

\begin{eqnarray}
{\rm CO_2  + h\nu } & \longrightarrow & {\rm CO + O(^1D)   } \:\:\:\:  z >  200 {\rm km} \\ 
& & {\rm CO + O(^3P)   } \:\:\:\:  z < 200 {\rm km} 
\end{eqnarray}

\noindent
and also through condensation in the lower stratosphere.

{\em Future directions: } \coo\ remains the most significant atmospheric oxygen repository after CO, but is shorter-lived and more reactive. It exhibits surprisingly little spatial and seasonal variation in Titan's atmosphere \cite{teanby19,vinatier15} which requires further astronomical monitoring to confirm.

%%%%%%%%%%%%%%%%%%%%%%%%%%%%%%%%%%%%%%%%%%%%%%%%%%%%%%%%%%%%%%%%%%%%%
\subsection{Growth of large particles}
\label{sect:haze}

\begin{figure}
\includegraphics[scale=0.55]{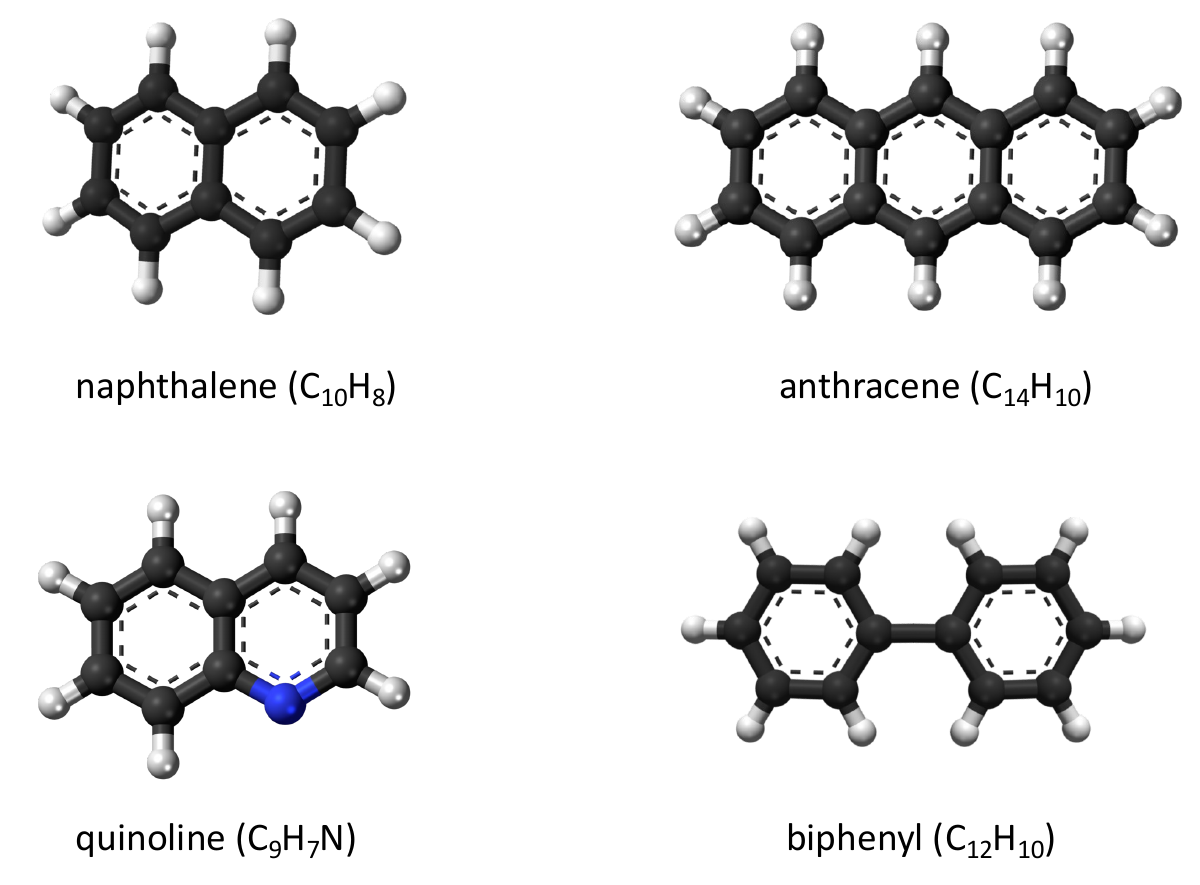}
  \caption{Cyclic or `ring' molecules. Naphthalene - two fused carbon rings. Quinoline - two fused carbon rings, with one nitrogen substitution (blue). Anthracene - three fused carbon rings. Biphenyl - two bonded but not fused carbon rings.  Image credit: individual PAH graphics from wikimedia commons. }
  \label{fig:rings}
\end{figure}

As hydrocarbon molecules grow to ever-larger sizes, they may take several forms: long chains, fused rings (PAHs - polycyclic aromatic hydrocarbons), or rings connected by hydrogen bonds (poly-phenyls) (see Fig.~\ref{fig:rings}). Nitrogen incorporation is also likely, for example in the form of PANHs (polycyclic aromatic nitrogen heterocycles). It is thought that eventually larger molecules clump together due to electrostatic forces to form fractal aggegrates \cite{cabane92,cabane93,rannou95,rannou00} - Titan haze particles which form the well-known golden haze at visible wavelengths. These in turn become the nuclei for stratospheric hydrocarbon ice particles or tropospheric methane raindrops \cite{lorenz93,lorenz95,mckay01,karkoschka09}, and fall to the surface where they form Titan's dune fields \cite{lorenz06}.

\begin{figure}
\includegraphics[scale=0.55]{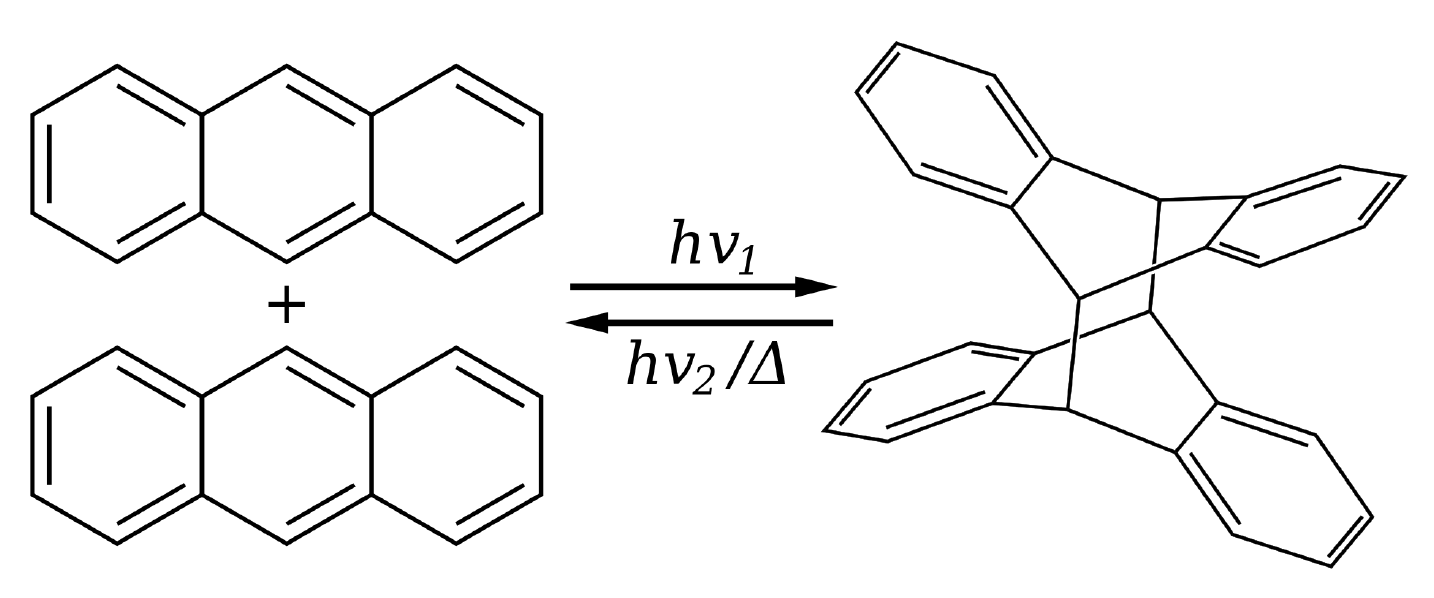}
  \caption{The anthracene dimer (wikimedia). }
  \label{fig:dimer}
\end{figure}

The presence of PAHs on Titan has been studied as far back as the 1990s \cite{sagan93} in laboratory experiments. However despite these predictions, detection of specific PAHs has remained elusive. The closest we have come so far to identification of a unique PAH in Titan's atmosphere was the sighting of a peak at $m/z=178$ in the CAPS spectrum by \citeauthor{waite07}\cite{waite07}, along with another peak at twice the mass: $m/z = 356$. These were tentatively identified as due to anthracene and its dimer (Fig.~\ref{fig:dimer}), although non-aromatic structures could not be ruled out. Note that the dimer itself may have formed by wall reactions in the instrument, but this would not rule out the fact that anthracene was entering the instrument. Likewise, no polyphenyls or N-heterocycles have been uniquely identified. 

{\em Production:} The formation of PAHs by addition to benzene rings has been a topic of debate for considerable time. Recently, \citeauthor{kaiser21}\cite{kaiser21} have categorized possible pathways into five principal mechanisms:

\begin{itemize}
\item
HACA - "hydrogen abstraction- \acet\ addition"\cite{frenklach91} 
\item
HAVA - "hydrogen abstraction- vinylacetylene addition" 
\item
PAC - "phenyl addition-dehydrocyclization"
\item
RRR - "radical-radical reactions"
\item
MACA - "methylidene addition-cyclization aromatization"
\end{itemize}

Each of these offers potential pathways to larger molecules. In brief, the HAVA mechanism, in which vinylacetylene ($\rm C_4H_4$) adds to aromatic rings (such as benzene) in a barrierless reaction is thought to be principle mechanism by which additional six-membered rings are added to existing rings at low temperatures, such as in planetary atmospheres. The other mechanisms offer alternate routes to addition of five and six-sided rings, predominantly at high temperatures of 1000s of K (HAVA, PAC, RRR), although MACA may operate at low temperatures to form indene. The reader is directed to the review paper by \citeauthor{kaiser21}\cite{kaiser21} for a full description, which is beyond the scope of this article.

No discussion of aerosol particle growth would be complete without mention of negative ions. The discovery of large negatively charged ions at high altitudes by Cassini's CAPS instrument was one of the major surprises about Titan's atmosphere early in the mission \cite{waite07, coates07, coates09a, coates09b, vuitton09b, crary09}. Small negative ions may be formed by several processes, including dissociative electron attachment, e.g.:

\begin{eqnarray}
{\rm CH_4 + e^- } & \longrightarrow & {\rm CH_2^- + H_2 } \\ 
{\rm CH_4 + e^- } & \longrightarrow & {\rm H^- + CH_3 } 
\end{eqnarray}

\noindent
and by radiative attachment to a radical species already formed through photochemistry:\cite{vuitton19}

\begin{eqnarray}
{\rm CH_3 + e^- } & \longrightarrow & {\rm CH_3^- + h\nu } \\
{\rm C_4H + e^- } & \longrightarrow & {\rm C_4H^- + h\nu } \\
{\rm C_6H + e^- } & \longrightarrow & {\rm C_6H^- + h\nu } \\
{\rm HC_5N + e^- } & \longrightarrow & {\rm HC_5N^- + h\nu } 
\end{eqnarray}

\noindent
Once H$^-$ is produced, it leads to the creation of some larger negative ions through proton abstraction:\cite{vuitton19}

\begin{eqnarray}
{\rm C_2H_2 + H^- } & \longrightarrow & {\rm C_2H^- + H_2 } \\
{\rm HC_3N + H^- } & \longrightarrow & {\rm C_3N^- + H_2 } 
\end{eqnarray}

Successively larger aerosol particles are produced through a variety of ion-neutral and ion-ion reactions \cite{wellbrock19, dubois19, mukundan18}. The largest charged particles tend to be predominantly negative ions, due to the larger to their higher mobility. 

{\em Loss:} Since aromatic rings, once formed, are very stable, with many possibilities to disperse absorbed energy internally, their principle loss channels will be either (a) to form radicals and then larger molecules; (b) to agglomerate or (c) to condense.

{\em Future directions:} Much work is still required to further elucidate the growth of large particles, especially the relative important of ion vs neutral chemistry at different altitudes.
See also the later Section on the topic of PAHs.

%%%%%%%%%%%%%%%%%%%%%%%%%%%%%%%%%%%%%%%%%%%%%%%%%%%%%%%%%%%%%%%%%%%%%
%%%%%%%%%%%%%%%%%%%%%%%%%%%%%%%%%%%%%%%%%%%%%%%%%%%%%%%%%%%%%%%%%%%%%
\section{Gaps in our knowledge}
\label{sect:future}
%%%%%%%%%%%%%%%%%%%%%%%%%%%%%%%%%%%%%%%%%%%%%%%%%%%%%%%%%%%%%%%%%%%%%
%%%%%%%%%%%%%%%%%%%%%%%%%%%%%%%%%%%%%%%%%%%%%%%%%%%%%%%%%%%%%%%%%%%%%

In this section we consider where the gaps are in our current knowledge and understanding of Titan's atmospheric composition and chemistry, and how these gaps might be addressed in the near future through combinations of astronomy, laboratory and theoretical work.

%%%%%%%%%%%%%%%%%%%%%%%%%%%%%%%%%%%%%%%%%%%%%%%%%%%%%%%%%%%%%%%%%%%%%
\subsection{Aliphatic Species}

%%%%%%%%%%%%%%%%%%%
\subsubsection{Hydrocarbons}
\label{sect:futurec}

At present, \diacet\ remains the only $\rm C_4$ hydrocarbon species definitively detected in the neutral atmosphere, while no aliphatic $\rm C_n$ species with $n \geq 5$ have been detected (benzene, \benzene , being detected as a ring). 

%Figure~\ref{fig:missing} shows the gaps in our knowledge of small Titan hydrocarbons: while we have relatively complete detections of C1--C3 long-lived hydrocarbons (green, yellow) with a a few exceptions (such as cyclopropane), only two hydrocarbons have been detected with C4--C6 (diacetylene and benzene) leaving many gaps in the 'hydrocarbon periodic table'.   

%\begin{figure}
%\includegraphics[scale=0.55]{missing-hydrocarbons.pdf}
%  \caption{Hydrocarbon molecular `periodic table' for Titan's atmosphere. Green  = compounds where expected stable isomers have been detected. Yellow = some isomers missing (e.g. cyclopropane for C$_3$H$_6$). Orange = compounds for which no detections have been made in the neutral atmosphere. Grey = no compound possible.}
%\label{fig:missing}
%\end{figure}

Some example reactions that are producing $\rm C_4H_x$ species are thought to occur include\cite{knyazev01, laufer04, loison06, lavvas08a}: 

\begin{eqnarray}
{\rm CH_3 + C_3H_3 +M}  & {\longrightarrow} & {\rm C_4H_6 + M^{\ast} } \\ 
{\rm CH + C_3H_8 }  & {\longrightarrow} & {\rm C_4H_8 + H } \\ 
{\rm C_2H_3 + C_2H_5 + M}  & {\longrightarrow} & {\rm C_4H_8 + M^{\ast} } 
\end{eqnarray}

\begin{figure}
\includegraphics[scale=0.4]{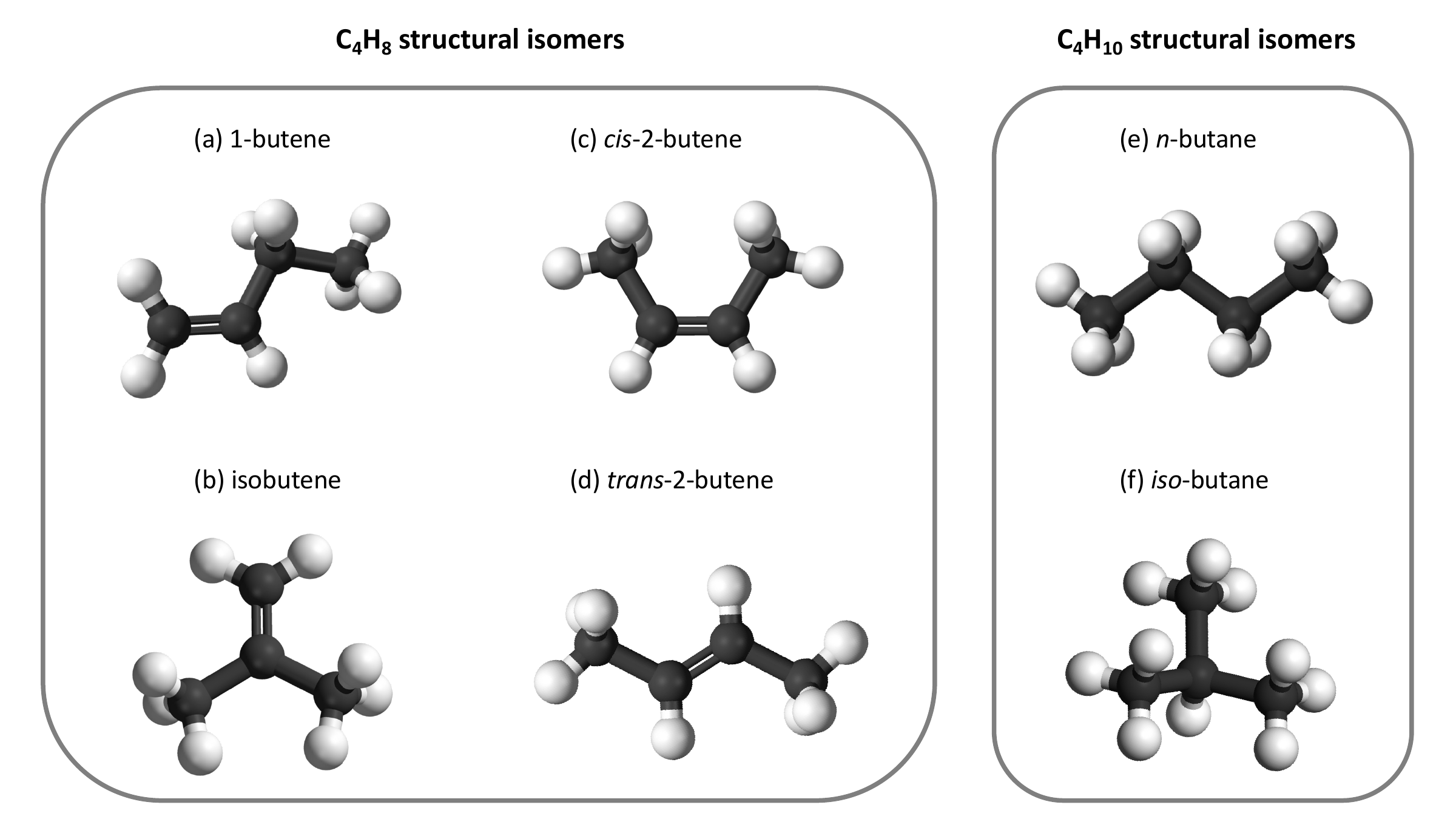}
\caption{Structural isomers of butene (C$_4$H$_8$) and butane (C$_4$H$_{10}$), showing that branched chains become possible when four or more carbon atoms are present. Not all possible isomers are shown - additional cyclic forms are also possible (e.g. cyclobutane, C$_4$H$_8$)}
\label{fig:c4s}
\end{figure}

Species with four or more carbon atoms show increased possibilities for structural isomerization, as illustrated in Fig.~\ref{fig:c4s}. This results in both increased challenges for observational detection, as well as new and interesting chemical possibilities such as branched chain molecules.

Upper limits for $n$-butane and $i$-butane (\butane ) have been calculated as $5\times 10^{-7}$ and $4\times 10^{-8}$ respectively, from CIRS spectra at 200--250~km, 30\dg N--30\dg S\cite{hewett20}.

{\em Future Work:} Astronomical detection of species such as $\rm C_4H_x$, $x = 4, 6, 8, 10$ (i.e. butynes and butenes) and even $\rm C_5H_x$ would greatly help to improve constraint on photochemical models, which are currently lacking in data to model for this regime. Building our knowledge of aliphatic species such as C4's and C5's may provide a better understanding of the pathways to benzene or other cyclic species. 

%In addition, specific molecules are of interest such as vinyl acetylene ($\rm C_4H_4$) which is implicated in the addition of further rings to benzene, and isoprene ($\rm C_5H_8$) which is an important biomarker on Earth, being produced by plants \cite{seager16}.

In parallel, photochemical models that currently treat multiple isomers under single formulae such as $\rm C_4H_8$ and $\rm C_4H_{10}$ must continue to expand treatment of separate isomers. To date this has been sparse and primarily for a few specific cases for which isomeric data exists: especially HCN and HNC, and \propadiene\ and \propyne , which are considered separately in current models \cite{hebrard12, hebrard13, lic15, vuitton19}. A good reason for lack of inclusion of separate isomers is lack of knowledge of branching ratios and isomer-specific reaction rates. Monte Carlo simulations have proved useful at showing which reaction rate uncertainties have the biggest effects on the uncertainties in the solution \cite{hebrard07, hebrard09}. These researches thereby provide important prioritization of where resources such as laboratory time and theoretical effort (e.g. TST calculations) can best be spent to most rapidly improve our knowledge.
 
%%%%%%%%%%%%%%%%%%%  
\subsubsection{Nitrogen Species}
\label{sect:futuren}

All of the nine detected nitrogen compounds in Titan's neutral atmosphere, formed from dissociation products of \nitrogen\ and \methane , are triple bonded. These include \nitrogen\ itself, and eight known cyanides, which include nitrogen in a terminal ${\text -}$C${\equiv}$N functional group. 

Nitriles (cyanides) appear stable and plentiful in Titan's atmosphere, and further examples are sure to be found. In 1985 in the wake of the {\em Voyager} IRIS discoveries, \citeauthor{cerceau85}\cite{cerceau85} studied the infrared spectra of seven undetected nitriles to facilitate further new detections. 35 years later, four of these seven species had been detected on Titan (\acetonitrile , \vinylcyn , \ethylcyn\ and \butynenitrile ), although ironically all these detections were made through sub-millimeter wave astronomy, not infrared spectroscopy \cite{marten02, cordiner15, palmer17, thelen20}. Upper limits for the three remaining undetected species, plus one other have been calculated by \citeauthor{coustenis93b}\cite{coustenis93b}: 

\begin{tabbing}
\hspace*{1cm} \= crotonitrile \hspace*{1cm} \= ($\rm CH_3(CH)_2CN$)  \hspace*{1cm} \= $2.5\times 10^{-7}$ \\
\> butanenitrile  \> ($\rm CH_3(CH_2)_2CN$) \>  $5.0 \times 10^{-7}$ \\
\> isobutyronitrile \> ($\rm (CH_3)_2 CHCN$) \> $2.0 \times 10^{-7}$ \\
\> cyanocycloproane \> ($\rm \Delta -CN$) \> $1.5\times 10^{-7}$.
\end{tabbing}

\begin{figure}
\includegraphics[scale=0.55]{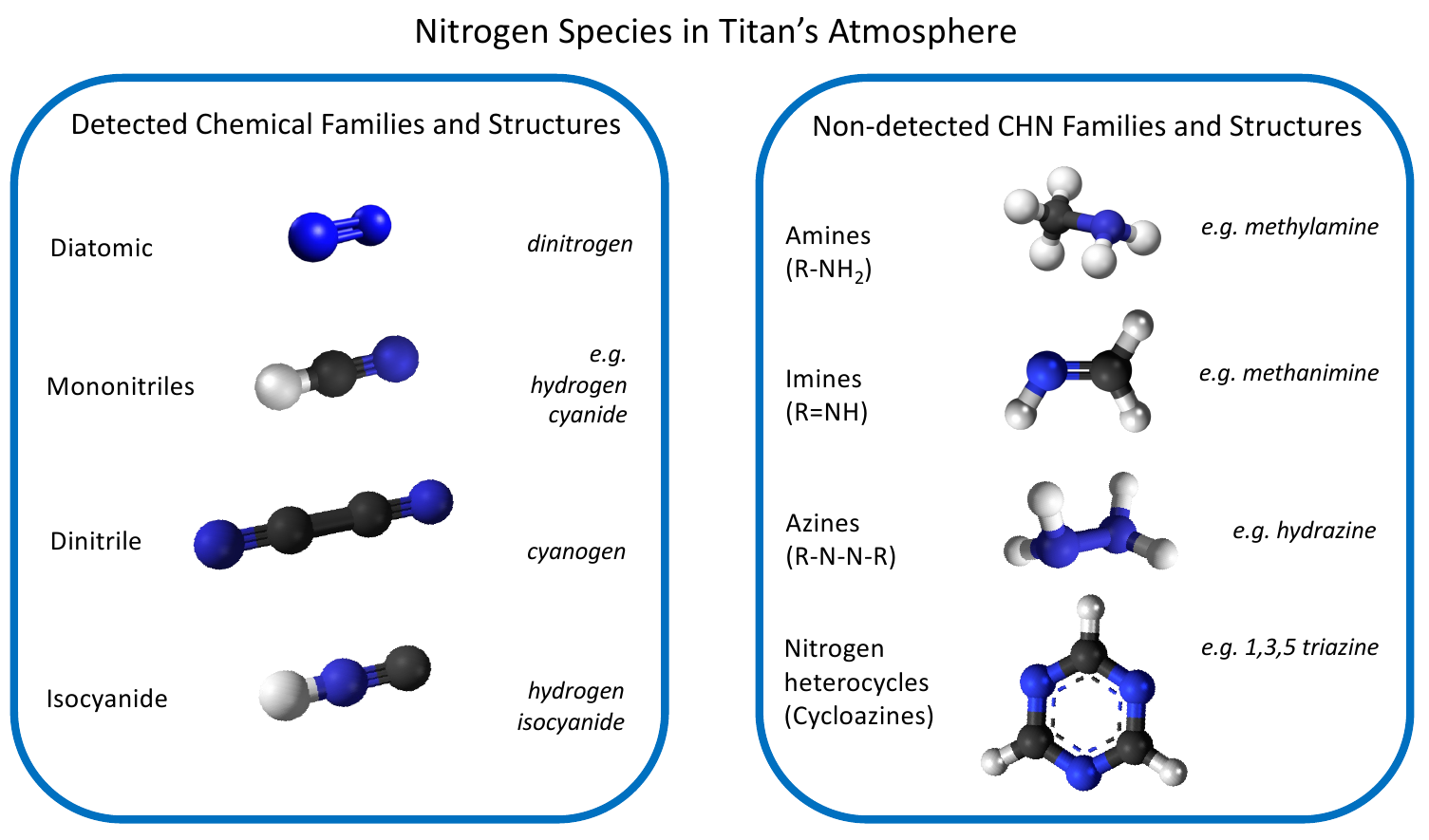}
  \caption{Examples of detected and non-detected nitrogen molecular families in Titan's atmosphere.}
  \label{fig:missingn}
\end{figure}

Nitrogen has yet to be detected in other types of bonding - {\em i.e.} where it is not in a triple bond, such as in amines, imines (beyond HNC), azines and nitrogen heterocycles (see Fig.~\ref{fig:missingn}). A simple example is ammonia (\ammonia ) which was tentatively inferred from {\em Cassini}'s mass spectra at high altitudes, at mixing fractions of ${\sim}3{\text -}4{\times}10^{-5}$, although not yet uniquely detected due to the barometric degeneracy problem of unit resolution mass spectroscopy.\cite{vuitton06a,vuitton07}

The major channel for ammonia production is thought to be via NH$_2$ + CH$_2$N, as follows:\cite{vuitton19}

\begin{eqnarray}
{\rm NH + C_2H_4 }  & {\longrightarrow} & {\rm NH_2 + C_2H_3  } \\ 
{\rm CH_2NH_2^+ + e^- }  & {\longrightarrow} & {\rm NH_2 + {^3}CH_2 } \\ 
{\rm N(^4S) + CH_3 }  & {\longrightarrow} & {\rm H_2CN + H }  \\
{\rm NH_2 + H_2CN }  & {\longrightarrow} & {\rm NH_3 + HCN }  
\end{eqnarray}

Ammonia is mostly lost through photodissociation.

Upper limits for some molecules in the lower atmosphere have been estimated, including for ammonia (\ammonia ) of 1.3 ppb (3-$\sigma$) at 107 km, 25\dg S \cite{nixon10b}, and methanimine (\methanimine ) at 0.35 ppb (3-$\sigma$) in the stratosphere \cite{teanby18}.

{\em Future Work:} No molecules having both oxygen and nitrogen, e.g. HNO, and more complex amino acids, have been detected. Detection of such functional groups and molecules, important for biological activity on Earth, is a key area of future research. Several nitriles have been detected only through sub-millimeter astronomical techniques as previously mentioned, but not using infrared techniques despite several decades of attempts \cite{cerceau85, coustenis93b, coustenis03, nixon10b}. In fact  laboratory line lists with intensities do not exist for many of the nitriles sought in the infrared, except \acetonitrile\ \cite{rinsland05, rinsland08}, hampering the search and implying a need for new laboratory work to obtain cold temperature spectra. 

%%%%%%%%%%%%%%%%%%%%
\subsubsection{Oxygen species}
\label{sect:futureo}

Few oxygen compounds have been detected in Titan's atmosphere: only CO, \coo\ and \water\ to date. Diatomic molecular oxygen, $\rm O_2$, readily found in the atmospheres of the inner planets, is absent, allowing organic chemistry to proceed to great complexity. The oxygen compounds detected appear to be attributable to an external source of oxygen (both O and OH), most likely originating from Enceladus \cite{hartle06, horst08}. Prior to the discover of the Enceladus plumes a meteoritic source was favored \cite{wong02}, and some meteoritic contribution may still be present \cite{dobrijevic14}.

Other than CO, which is present at a relatively high abundance ($\sim 50$~ppm \cite{serigano16}), oxygen is a minor although potentially important ingredient of Titan's atmosphere. This is because many molecules of biological importance require oxygen \cite{seager16}. As early as the 1980s it had already been demonstrated that hydrolysis of Titan tholins ($\rm H_xC_yN_z$) added oxygen to form amino acids of biological relevance \cite{khare84b} - subsequently confirmed in many similar experiments. More recently, \citeauthor{horst12}\cite{horst12} showed that even in the gas phase, amino acids may be synthesized in a Titan-like atmosphere when CO is added to mixtures of \methane\ and \nitrogen\ in RF discharge experiments.

Photochemical models \cite{dobrijevic14} predict that trace amounts of molecules such as HNO, HNCO, \formaldehyde\ and \methanol\ should be present in the atmosphere, and perhaps detectable at high altitudes by observatories such as ALMA. To date, few published studies have attempted to directly identify further oxygen species. The {\em Cassini} INMS team published upper limits for methanol ($\rm CH_3OH$) and acetaldehyde ($\rm CH_3CHO$) of 30 ppb and 10 ppb in the ionosphere at 1100 km \cite{vuitton07}. In the stratosphere, upper limits have been determined for methanol and formaldehyde of 6 ppb and 2 ppb respectively at at 107 km, 25\dg S \cite{nixon10b} from infrared spectroscopy with {\em Cassini} CIRS.

\begin{figure}
\includegraphics[scale=0.55]{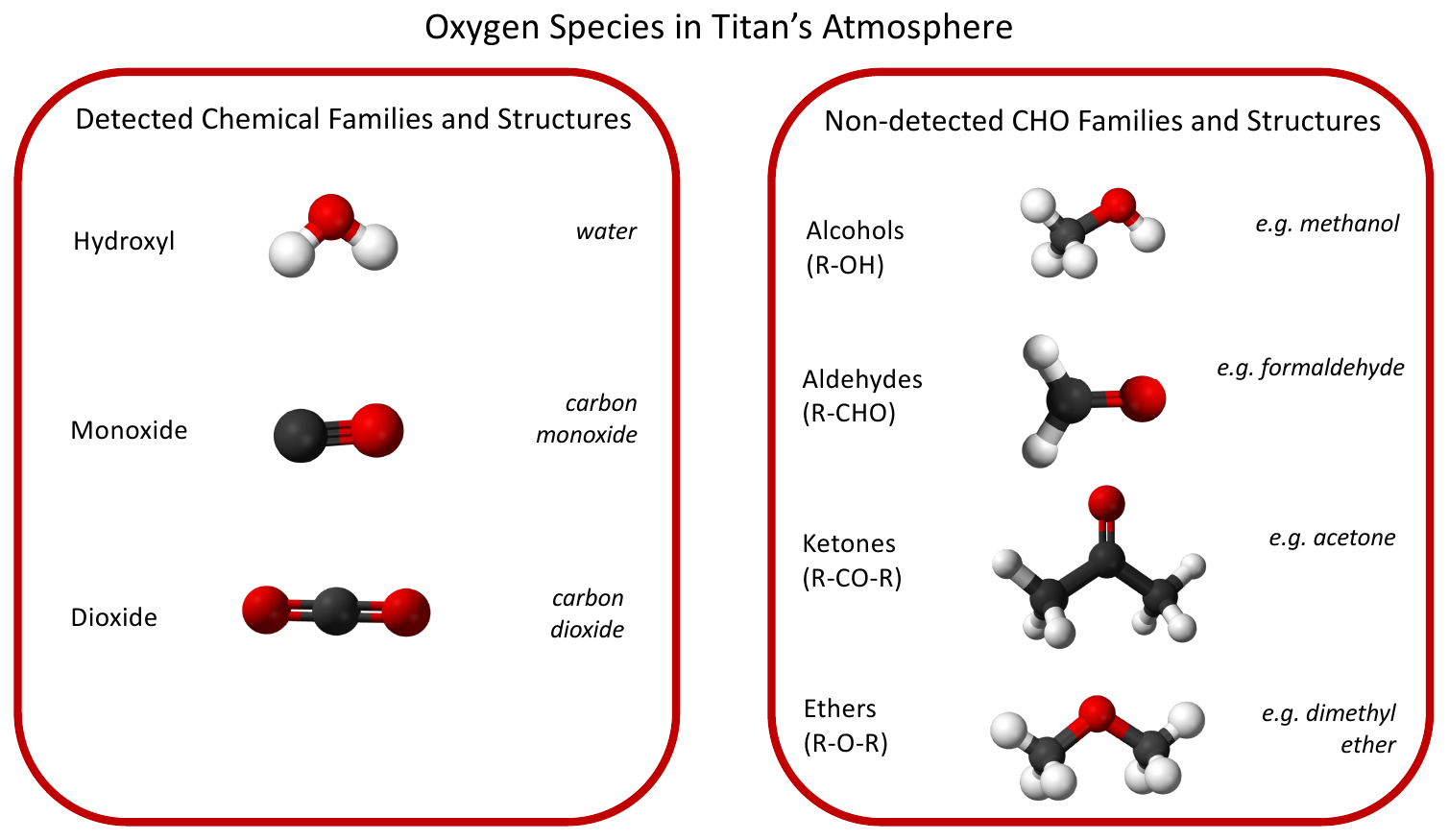}
  \caption{Examples of detected and non-detected oxygen molecular families in Titan's atmosphere.}
  \label{fig:missingo}
\end{figure}

{\em Future Work:} Vibrational (IR) and/or rotational (sub-mm) line lists exist for most small oxygen compounds currently undetected in Titan's atmosphere (see Fig.~\ref{fig:missingo}), making astronomical searches viable.\cite{endres16, delahaye21, gordon22}

%%%%%%%%%%%%%%%%%%%%%%%%%%%%%%%%%%%%%%%%%%%%%%%%%%%%%%%%%%%%%%%%%%%%
\subsection{Cyclic molecules}

%%%%%%%%%%%%%%%%%%%%%
\subsubsection{Single hydrocarbon rings}

Two small cyclic molecules have been definitively detected in Titan's neutral atmosphere: cyclopropenylidene (\cpld ) \cite{nixon20} and benzene (\benzene ) \cite{coustenis03}. Other small cyclic molecules likely to exist include the saturated cycloalkanes - cyclopropane ($\rm c{\text -}C_3H_6$), cyclobutane ($\rm c{\text -}C_4H_8$) and larger - and possibly cycloalkenes such as cyclopropene ($\rm c{\text -}C_3H_4$),  cyclobutene ($\rm c{\text -}C_4H_6$), cyclobutadiene ($\rm c{\text -}C_4H_4$), and others. Substituted rings are also possible: cyanocyclopropane ($\rm c{\text -}C_3H_5CN$).

In the ISM, several single ring molecules have been detected. These include the small 3-carbon rings cyclopropenylidene (\cpld\  \cite{thaddeus85}) and its related radical $\rm c{\text -}C_3H$ \cite{yamamoto87}, and a substituted species, ethynyl cyclopropenylidene ($\rm c{\text -}C_3HCCH$) \cite{cernicharo21a}. The five-sided ring cyclopentadiene ($\rm c{\text -}C_5H_6$), and the six-sided rings benzyne ($\rm o{\text -}C_4H_6$) and benzene (\benzene ) have also been detected \cite{cernicharo01, cernicharo21a, cernicharo21b}.

\begin{figure}
\includegraphics[scale=0.55]{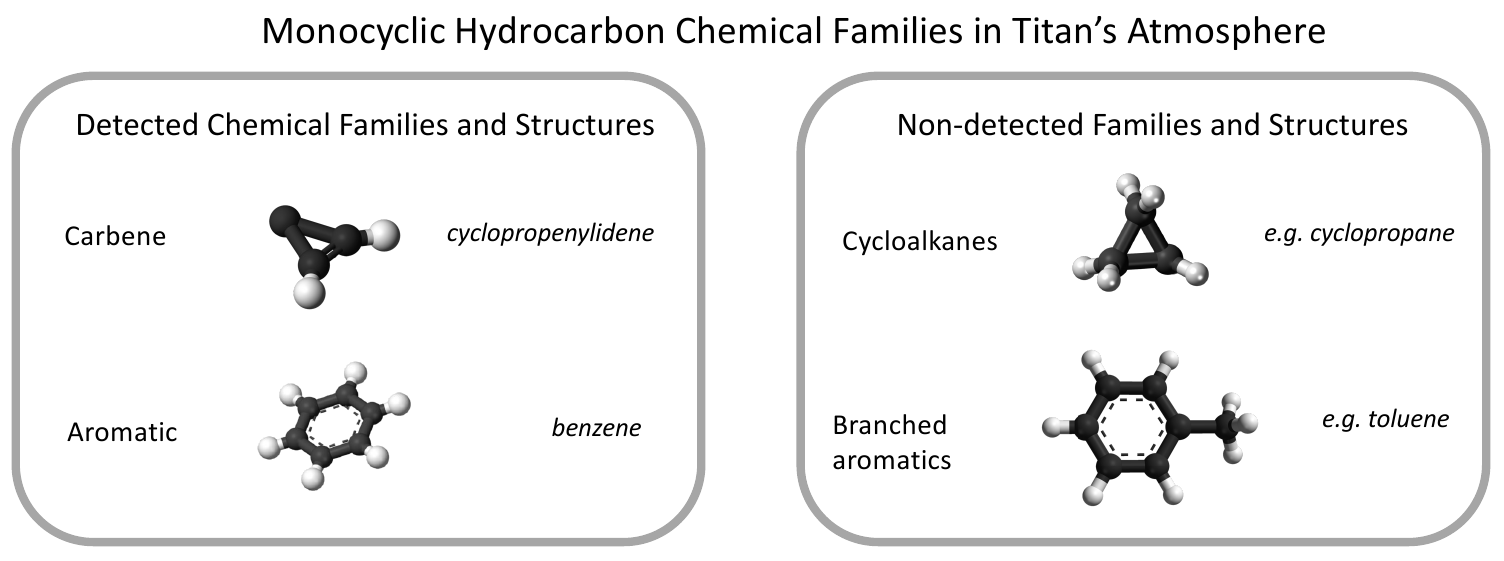}
  \caption{Examples of detected and non-detected monocyclic hydrocarbons in Titan's atmosphere.}
  \label{fig:missingmonos}
\end{figure}

{\em Future work:} Work is needed on all fronts to advance our understanding of the formation pathways, stability and prevalence of small cyclic rings in Titan's atmosphere (see Fig.~\ref{fig:missingmonos}). These include astronomical observations, laboratory work and photochemical modeling. In the 2030s, we anticipate direct {\em in situ} measurement of such molecules by the {\em Dragonfly} probe DraMS ({\em Dragonfly} Mass Spectrometer) instrument, which, unlike {\em Cassini} INMS will have ability to definitively identify molecular structure through a combination of tandem mass spectrometry (MS/MS) and Gas Chromatograph Mass Spectrometry (GCMS) \cite{trainer21}. 

%%%%%%%%%%%%%%%%%%%%%
\subsubsection{Multi-ring hydrocarbons: PAHs and polyphenyls}
\label{sect:pahs}

\begin{figure}
\includegraphics[scale=0.55]{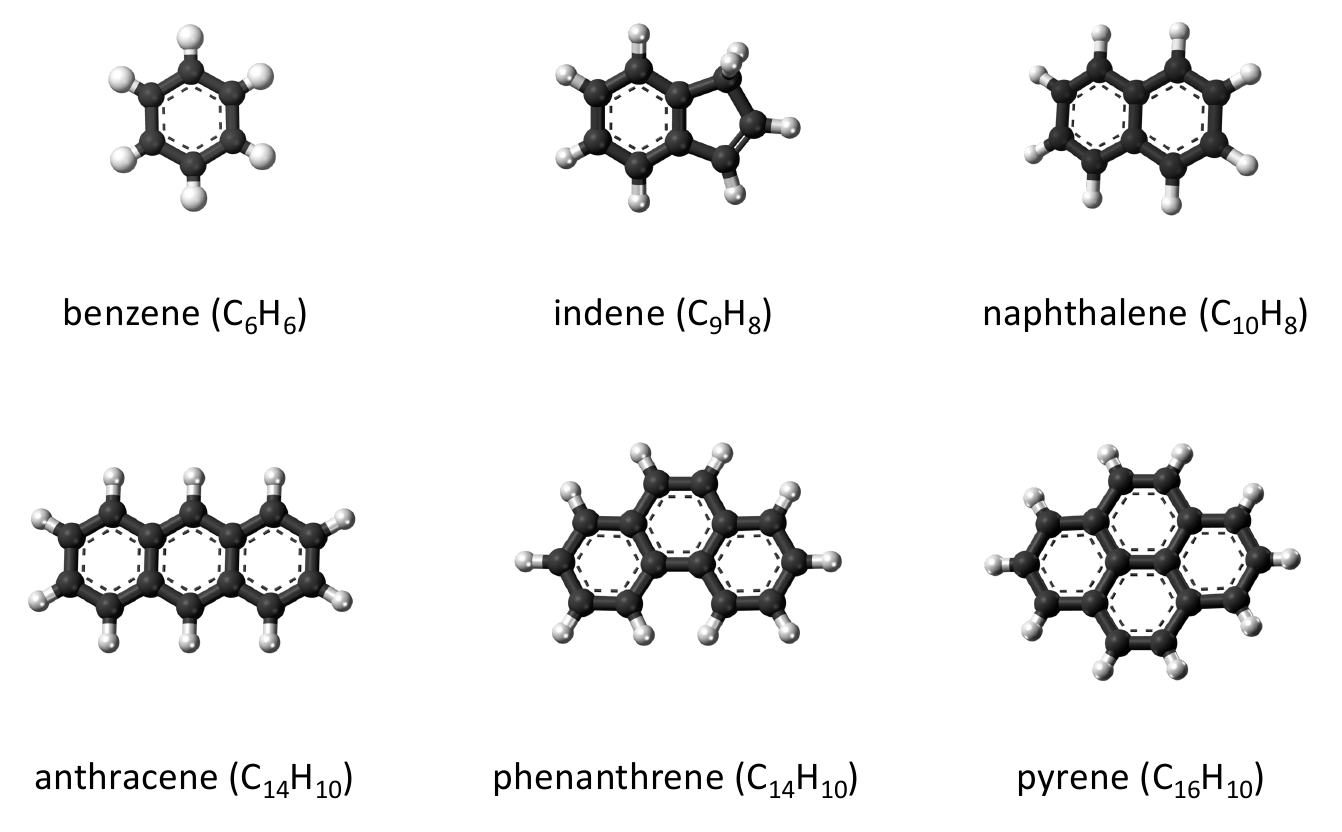}
  \caption{Benzene and polycyclic aromatic hydrocarbon molecules (PAHs). Image credit: individual PAH graphics from wikimedia commons. }
  \label{fig:pahs}
\end{figure}

Polycyclic aromatic hydrocarbons (PAHs) are multi-ring molecules composed of carbon and hydrogen that exhibit aromatic character, that is to say they have delocalized $\pi$ electron bonds. Example include naphthalene (two 6-membered rings), indene (one five-sided and one six-sided ring), anthracene and phenanthrene (three 6-membered rings), and larger examples (see Fig.~\ref{fig:pahs}). Benzene is not considered to be a PAH, since it has only one ring.

PAHs have long been suspected to exist in interstellar space\cite{sagan72}, and have been implicated as culprits responsible for the so-called `diffuse interstellar bands' (DIBs)\cite{donn68} (see review by \cite{herbig95}). To date, only a single non-functionalized PAH has been uniquely identified in space (indene - $\rm C_9H_8$ \cite{cernicharo21a}), although a greater number of CN-substituted single and double rings (cyano-PAHs) have been identified, assisted by their strong rotational lines due to the cyanide group - see review by \citeauthor{mccarthy21a}\cite{mccarthy21a}. 

Near-IR emission at 3.28 \micron\ has also been seen in Titan's dayside spectrum by {\em Cassini VIMS} \cite{dinelli13}. In a model by \citeauthor{lopezpuertas13}\cite{lopezpuertas13} this emission was attributed to a combination of PAHs with 9-96 carbons (up 11 fused rings), using laboratory cross-section as measured in the Ames PAH database \cite{bauschlicher10}, however unique identification of individual PAHs was not possible. 

Laboratory experimental work on haze formation by UV photolysis has shown that the inclusion of benzene in initial reagent mixtures along with \nitrogen\ and \methane\ leads to the formation of significantly larger molecules than when a \nitrogen /\methane\ mixture is used \cite{trainer13}.  More recently, naphthalene has also be used as a starting reagent in lab tholin experiments \cite{gautier17} showing similar results. It should be noted that, on Titan, both benzene and naphthalene would presumably first have to form from methane, so their use in lab experiments may be considered an acceleration of a natural process that could in principle start from a pure \nitrogen /\methane\ mixture and arrive at the same result over long time periods.

Multi-ring organic molecules are not constrained to form as `fused' ring PAHs such as naphthalene, but may instead form as polyphenyls (see Fig.~\ref{fig:rings}). It has been argued that a significant amount of carbon rings in Titan's atmosphere may be in the form of polyphenyls rather than as fused rings \cite{delitsky10}. Titan's aerosols are likely to be a mixture of fused and unfused rings, forming monomers and then fractal aggregates \cite{tomasko08a}.

{\em Future work:}  Future laboratory, and eventual in situ experimental work is required to determine the relative importance of fused vs non-fused rings.

%%%%%%%%%%%%%%%%%%%%%
\subsubsection{Fullerenes}
\label{sect:fullerenes}

\begin{figure}
\includegraphics[scale=0.55]{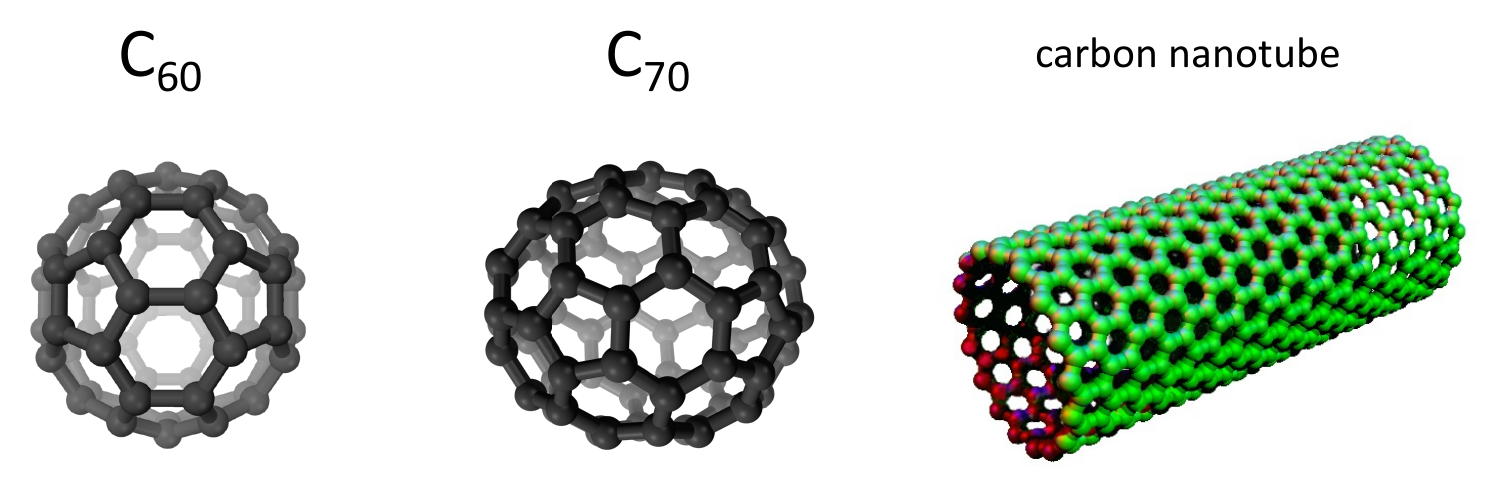}
  \caption{Fullerenes: $\rm C_{60}, C_{70}$ and carbon nanotube. Image credit: individual graphics from wikimedia commons. }
  \label{fig:fullerenes}
\end{figure}

Fullerenes are carbon allotropes formed of rings of 5 to 7 atoms in closed or partially closed mesh structures. These can include `buckyballs' (or buckminsterfullerenes), such as the spherical ($\rm C_{60}$) and ellipsoidal ($\rm C_{70}$) molecules \cite{kroto85, kroto91}, but also as sheets (graphene) and cylinders (carbon nanotubes) (see Fig.~\ref{fig:fullerenes}). Despite their large size and apparent complexity, buckyballs have been detected in space both as neutrals \cite{cami10} and ions \cite{cordiner19b}, and in meteorites \cite{becker99}. Fullerenes have been hypothesized to exist in Titan's atmosphere \cite{sittler20} although a recent attempt to detect them in {\em Spitzer} data proved unsuccessful.\cite{coy23}

{\em Future work:} Greater sensitivity with observatories such as JWST \cite{nixon16a} may enable more sensitive searches for fullerenes in Titan's atmosphere. In addition, little lab work has been done at present to determine what effect the presence of fullerenes could have on Titan's atmospheric chemistry, aerosol formation and surface geology.

%%%%%%%%%%%%%%%%%%%%%
\subsubsection{PANHs}
\label{sect:panhs}

\begin{figure}
\includegraphics[scale=0.55]{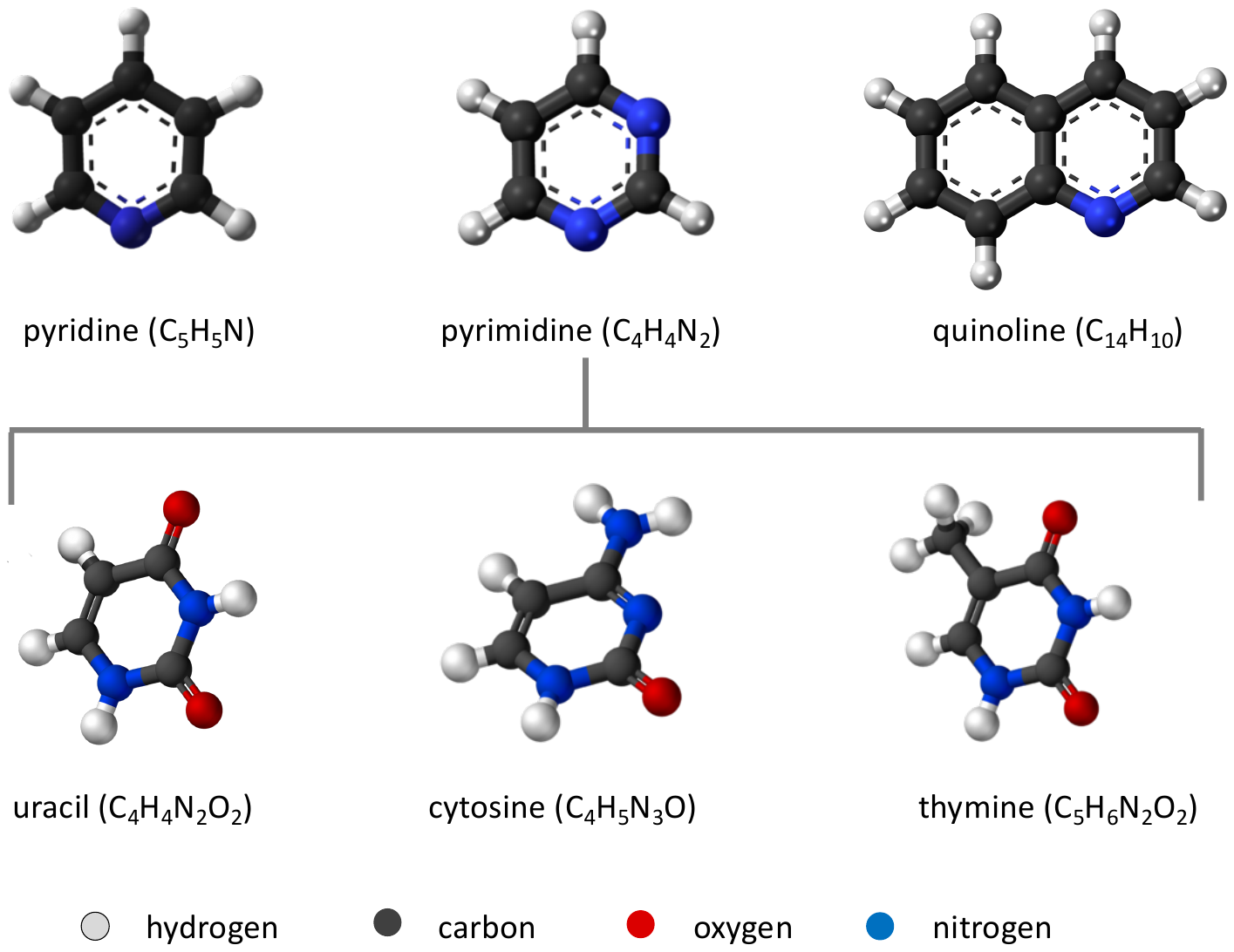}
  \caption{Nitrogen heterocycles and PANHs (polycyclic aromatic nitrogen heterocycles). Upper row: simpler N-heterocycles with one or two rings, and one or two nitrogen atoms incorporated. The importance of the search for pyrimidine is illustrated by the lower row: pyrimidine forms the backbone ring for two of the four nucleobases in DNA (cytosine, thymine) and one in RNA (uracil). Image credit: individual PAH graphics from wikimedia commons. }
  \label{fig:panhs}
\end{figure}

Nitrogen heterocycles and PANHs (polycyclic aromatic nitrogen heterocycles) are similar to PAHs, but with nitrogen incorporation into the ring structure (see Fig.~\ref{fig:panhs}, \cite{nixon20}). Nitrogen-heterocycles have been sought unsuccessfully in interstellar space, with upper limits for molecules such as pyridine and quinoline derived \cite{charnley05}. Mass spectroscopy of Titan's atmosphere with {\em Cassini} INMS has identified peaks at masses that could correspond to N-heterocycles, such as $\rm C_5H_5NH^+$ (could be protonated pyridine) at mass 80 and $\rm C_4H_4N_2H^+$ at mass 81 (possibly protonated pyrimidine) \cite{vuitton07}. However, aliphatic variants are possible making the PANH ion identification uncertain.

Recently, \citeauthor{nixon20}\cite{nixon20} made the first astronomical search for pyridine and pyrimidine in Titan's atmosphere using ALMA, deriving upper limits on their disk-averaged (global) abundances: pyridine (\pyridine ) at 1.15 ppb (2-$\sigma$) above 300 km \cite{nixon20}; and similarly pyrimidine (\pyrimidine ) at 0.85 ppb (2-$\sigma$) also above 300 km.

Laboratory work has examined how tholin (Titan haze analog) formation in UV photolysis experiments is affected by the inclusion of N-heterocycles such as pyridine and quinoline \cite{sebree14, gautier14}. Changes in the spectrum show similarities to features in Titan's haze spectrum \cite{sebree14}, and the structures formed show a mixture of polymeric and random co-polymeric structure \cite{gautier17}.

{\em Future work:} 
In future, more sensitive astronomical searches may be undertaken, for example using ALMA and JWST. Photochemical models at present do not include detailed description of N-heterocycle formation and growth, which must be included in future generations to adequately model processes leading to haze formation.

%%%%%%%%%%%%%%%%%%%%%%%%%%%%%%%%%%%%%%%%%%%%%%%%%%%%%%%%%%%%%%%%%%%%
\subsection{Sulfur and Phosphorous Chemistry}

To date, no compounds of sulfur or phosphorus have been found in Titan's atmosphere. \citeauthor{nixon13b}\cite{nixon13b} made the first quantitative assessment of upper limits for the most simple reduced species - \phosphine\ and \hsulfide\ - of 1 ppb and 300 ppb respectively in the stratosphere ($\sim$250~km). 
Phosphine gas has even been mooted as a biosignature gas in planetary atmospheres \cite{sousa-silva20}, although this conclusion has been debated.\cite{cockell21}
A sensitive search for CS was made by \citeauthor{teanby18}\cite{teanby18} with ALMA, yielding an upper limit of either 7.4 ppt (uniform profile above 100~km) or 25.6 ppt (uniform profile above 200~km).

The first (and only so far) photochemical model to include sulfur chemistry was published by \citeauthor{hickson14}\cite{hickson14}, which predicts that CS and H$_2$CS should be the most abundant sulfur-bearing species in the upper atmosphere, transitioning to C$_3$S, H$_2$S and CH$_3$SH in the lower atmosphere. However in the absence of constraints, predictions remain highly uncertain.

In principle the O/S ratio should allow further constraint of the source of Titan's oxygen flux, since the O/S ratio is predicted to be some 1000$\times$ less for a cometary source (O/S$\sim$100)\cite{crovisier09} than an Enceladus source (O/S$\sim 10^5$)\cite{waite09}. Detection of sulfur in Titan's atmosphere may also provide evidence for cryovolcanic activity \cite{fortes07}.

Both phosphorus and sulfur are among the six most essential elements for biochemistry on Earth (the so-called CHNOPS elements). With four of the six already detected on Titan, it is therefore of considerable interest to seek the remaining two, to further assess Titan's potential for astrobiology. Recent reports that all six CHNOPS elements have now been detected in the plume material of Enceladus \cite{waite09, postberg23}, make it feasible that some trace amounts of P and S arrive at the top of Titan's atmosphere, as is the case apparently with O\cite{hartle06}.

{\em Future work:} Future work is required on both the direct detection of P and S-containing substances by both astronomical and in-situ techniques - but also in laboratory work, computer photochemical modeling, and clarification of reaction pathways and rates, especially at low temperatures.

%%%%%%%%%%%%%%%%%%%%%%%%%%%%%%%%%%%%%%%%%%%%%%%%%%%%%%%%%%%%%%%%%%%%%
\subsection{Radicals}

For completeness, we include consideration of radical species - atoms and molecules with unpaired free electrons - even though these are highly reactive and unlikely to be found in significant numbers, or significantly deep in Titan's neutral atmosphere. Radicals include fragments of \methane , such as CH (methylidene), CH$_2$ (methylene), and CH$_3$ (methyl), as well as ground state and excited nitrogen atoms formed from break up of \nitrogen , such as as N($^2$D) and N($^4$S).

Many radicals (CN, OH etc) have been observed in the ISM \cite{mcguire18b}, including at least one cyclic radical ($\rm c{\text -}C_3H$)\cite{yamamoto87}; also in comets \cite{mumma11} and tenuous satellite exospheres \cite{smyth06}. However, to date  only the methyl radical has been detected by astronomical techniques in a planetary atmosphere \cite{bezard98, kunde04}.

{\em Future Work:} ALMA observations are particularly sensitive to Titan's upper atmosphere, and can sense HCN at altitudes of up to 1200~km \cite{lellouch19,cordiner20}. Future investigations with ALMA may allow the detection of radicals with dipoles, and more sensitive infrared observatories such as JWST may prove effective at detecting radicals such as methyl in the infrared \cite{nixon16a}.

%%%%%%%%%%%%%%%%%%%%%%%%%%%%%%%%%%%%%%%%%%%%%%%%%%%%%%%%%%%%%%%%%%%%%
\section{Conclusions}
%%%%%%%%%%%%%%%%%%%%%%%%%%%%%%%%%%%%%%%%%%%%%%%%%%%%%%%%%%%%%%%%%%%%%

The organic-rich atmosphere of Titan constitutes the most complex atmospheric chemical network known outside of Earth. This provides a unique natural laboratory for understanding the synthesis of organic compounds,  processes that may have been important early in the history of the Solar System \cite{ehrenfreund00, sandford20}, and may have seeded the origins of life on Earth \cite{chyba90, chyba92}. 

Therefore it is of substantial scientific importance to better understand these processes and chemical results. This field of enquiry brings together astronomers, laboratory chemists, theoretical chemists and atmospheric modelers whose combined approaches are needed to unravel the entire picture. Substantial progress has been made, especially stimulated by the recent wealth of data from the {\em Cassini-Huygens} mission \cite{lebreton02, matson02}, and the selection of the {\em Dragonfly} mission\cite{barnes21} that will arrive in the 2030s. 

{\em Astronomy:} Titan is a distant object and gathering robust information is difficult in the absence of spacecraft: astronomy is currently the only means to gather data about Titan directly. Currently active ground and space-based observatories such ALMA, IRTF and JWST are providing continuity of data collection since the end of {\em Cassini-Huygens} regarding seasonal changes in Titan's atmosphere,\cite{cordiner20, thelen22} and new information on the chemistry, composition and isotopic ratios.\cite{lombardo19a, lombardo19b, lombardo19c, lai17, palmer17, thelen20, nixon20, molter16, serigano16}.

{\em Laboratory Studies:} Whilst data collection through astronomy and remote sensing continues, a robust campaign of laboratory experiments and theoretical work is continuing in parallel to understand the origins and chemical evolution of the atmosphere, and interaction with the surface and subsurface. These diverse enquiries include 
spectroscopy of gases,\cite{jolly10, sung13, vanderauwera14, jolly15, sung18, hewett19, sung20, hewett20, bernath21, sorensen22, bernath23}
measurement of reaction rates,\cite{gu09a, gu09b, morales10, balucani10, balucani12, dutuit13, fleury14, mancini21, kaiser21b, vanuzzo22}
experimental work on co-crystals,\cite{cable14, cable18, cable19, cable20, cable21a, czaplinski23}
ices,\cite{moore10, hudson14a, hudson14b, anderson18, abplanalp19, nna19, materese20, hudson20, hudson22, hudson22b, gerakines22, hudson23}
 and hazes (tholins).\cite{cable12, imanaka12, gautier12, carrasco12, sciamma14, sciamma17, he18, yu20, dubois20, nuevo22, he22, li22}

{\em Modeling:} The chemistry of Titan's atmosphere cannot proceed without disturbing the medium in which it takes place: the minor gases and haze generated have a substantial effect on the heating and cooling of the atmosphere \cite{bezard18}, and in turn changes to the thermal structure of the atmosphere leads to dynamical motions including vertical eddy mixing, and meridional transport. As gases are transported, they enter atmospheric regions that may have greater or lesser photon flux from the Sun, have fewer or greater opportunities to interact with electrons, and encounter differing densities of radicals and other reagents. Therefore, decoupling chemistry from dynamics is not possible. Combining the current generation of Global Circulation Models (GCMs) \cite{lora15, faulk17, lora19, newman11, newman16, lombardo23} and photochemical models \cite{lic15, willacy16, loison19, vuitton19, dobrijevic21} to create 2D and 3D coupled chemical-GCMs is a challenging but important task that must occupy the next generation of modelers.

{\em Future Missions:} The {\em Dragonfly} mission estimated to land on Titan $\sim$2034 will provide a wealth of new data about Titan's surface and atmospheric boundary layer. However {\em Dragonfly} will only investigate the low latitudes, including dunes and the crater Selk. In the future an orbiter, as envisaged by several published studies \cite{coustenis09, nixon16b, rodriguez22}, would provide the important benefit of a truly global picture, including potentially complete global mapping of the atmosphere and surface at uniform resolution, with time-domain information to search for changes occurring. Other elements such as a balloon, airplane and/or floating lake probe could provide valuable in situ information about other environs \cite{lorenz08c, barnes11c, stofan13}.

There is no doubt that Titan still offers many challenges to our understanding that will provide fertile areas of study for future generations of scientists \cite{nixon18, mackenzie21}, and offer rewards through important insights into the chemical evolution of the Solar System and origins of life in the universe.


\begin{acknowledgement}

Funding for this work was through NASA's Astrobiology program. The author wishes to thank the Astrobiology Program manager Dr Mary Voytek (NASA HQ) and the Principal Investigator of the CAN-8 Project "Habitability of Hydrocarbon Worlds: Titan and Beyond" Dr Rosaly Lopes (JPL/Caltech) for their support of this work. Thanks is also due to Nicholas Lombardo and Alexander Thelen for providing text files of retrieved gas profiles from prior publications. The author sincerely thanks two anonymous reviewers and the guest editors of the ACS Earth and Space Chemistry special edition on astrochemistry, Martin Cordiner and Christopher Bennett for their thoughtful comments and feedback that helped to improve the manuscript; and to the Journal Editor Eric Herbst, and staff for their assistance in the review and publication process.  Last but not least, the author is very grateful to a cadre of early-career students and postdocs who proof-read parts of the final submitted version of the manuscript: Brandon Coy (UCLA), Nicholas Kutsop (Cornell University), Paige Leeseberg (University of Iowa/SURA), Siobhan Light (University of Maryland/SURA), Nicholas Lombardo (Yale University), Edward Molter (University of California Berkeley), Jonathon Nosowitz (Catholic University), Alexander Thelen (Caltech). Any remaining errors or inaccuracies are the responsibility of the author.

\end{acknowledgement}

%%%%%%%%%%%%%%%%%%%%%%%%%%%%%%%%%%%%%%%%%%%%%%%%%%%%%%%%%%%%%%%%%%%%%

\clearpage
\begin{appendix}

\section{Acronyms and Abbreviations}

\begin{table}
\begin{tabular}{ll}
ALMA & Atacama Large Millimeter/submillimeter Array \\
CAPS & {\em Cassini} Plasma Spectrometer \\
CDA & Cosmic Dust Analyzer ({\em Cassini} instrument)  \\
CIRS & Composite Infrared Spectrometer ({\em Cassini} instrument)  \\
DR & Dissociative Recombination \\
DSMC & Direct Simulation Monta Carlo \\
ESA & European Space Agency \\
GCM & Global Circulation Model \\
GCMS & Gas Chromatograph / Mass Spectrometer ({\em Huygens} instrument) \\
HASI & Huygens Atmospheric Structure Instrument \\
INMS & Ion and Neutral Mass Spectrometer ({\em Cassini} instrument) \\
IRAM & Institut de Radioastronomie Millimetrique \\
IRIS & Infrared Interferometer Spectrometer ({\em Voyager} instrument) \\
IRTF & {\em Infrared Telescope Facility} \\
ISM & InterStellar Medium \\
ISO & {\em Infrared Space Observatory} \\
IUPAC & International Union of Pure and Applied Chemists \\
JWST & {\em James Webb Space Telescope} \\
LTE & Local Thermodynamic Equilibrium \\
NASA & National Aeronautics and Space Administration \\	
PAH & Poly-Aromatic Hydrocarbon \\
PANH & Poly-Aromatic Nitrogen Heterocycle \\
RPWS & Radio and Plasma Wave Spectrometer ({\em Cassini} instrument) \\
RSS & Radio Science Subsystem ({\em Cassini|} instrument) \\
SNR & Signal to Noise Ratio \\
TEXES & Texas Echelon Cross Echelle Spectrograph \\
TMC & Taurus Molecular Cloud \\
TST & Transition State Theory \\
UVIS & Ultraviolet Imaging Spectrometer ({\em Cassini} instrument) \\
VIMS & Visible and Infrared Mapping Spectrometer ({\em Cassini} instrument) \\
VMR & Volume Mixing Ratio 
\end{tabular}
\end{table}

\end{appendix}


\clearpage
\providecommand{\latin}[1]{#1}
\makeatletter
\providecommand{\doi}
  {\begingroup\let\do\@makeother\dospecials
  \catcode`\{=1 \catcode`\}=2 \doi@aux}
\providecommand{\doi@aux}[1]{\endgroup\texttt{#1}}
\makeatother
\providecommand*\mcitethebibliography{\thebibliography}
\csname @ifundefined\endcsname{endmcitethebibliography}
  {\let\endmcitethebibliography\endthebibliography}{}

\clearpage
\Large{*** For TOC Only ***}

\begin{figure}
\includegraphics[scale=0.40]{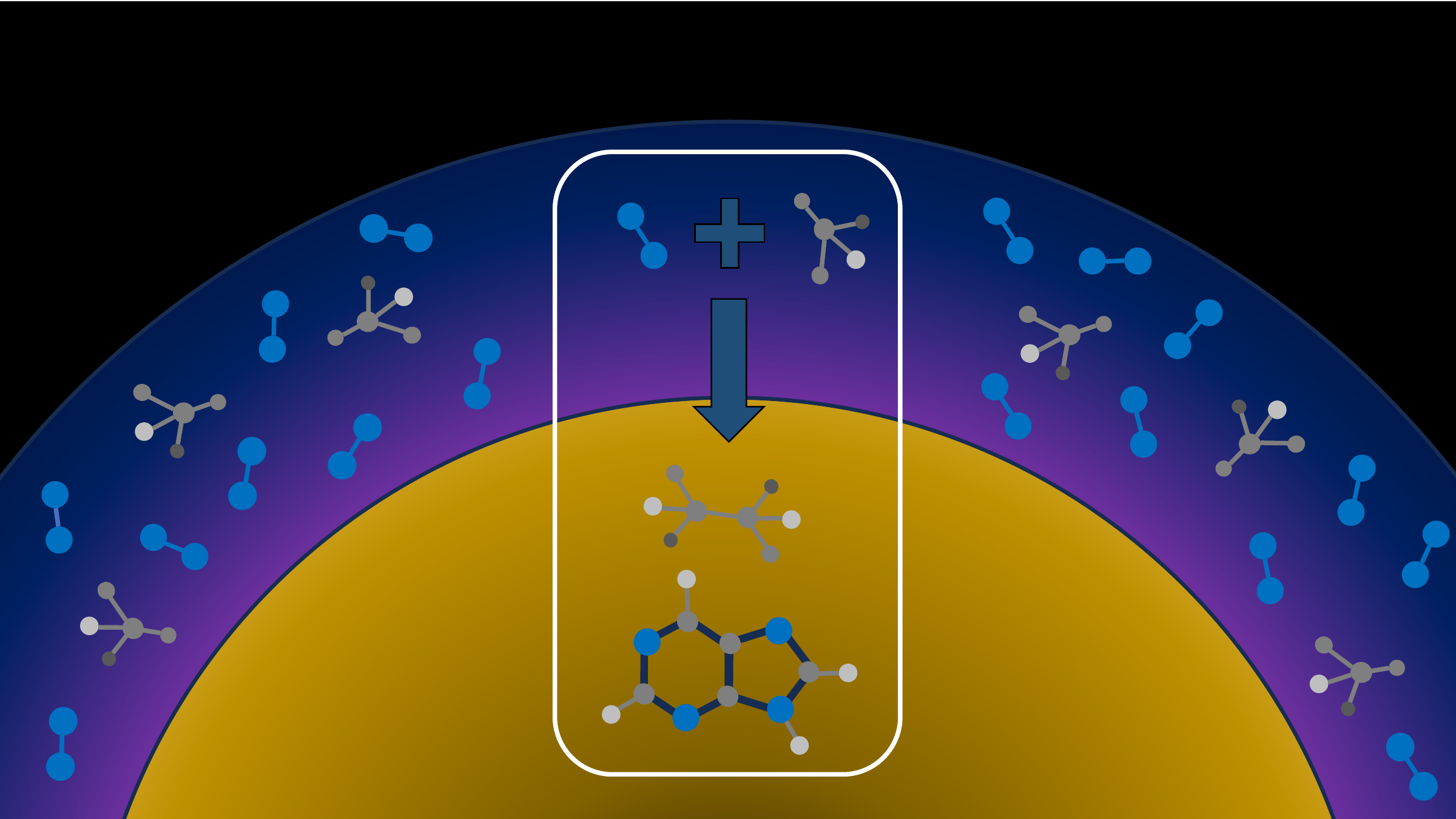}
  \caption{Figure for contents page only. }
  \label{fig:contents-fig}
\end{figure}

\end{document}